\documentclass[10pt, preprint]{aastex}
\usepackage{epsfig}

\shorttitle{Stellar populations of $z~1$ AGN with HST and LGSAO}
\shortauthors{Ammons et al.}

\begin{document}

\title{Spatially Resolved Stellar Populations of Eight GOODS-South AGN at $z \sim 1$} 

\author{S. Mark Ammons\altaffilmark{1}, Jason Melbourne\altaffilmark{2}, Claire E. Max\altaffilmark{1}, David C. Koo\altaffilmark{1}, David J. V. Rosario\altaffilmark{1}}

\altaffiltext{1}{present address:  Astronomy Department, University of California, Santa Cruz, 1156 High St., Santa Cruz, CA  95064, ammons@ucolick.org, koo@ucolick.org, max@ucolick.org}
\altaffiltext{2}{present address:  California Institute of Technology, MS 320-47, Pasadena CA 91125, jmel@ucolick.org}

\begin{abstract}
We present a pilot study of the stellar populations of 8 AGN hosts at $z\sim1$ and compare to (1) lower
redshift samples and (2) a sample of nonactive galaxies of similar redshift.  We utilize K$'$ images
in the GOODS South field obtained with the laser guide star adaptive optics (LGSAO) system at Keck
Observatory.  We combine this K$'$ data with B, V, i, and z imaging from the Advanced Camera for Surveys
(ACS) on HST to give multi-color photometry at a matched spatial resolution better than 100 mas in all bands.
The hosts harbor active galactic nuclei (AGN) as inferred from their high X-ray luminosities
($L_X > 10^{42}$ ergs s$^{-1}$) or mid-IR colors.  We find a correlation between the presence of younger
stellar populations and the strength of the AGN, as measured with [OIII] line luminosity or X-ray (2-10
keV) luminosity.  This finding is consistent with similar studies at lower redshift.  Of the three Type II galaxies, two are disk galaxies and one is of irregular type, while in the Type I sample there only one disk-like source and four sources with smooth, elliptical/spheroidal morphologies.  In addition, the mid-IR SEDs of the
strong Type II AGN indicate that they are excited to LIRG (Luminous InfraRed Galaxy) status via galactic starbursting, while the
strong Type I AGN are excited to LIRG status via hot dust surrounding the central AGN.  This supports
the notion that the obscured nature of Type II AGN at $z\sim1$ is connected with global starbursting
and that they may be extincted by kpc-scale dusty features that are byproducts of this starbursting.

\end{abstract}


\section{INTRODUCTION}
It is observed that the evolution of active galactic nuclei (AGN) and quasi-stellar objects (QSOs) is connected with the properties of the bulges in their host galaxies \citep{mag98, fer00, geb00} at low redshifts.  These correlations between black hole mass and various bulge properties (e.g., velocity dispersion, half-light radius, total mass) hint at simultaneous evolution of the AGN and its host galaxy.  While this connection between activity at sub-parsec and kiloparsec scales is observable in snapshots of an AGN's life history, a complete understanding of its evolution through time remains elusive.  It is also unclear how the conditions and morphology of the galaxy at large may be related to the canonical dust torus on sub-parsec scales \citep{ant93}.

Several scenarios explaining these observations have arisen in recent years from simulations of galaxy formation.  In the major merger scenario, two colliding gas-rich galaxies trigger central, dusty star formation and black hole fueling, followed by an unobscured quasar (QSO) phase resulting as AGN feedback expels the gas and dust \citep[e.g.,][]{san88, hop05, spr05}.  This quenching of star formation may be a mechanism by which AGN activity controls the evolution of galaxies from the quiescent blue cloud to the red sequence \citep{hop06, cro06}, serving as a component of the ``mixed'' evolution scenario of \citet{fab07}.  An alternate picture is that the triggering of AGN activity simply depends on the conditions at the time.  Both the condition of existence of a supermassive black hole (SMBH), met in most if not all ellipticals, and the condition of presence of a large reservoir of cold gas, are required for AGN activity to occur \citep{kau03, kau07, sil08}.

Large, carefully controlled studies of the morphological properties, emission properties, and stellar populations of AGN hosts at low and high redshift have shed light on these theories.  Several studies have found that luminous quasars and radio galaxies inhabit massive ellipticals formed at or before $z\geq2.5$ \citep{mcl99, nol01}.  Contrastingly, other studies have found that intermediate-luminosity AGN ($10^{42} < L_{X, 0.5-10 keV} < 10^{44}$ ergs s$^{-1}$) at low redshift tend to reside in a unique class of objects that possess elliptical morphologies, but additionally display the signatures of young stellar populations (YSPs), which places them in the green valley region between the blue cloud and red sequence.  \citet{kau03} find that narrow-line AGN (Type II, $0.02 < z < 0.3$) of all luminosities reside almost exclusively in massive galaxies and have distributions of sizes, stellar surface mass densities and concentrations that are similar to those of ordinary early-type galaxies.  In this Sloan Digital Sky Survey (SDSS) study, the hosts of high-luminosity AGN have much younger mean stellar ages than those of the low-luminosity AGN, which are similar in stellar makeup to old ellipticals.  In addition, the young stars are not preferentially located near the nucleus of the galaxy, but are spread out over scales of at least several kiloparsecs.  The interpretation put forth is that these ``young bulges'' simply possess the necessary conditions for AGN activity - a SMBH and a reservoir of cold gas.  Many other studies have uncovered evidence of young stellar populations in local Type II AGN \citep{sch99, cid01, jog01, gon01, sto01}.  The tendency of AGN to reside in hosts with elliptical-like morphologies has been seen in previous studies \citep{sto01, ho03}.  Additional information on the spatial distribution of YSPs in local AGN hosts is presented in \citet{kau07}.  Hosts with young bulges and strongly accreting black holes often display outer disks ($r > 10$ kpc) with blue near-UV - r colors, suggesting that outer star formation is connected with AGN growth, according to near-ultraviolet (UV) fluxes from the Galaxy Evolution Explorer (GALEX).  The authors hypothesize again that these AGNs possess reservoirs of cold gas, the raw material for star formation.  Galaxy interactions \citep{gun79, her89,her95} or bars \citep{shl92} must then be invoked to funnel this gas into the nucleus and thence onto the black hole.

Although the quantity of work on these issues at low redshift is substantial, AGN evolution theories additionally make predictions about the properties of their progenitors at higher redshift ($z \sim 1-2$).  In particular, it was once thought that the major merger - AGN fueling scenario predicted that mergers and close companions would be more common in AGN hosts, but this has not been observed \citep{pie07, gro05}.   Direct observations of AGN feedback have been difficult, as the timescale for this process is short \citep[$< 1$ Gyr,][]{sil08}, cutting down on number statistics.  Several studies have focused on the morphological properties of high redshift AGN hosts.  For example, \citet{has07} finds that the hosts of intermediate luminosity AGN ($10^{42} < L_X < 10^{43.5}$ ergs s$^{-1}$) at $z\sim1$ are more likely to be massive ellipticals or bulge-dominated disks, just as at low-redshift.  \citet{sil08} analyzes the rest-frame colors of the hosts of intermediate luminosity AGN at $z \sim 1$ and arrives at a similar conclusion.   

Studies of the stellar populations of AGN hosts at high redshift can impart independent information about their individual histories than can be inferred from their structural or morphological properties alone.  For example, the presence of YSPs in high-redshift AGN hosts would signal that the processes controlling AGN fueling were not vastly different from today.  
A comparison of the ages of dominant stellar populations for a sample of AGN against a sample of nonactive systems could indicate a delay time between AGN activity and starbursting.  

In this paper we search $z \sim 1$ AGN hosts for young stellar populations, using adaptive optics imaging together with data from the Hubble Space Telescope Advanced Camera for Surveys (HST ACS) to enable stellar population modeling via 5-band Spectral Energy Distribution (SED) fitting.  We investigate AGN in the Great Observatories Origins Deep Survey South (GOODS-South) field, taking advantage of the large number of AGN revealed by the 1 Megasecond Chandra exposure in this field \citep{gia01}.  We use the Laser Guide Star Adaptive Optics (LGSAO) system on the Keck Telescope \citep{wiz06} with the Near Infrared Camera-2 (NIRC2) at K$'$-band and ACS observations at shorter wavelengths.   This wavelength baseline of rest-frame UV to rest-frame near-IR is optimal for constraining stellar age independent of dust extinction \citep{gil02}.  We approach these multicolor data with several goals:  (1)  Understand the comparison of the stellar populations and color gradients of high-redshift AGN hosts with those of a non-active population of similar morphologies and mean redshift.  Interpret this comparison in light of similar analyses at low redshift that find correlations between AGN strength and the presence of YSPs.  (2)  Characterize the morphologies of the hosts with diffraction-limited near-IR imaging and determine if the morphological classification is consistent with the stellar population distribution.  (3)  Compare the stellar populations of Type I and Type II AGN to shed light on the nature of AGN obscuration at high-redshift.  

To achieve these goals, we have chosen observational modes that deliver excellent spatial resolution ($0\farcs1 \sim 600-800$ pc) to separate bulge from disk and to distinguish nuclear star formation from the central AGN.  Particularly for Type I AGN hosts, good resolution is necessary to subtract the point-like non-stellar continuum of the AGN itself, revealing the underlying host.  At these redshifts and magnitudes ($19 < R < 23.5$), stellar age determination via photometric SED fitting is preferable to spectroscopic analysis from the point of view of observing efficiency; AO resolution is necessary to study the bulge separately from the central point source and from the galactic disk, if either are present.

We describe the data and analyze selection biases in the AGN sample in \S 2.  In \S 3, we utilize multiple methods of AGN/starburst discrimination to demonstrate that the sample galaxies are indeed AGN.   \S 4 explains how annulus photometry and point source subtraction is performed on all imaging data.  Section 5 includes morphological analysis via multi-component GALFIT \citep{pen02} models of each band, with arguments for the presence or non-presence of a central point source in each band.  In \S 6, we present B-i and V-K$'$ color maps and multi-color surface brightness gradients for all AGN galaxies.  Section 7 gives detailed stellar populations analysis of the color maps for the AGN sample.  We make conclusions in \S 8 and summarize in \S 9.  We outline the method used to estimate the Keck Laser Guide Star AO (LGSAO) point spread function (PSF) in Appendix A.  We give morphological and stellar populations analyses for a comparison sample of normal (non-active) galaxies in Appendix B and C.  

We assume a flat cosmology throughout, with $H_0 = 70$ km/s/Mpc, $\Omega_M = 0.3$, and $\Omega_{\Lambda} = 0.7$.

\section{MULTIBAND DATA}

\subsection{GOODS-South Optical, X-ray, and Ground-Based IR Data}

We investigate AGN in the Great Observatories Origins Deep Survey (GOODS) field South, or Chandra Deep Field South (CDF-S), an ideal field in the optical/IR both for depth and wavelength coverage.    The wide wavelength baseline in this field enables the use of SED-fitting to estimate fundamental properties of the underlying stellar populations of galaxies.  The deep, 1 Megasecond Chandra exposure in this region has revealed hundreds of AGNs at $0.5 < z < 1.5$ \citep{fan01, fan03, bar03}.  HST ACS images of the GOODS fields \citep{gia04} in the F435W (B), F606W (V), F775W (\textit{i}), and F850LP (\textit{z}) filters are the most sensitive optical images to date in these regions.  We obtain spectroscopic redshifts in GOODS-S/CDF-S from \citet{szo04} and photometric redshifts from COMBO-17 \citep{wol04, zhe04}.  In this paper, all source ID numbers are in the form ``XID \#'' as reported in \citet{gia02}.  

We also make use of deep Ks imaging in the CDFS using the ISAAC instrument mounted at the Antu Unit Telescope of the VLT.  All sources are detected in this Ks imaging.  The data reduction and depths are described at the GOODS/ISAAC Final Data Release (V2.0) webpage \citep{ret07}.

\subsection{K$'$ Band Adaptive Optics Imaging}

All K$'$-band adaptive optics imaging has been obtained as part of the CfAO Treasury Survey \citep[CATS,][]{mel05, koo07}, with three nights of laser and natural guide star imaging on the Keck II telescope. Table 1 lists pointings and tip/tilt star coordinates.   These pointings were specifically chosen to optimize correction of the Chandra sources in the available field of view (FOV) of the adaptive optics system. Field centers were chosen to maximize the number of X-ray sources as well as serendipitous field galaxies in the FOV; the latter of which are used as galaxies for comparison in this study.  The $KÕ$ images were obtained with the NIRC2 wide camera ($40\farcs \times 40\farcs$) with a pixel sampling of $0\farcs04$.  This provides good sampling of the Keck telescope diffraction-limited PSF at 2.2 microns, whose full width at half maximum (FWHM) was $0\farcs054$.  

Imaging and reduction strategies are detailed in \citet{mel05, mel07}, with a cursory summary here.  The tip/tilt star, varying in brightness between $R = 15$ and $R = 17$ for different fields, is included in the $40\farcs \times 40\farcs$ corrected field of view to enable good temporal sampling of the PSF.  The spatial variation of the PSF is captured with nightly observations of a globular or open cluster.  For dim PSF stars ($R < 16$), we obtain 2 exposures of 2 coadds of 30 seconds at each dither position.  The typical exposure of one hour consists of 30 dither positions marking random locations on a circle of radius  $2\farcs$  The reduction entails the creation of super-sky and flat-field frames by median-combining the object frames and masking out sources.  UKIRT IR standards were used to determine the photometric zeropoint.

\subsection{Sample Selection}

GOODS-South is an ideal field from which to choose targets for purely photometric stellar populations analysis, as deep ACS photometry is available in four bands, covering a wide wavelength baseline.  The sample of AGN in this study is X-ray selected, using the catalog of sources extracted from the 1 Ms \textit{Chandra} Deep Field South (CDFS) observation as described in \citet{gia02}.  The flux limits of this final catalog for the soft (0.5 - 2.0 keV) and the hard (2-10 keV) bands are $5.5\times10^{-17}$ ergs s$^{-1}$ cm$^{-2}$ and $4.5\times10^{-16}$ ergs s$^{-1}$ cm$^{-2}$, respectively.  Sources were extracted with a modified version of the SExtractor algorithm \citep{ber96} with a S/N detection threshold of 2.1, where single-band detection was sufficient for inclusion in the catalog.  The extraction methods and accompanying selection biases are described in \citet{gia02}.   Throughout this paper, the hardness ratio is defined as $HR = (H-S) / (H+R)$ as in \citet{gia02} and \citet{szo04}, where H and S
are the net instrument count rates in the hard (2-10 keV) and the soft (0.5-2 keV) 
bands, respectively.

The Keck LGSAO system requires a point source in the field with $R < 17$ to serve as a tip/tilt reference.  Unfortunately, few stars are available in GOODS-South to use as tip/tilt references.  Therefore, our sample of galaxies is limited by the number of available tip/tilt stars in the field.  Fields with bright tip/tilt stars were observed first to maximize science return, as the quality of adaptive optics correction suffers for dim tip/tilt reference stars.  Little selection bias is introduced by concentrating on targets with nearby bright stars.  Of the 346 sources in the CDFS X-ray catalog, $\sim30$ sources are present within the ACS fields, possess suitable tip/tilt stars nearby, and have photometric or spectroscopic redshifts measured \citep{zhe04, szo04}.  

We also impose an R band magnitude cutoff for each galaxy of 23.5.  X-ray sources with optical counterparts dimmer than this value do not possess sufficient signal to perform stellar populations analysis in the outer regions of the host ($r > 1\farcs0$) in any band.  Of the 12 AGN sources imaged thus far by the CATS survey, this magnitude cut removes three sources:  XID 629 (HR = -1.00, R = 25.26), XID 532 (HR = -0.06, R = 24.59), and XID 61 (HR = -0.45, R = 23.98).    XID 629 appears in the ACS bands as a compact, asymmetric structure on the order of 1-2 kpc in half-light radius (assuming $z > 0.5$), with a soft X-ray spectrum that would classify it as a starbursting galaxy or Type I AGN.   XID 629 does not appear in the LGSAO K$'$ image.  XID 532 appears in the ACS bands as a powerful point source surrounded by a streak of compact, asymmetric star formation and would be classified as a Type II object or starbursting galaxy from hardness ratio alone.  XID 61 is a pure optical point source, which when taken with the hardness ratio of -0.45, classifies it as a likely Type I QSO.  XID 61 is a dim source located $4\farcs$ away from a bright star, which prevented good spectroscopy in \citet{szo04}, although a photometric redshift is available.    

Apart from the R = 23.5 magnitude cut, we also remove sources that are clearly dominated by a point source.  It is challenging to isolate and analyze the hosts of these compact or point-source dominated sources.   XID 30 (HR = -0.53, R = 20.42, z = 0.837), which has been classified as a Type I broad-line AGN (BLAGN) from optical spectra and as a Type I QSO from X-ray spectra \citep{szo04}, appears as a pure point source in the ACS bands.  A detailed two-component fit to the 2D image (see \S 5.1 for more information on GALFIT) reveals that the central point source is more than 3 times brighter in \textit{i}-band than the total luminosity of any underlying host.  

Properties of the remaining sample are enumerated in \S 3.1.

\section{AGN AND COMPARISON SAMPLES}
Following the magnitude and morphology cuts detailed in \S 2.3, eight AGN remain in the sample.  Coordinates, redshifts, and broad photometric properties are given for these sources in Table 2.  

\subsection{Verifying AGN identification}

Identifying AGN in high-redshift surveys of any wavelength has been historically difficult due to their widely variant morphologies and spectral properties.  In particular, galaxies that are undergoing vigorous star formation may display X-ray or IR luminosities that mimic galaxies that harbor nuclear activity.  Multiple historical methods of distinguishing AGN from starbursts exist, including (1) total X-ray luminosity or flux ratio of x-ray to optical; (2) particular optical narrow-line ratios like $[OIII] \lambda5007 /H\beta$ and $[NII] \lambda6583/H\alpha$; (3) flux ratios in the near- and mid-IR; and (4) presence of broad-lines in the optical spectrum.  In this section, we use combinations of evidence from all four categories above to demonstrate that all the objects are truly AGN.  These methods of discrimination are described in more detail in the following paragraphs.  Shown in Table 3 are X-ray, infrared, and optical line luminosities and ratios for each of the sample galaxies.  

Sources with total X-ray luminosity $L_{X, 0.5-10 keV}$ greater than $10^{42}$ ergs s$^{-1}$ are likely to be AGN, as few local starbursting galaxies have luminosities greater than this value \citep{ros02, ale01}.  The X-ray classification shown in Table 2 is derived from a two-dimensional scheme originally presented in \citet{szo04}, in which hardness ratio purely determined AGN Type and X-ray luminosity purely determines strength.  Objects with hardness ratio less than -0.2 are Type I objects; those with hardness ratios greater than -0.2 are Type II objects.  X-ray sources are classified as ``AGN'' if $10^{42}$ ergs s$^{-1} < L_{X,0.5-10 keV} < 10^{44}$ ergs s$^{-1}$ for Type I objects and $10^{41}$ ergs s$^{-1} < L_{X,0.5-10 keV} < 10^{44}$ ergs s$^{-1}$ for Type II objects, or classified as ``QSOs'' if $L_{X,0.5-10 keV} > 10^{44}$ ergs s$^{-1}.$  Local hard X-ray surveys find very good correspondence \citep{tue08} between X-ray typing by hardness ratio or neutral hydrogen column density and traditional typing by characterizing widths of lines in the optical spectra as in \citet{ost89}.  

The X-ray-to-optical flux ratio is a good discriminator between AGN and normal galaxies.  The original formulation for local AGN in \citet{mac88} is a function of the 0.3-3.0 keV soft band flux and the V-band magnitude.  Since modified by \citet{szo04} for higher redshift objects, the X-ray-to-optical flux ratio (X/O hereafter) in Table 3 uses the R-band magnitude and the 0.5-2.0 keV Chandra soft band flux.  AGN at higher redshift tend to satisfy $-1.4 < X/O < 1.1$ and normal galaxies have $-2.2 < X/O < 0.0$.   Table 3 plots values of the X/O flux ratio for all systems, including a correction in the Type II systems for soft-band absorption by dust.  

An additional tool we utilize is the 24 micron to 8 micron flux ratio (24/8 hereafter), as discussed in \citet{bra06}.  At high redshift a histogram of 24/8 ratios for a mid-IR selected sample is bimodal, centered about 24/8 = 0 and 0.5.  Sources clustered about 0 are identified with AGN that produce mid-IR SEDs similar to power laws; those clustered about 24/8 = 0.5 are identified with starbursting galaxies that possess SEDs characteristic of warm dust.  

\subsubsection{Type II AGN}
XID 155, 56 and 266 are identified as Type II AGN by hardness ratio and X-ray luminosity.  These systems also have X/O ratios between -0.1 and -0.7, which places them in the AGN range.  XID 56 displays optical spectral signatures of AGN, including [Ne V].  [OIII] $\lambda5007$ and $H\beta$ line intensities are taken directly from \citet{net06}, with an upper limit on $H\beta$ estimated for XID266.  The [OIII] $\lambda5007 / H\beta$ line ratio is between 4 and 8 for all systems, placing them in the upper half of a BPT diagram \citep{bal81}.  Although the X-ray properties of these systems clearly classify them as AGN, their 24/8 flux ratios are all greater than 0.3, which implies that their mid-IR SED includes some contribution from a starburst.  

\subsubsection{Type I BLAGN}
XID 15 and 32 are broad-line AGN based on their optical spectra.  Their X-ray luminosities ($L_{X,0.5-10 keV} = 63\times10^{42}$ ergs s$^{-1}$ and $11.1\times10^{42}$ ergs s$^{-1}$, respectively), X/O ratios (0.2 and -0.1), and hardness ratios additionally classify them as X-ray Type I AGN.  We estimate the [OIII] $\lambda5007$ line luminosity of XID 32 with spectra from \citet{szo04}.  XID 15, however, is at too high of a redshift (z = 1.23) to measure the [OIII] $\lambda5007$ strength with an optical spectrum, so we estimate its strength via a log-linear calibration to hard x-ray luminosity presented in \citet{net06} for Type I AGN and QSOs.  It appears from Figure 2 of that work that the precision of this calibration is half a decade.

\subsubsection{Type I AGN with no optical spectral signatures}
XID 536, 594, and 83 are X-ray sources that have no optical signatures of the presence of an AGN.  XID 594 is classified as a Type I AGN by its X-ray luminosity of $2.44 \times10^{42}$ ergs s$^{-1}$, its X/O ratio of -1.1, its 24/8 ratio of -0.2, and its hardness ratio of HR = -1.0.  XID 83 is similarly classified as a Type I AGN by these same properties ($L_{X, 0.5-10 keV} = 4.4\times10^{43}$ ergs s$^{-1}$, X/O = -0.1, and 24/8 of -0.18).  XID 83's redshift is derived from a photometric analysis \citep{zhe04} with a $2\sigma$ confidence interval of 1.68 to 1.95.  
A true redshift of less than 0.5 would imply that the true 0.5-10 keV X-ray luminosity were less than $10^{42}$ ergs s$^{-1}$, but such a scenario is unlikely for this red ($R_{AB} - K'_{Vega} = 5.81$) source.  XID 536 has an X-ray luminosity and X/O ratio below the AGN threshold ($L_{X, 0.5-10 keV} = 3.0\times10^{41}$ ergs s$^{-1}$ and X/O = -2.3) and is thus classified as a galaxy in Table 2.  This X-ray luminosity, however, would still require a substantial starburst to explain.  As we show in \S 8, the stellar populations of this object are not consistent with a starburst.  In addition, the 24/8 flux ratio of -0.40 is quite low for a starburst, but can be better accounted for with low AGN activity.  We surmise that XID 536 is a weak AGN and therefore include it in the sample as such.  Further, its neutral hydrogen column density of less than $3.0\times10^{21}$ cm$^{-2}$ argues for an unobscured, Type I source.  

We estimate the [OIII] $\lambda5007$ line luminosity of XID 594 with spectra from \citet{szo04} and show the results in Table 3.  For XID 536 and 83, sources that do not have optical spectra in \citet{szo04}, we estimate the [OIII] $\lambda5007$ line luminosity using the X-ray luminosity calibration of \citet{net06} as applied on XID 32.

\subsection{The Comparison Sample}
Stellar populations analysis has been shown to be prone to systematic errors in various cases.   It is useful to compare the stellar populations statistics of the sample of interest to a control sample at the same redshift.  In this way, the systematic errors present in the stellar populations fitting will be present in both samples and the relative differences may be investigated.  

Nine non-active galaxies have been chosen from the $40\farcs \times 40\farcs$ AO corrected field of view to serve as a comparison sample.  All selection criteria imposed on the AGN sample have also been imposed on this sample.  In addition, the galaxies were selected to have a similar distribution of integrated K$'$ apparent magnitudes.  This necessitated choosing the brightest galaxies in the adaptive optics K$'$ images to match the AGN.  The system coordinates, R-band magnitudes, and redshifts are given in Table 2.  The mean R and K$'$ (AB) magnitudes for the comparison sample are 22.5 and 20.1, respectively, as compared to 21.65 and 19.3 for the AGN sample.  The mean R-K$'$ color for the comparison galaxies is 1.40, as compared to 2.35 for the AGN sample.  The mean redshift is 0.724 for the comparison sample and 0.836 for the AGN sample.  It is expected that the mean redshifts should be similar, and the small difference seen here does not affect the broad conclusions of the paper.  K$'$ is a rough proxy for stellar mass, and AGN have been shown to preferentially lie in massive galaxies (e.g., Kauffmann 2003).   Thus, choosing the most K$'$-bright (or most massive) non-active systems to compare to an active sample is suitable.  The mean K$'$ magnitude of the AGN sample is still 0.8 magnitudes brighter than that of the comparison sample, which reflects our inability to find sufficiently massive galaxies for the comparison sample in the limited survey field of view.  These minor differences will be taken into account in later analysis.

\section{PHOTOMETRY}  

\subsection{Estimation of LGSAO Point Spread Function}
Adaptive Optics systems tend to produce point spread functions (PSFs) that are the sum of two components, a diffraction-limited core and a wide halo with FWHM roughly the size of the seeing disk.  The \textit{Strehl} is a measure of the amount of power in the core of the PSF relative to a perfect PSF produced by the telescope and instrument.  The PSF in AO systems is also spatially and temporally variant, complicating precise photometry on spatial scales comparable to the diffraction limit.  The PSF must be well-known to permit the photometric accuracies we seek for the stellar populations analysis in this work.  We use a semiempirical method of determining the PSF, which utilizes an LGSAO image of a globular or open cluster taken once throughout night and the PSF of the tip/tilt star within the corrected FOV \citep{ste04, ste05}.  We present this method and the nuances of its application to this work in Appendix A.

\subsection{Subtraction of Central Point Source}

AGN and QSOs tend to have contamination by non-thermal continua in the central regions, due to the accretion disk.  This feature is expected to be unresolved and bluish.  Unfortunately, the feature we search for (circumnuclear starbursting) is typically less than 1 kpc in extent and is likely hidden within the non-thermal contamination.  The Keck diffraction limit at $KÕ$ is 54 mas, which corresponds to $\sim400$ pc at a redshift of $0.8$.  Thus, the principal strategy is to estimate stellar populations statistics surrounding the central point source ($r > 1$ kpc) and ignore the stellar populations of the central point source itself.

However, central unresolved point sources contribute scattered light to the outer regions of interest, so it is necessary to subtract the central point source before attempting aperture photometry.  We determine whether or not a point source is present in a particular band with GALFIT.  We first fit the most radially extended components of the galaxy with two sersic profiles (bulge and disk) on the i-band image.  Then, we introduce a Gaussian component with a range of FWHM between $0-3$ pixels, or less than 100 mas, and refit all other parameters (including the Gaussian magnitude).   If the variation in $\chi^2$ between the  zero-width Gaussian case and the 3 pixel Gaussian case is not significantly different at the 5-sigma level according to an F-test, the central peak is not judged to be a point source.  If the central peak is judged to contain a point-like component, it is subtracted from the images before annulus photometry is performed. 

By the criterium above, full subtraction of a central point source was necessary for two sources, XID 15
and XID 32, both Type I sources and BLAGN by optical spectrum.  To indicate that the results of the point source subtraction are sensible, we plot the best fit SED of the removed point source for each source in Figure 1.  These SEDs are plotted against quasar color models from \citet{fra00}.  It appears that a power-law SED ($F_{\nu} \propto \nu^{-0.3}$) with a small infrared emission component is a sensible fit to the point source SED of XID 32.  The RMS error to this fit is 0.06 mags, which is on the order of the expected point source subtraction error.  A power-law SED with a small dust absorption component, $E(B-V) = 0.3$, fits the SED of XID 15 to an RMS error of 0.07 mags, which is also on the order of the expected error.

Subtraction of central point sources is necessary to analyze host stellar populations in AGN and QSOs, but the point source SED in itself gives information about the extinction seen by the nuclear emission.  For those systems that lack a central point source in a particular band according to the criterion above, we derive an upper limit on the magnitude of a point source by subtracting the brightest possible PSF from the core of the galaxy that does not invert the radial profile in the central regions.  Typical galaxies are not expected to have ``brightness holes'' in the nucleus, so the brightest PSF that subtracts to produce a flat radial profile is an upper limit on the magnitude of a central point source.  

\subsection{Aperture Corrections for Annulus Photometry}

Knowledge of the intrinsic brightness profile of a galaxy is limited by PSF broadening in this study.  Here, we compose sersic-type models of the intrinsic distribution and convolve them with the known PSF in an attempt to match the data.  After finding the best model, we compute the annular ``aperture corrections'' that account for PSF broadening.  These corrections are then applied to the photometry measured on the reduced (but unprocessed) data.  We use GALFIT to perform this model-fitting, dividing the unconvolved model by the convolved model to yield the aperture corrections.

We perform annulus photometry over 10 successive, non-overlapping circular annuli with increasing radial widths of $0\farcs15$.  This gives a central core aperture with a diameter of 300 mas.  

This method is illustrated in Figure 2 as a series of radial profiles taken at intermediate steps during the calculation of the aperture corrections for XID 56 (K$'$).  Shown are the raw radial profile, the best GALFIT estimate of the intrinsic light distribution, and the corrected annular photometry.

\subsection{Errors in Annulus Photometry}

We choose a wide central aperture because the variance of the 300 mas encircled energy across the diameter is typically 2-3 times lower than the Strehl variance across the field, especially for our high airmass fields.  The errors in our PSF estimation method are manifest on small spatial scales, as they are due to misestimates of the isokinetic angle, so only the first radial point is affected in the annulus photometry.  We do not attempt to draw conclusions about morphology on spatial scales on the order of the Keck $KÕ$ diffraction limit ($\sim 54$ mas).  

Errors are calculated in each annulus by summing in quadrature the following error sources:  (1) photon noise, (2) read noise, (3) dark current noise, (4) sky noise, (5) PSF subtraction error, and (6) aperture correction error due to uncertainty of the PSF in $KÕ$.  Errors 2-4 are computed by the root mean square of an empty sky region.  Source (5) refers to the errors in estimating the photometry of the host galaxy caused subtracting a central point spread function.  In point source-dominated galaxies (like Type I QSOÕs), this error increases beyond 0.3 magnitudes in the central regions.  It is estimated with a Monte Carlo technique of simulating the accuracy of PSF recovery for a particular galaxy.  A central core PSF is added to the host, which is first obtained by subtracting the original central PSF.  GALFIT is then used to refit the galaxy components plus PSF, and the best-fit PSF magnitude recovered.  The GALFIT initial conditions are varied and the 1-sigma standard deviation of recovered PSF magnitudes about the true simulated magnitude is calculated.  This error source is strongly centered in the first annular core and drops sharply afterward.

Error source (6) is computed with a separate Monte Carlo technique to determine the effect of PSF misestimation on final host galaxy photometry.  As discussed above, the most uncertain factor in the PSF estimation is the isokinetic angle, which strongly affects the encircled energy in the first few annular rings.  This error plays into the host galaxy photometry through the aperture corrections.  We compute the error on the aperture corrections by simulating 100 PSFs with a distribution of isokinetic kernels convolved (i.e., Gaussian kernels with varying FWHM).  The aperture corrections are computed in each case and the one-sigma errors are derived from the resulting distributions for each annular ring.  

\section{MORPHOLOGICAL PROPERTIES OF INDIVIDUAL SOURCES}

We now describe the morphological properties of each of the sources in the AGN and comparison samples.  This section attempts to answer the following questions for each AGN host:
1.  What are the broad morphological properties of the system?
2.  What limits can be placed on the presence of a point source in each band?
The answer to question (2) may influence our interpretation of each system as an obscured or unobscured AGN.  Specifically, the SED of such a point source can be used to constrain the extinction seen by the central emission.  In this section, we answer these questions for each AGN host.  Figures 3-10 show B-i and V-K$'$ colormaps for each AGN source to explore trends in color gradients, as well as tiled imagery and model fits.  A summary is presented at the end of the section.  We display image tiles, GALFIT fits, model residuals, and morphological analyses for the normal, comparison sample in Appendix B.

Colormaps are generated with the following process.  For a particular galaxy, images from each of two bands are convolved with the PSF of the other band to match resolutions.  Then, to preserve signal in the noisy wings of the galaxy, the images are both convolved with a Gaussian kernel with spatially variant FWHM.  The FWHM increases linearly in a radial pattern, from zero in the center to $1\farcs0$ at a radius of $0\farcs75$.   Thus, good spatial resolution in the colormaps is only preserved in the central square arcsecond.  The V-K$'$ colormaps utilize near-IR data from LGSAO imaging in the central $2\farcs0$ and deep ISAAC imaging in the surrounding regions.  

\subsection{Type II AGN}

\subsubsection{XID 56}
XID56 is a clear merging system, with central radially symmetric emission closely surrounded by blue spiral arms.  Several point sources are seen in the central 10 kpc, including one blue source offset from the center that aligns with a line of star formation in the northeast.  This blue line of star formation may alternatively be the optical counterpart of a jet associated with this point source \citep{mel05}.  The GALFIT fit to the K$'$ band image possesses a large, extended spherical structure lying beneath the many features in this merger.  Both the B-i and V-K$'$ colormaps reveal this component to be redder than the bluer spiral arms and knot-like structures.  The central structure is not well-fit by a point source in any band.

\subsubsection{XID 155}
XID 155 has two major components:  A large vertical disk-like structure lying beneath a compact, central source is visible in the V-K$'$ colormap shown in Figure 4.  The central structure is not well-fit by a point source in any band.  The underlying inclined disk structure, seen in both B-i and V-K$'$ colormaps, is bluer  than the central source.  This appears to be a disk with a significant bulge component.  

\subsubsection{XID 266}
XID 266 is an inclined star forming disk with an appreciable central bulge component.  Small tidal tails and star-forming knots on the North and South ends of the disk, as well as a possible blue companion 1'' to the NW, indicate that minor interactions are probable.  The disk star formation is manifest in multiple star forming knots that are pinned to two major spiral arms.  Judging from the connection between the spiral arms and the bulge, there appears to be a central bar as well.  Both B-i and V-K$'$ colormaps indicate strong negative color gradients, or redder in the bulge and bluer in the outer regions, which is consistent with disks with large bulges.  Central point source models do not fit well to the bulge in any bands.    

\subsection{Type I AGN}	

\subsubsection{XID 83}
XID 83 is a spherically symmetric object with some hints of a disk and an appreciable spheroidal component.  Small spiral arms are apparent in the log-scale images and are seen clearly in model-subtracted residuals in all bands.  The nearby blue structure in the SE corner was masked for measuring photometry.  The B-i colormap indicates that the central core is bluer than the underlying spheroidal component.  The V-K$'$ colormap indicates this positive color gradient as well, but the overall system is quite red in V-K$'$ and is very bright in K$'$ out to a radius of $1.5-2\farcs0.$  Red observed frame V-K$'$ and bluer B-i colors are expected for such a high-redshift object ($z = 1.76$).  The very red, dominating spheroidal component underlying a bluer core may be indicative of a large elliptical with central star formation.  Central point sources models do not fit well to the central core in any optical band, although a small ($R_e < 1$ kpc) knot is possibly present in K$'$-band.

\subsubsection{XID 536}
XID 536 is a clear merger of two spheroidal galaxy cores, previously presented in \citet{mel05}.  Neither central core is a point source in any band, although individual sersic profiles fit well.  The underlying spheroid is well-fit by a single, radially symmetric sersic profile with no disk-like signatures.  Present in the optical model-subtracted residuals are tidal features and shells, which indicates kinematic interaction between the two cores.  These features may also be remnants of spiral arms, although they do not display knot-like star formation.  The B-i colormap indicates that the underlying spheroid is redder than the Western (right) core and bluer than the Eastern (left) core.  Both cores are bluer in V-K$'$ than the surrounding $\sim1''$, but about the same color as the outer regions ($1'' < r < 2''$).  

\subsubsection{XID 594}
XID 594 is a spheroidal galaxy with a bright central core and no disk-like features.  All bands display smooth morphology in the model-subtracted residuals.  Both the B-i and V-K$'$ colormaps indicate that the central core is very red, while the underlying spheroid is slightly bluer, but still red ($B-i \sim 4.0$, $V-K' \sim 3.5$).  The central core is larger than a point source in all bands.  

\subsubsection{XID 15}
XID 15 is a broad-line AGN (BLAGN) with a clear point source present at the center in all bands.  Surrounding the central, blue point source is an inclined ring of blue star formation (axis ratio = 0.77).  There do not appear to be any other significant morphological features.  The SED of the central point source is consistent with a dust-reddened QSO ($E(B-V) \sim 0.25$ by comparison with Figure 1 of \citet{fra00}) and a spectral index $\alpha = -0.3$ (i.e., $S_{\nu} \propto \nu^{\alpha}$).    

\subsubsection{XID 32}
XID 32 is a broad-line AGN with a clear, blue point source in all bands.  The SED of the point source is consistent with an unreddened QSO with some infrared emission (e.g., see Figure 2 of \citet{fra00}) and a spectral index $\alpha = -0.3.$  Unlike XID 15, this BLAGN sits on a spheroidal component which is redder than the point source.  This component is eight times dimmer than the point source in observed B band and three times brighter in observed z-band.

\subsection{Summary of Morphological Properties}    
    
The AGN in this sample represent a range of morphologies, including irregulars/mergers, normal star forming disks with bulges, and ellipticals.  All of the AGN sources except one (XID 15) have an appreciable bulge or spheroidal component.  Although there appears to be no preferred morphology for AGN hosts, a correlation between morphology and AGN Type is supported.  All of the Type II AGN either possess disks (XID 155 and XID 266) or are irregular mergers (XID 56).  One of the broad-line Type I AGN is a star forming ring, which indicates disk-like morphology, but four of five of the Type I AGN are ellipticals (or merging ellipticals, in the case of XID 536).  This is consistent with a previous study of morphology of AGN hosts at high redshift in the Extended Groth Strip (EGS) survey by \citet{pie07}.  The authors utilized Gini-M20 non-parametric analysis on Chandra sources in EGS, finding that the median hardness ratio of E/S0/Sa galaxies is -0.46 (Type I unobscured) and the median hardness ratio of Sc/Irr/d hosts is 0.55 (Type II obscured).  

The normal galaxies also represent a range of morphologies, including 6/9 disks, spirals and irregulars and 3/9 ellipticals.   Ellipticals are 50\% more common in the AGN sample (4/8) than in the normal sample (3/9), but this is not significant for this small sample.  No conclusions about the comparison between AGN hosts and nonactive samples can be drawn from the morphologies alone.  However, the sample of normal galaxies will be helpful in the following stellar populations analysis in that they may reveal systematic errors in this fitting process.
    
\section{STELLAR POPULATIONS ANALYSIS}

For each annulus, aperture-corrected photometry is available in observed B, V, i, z, and KÕ.  These are compared to \citet{bru03} tau models at the redshift of interest.  Bruzual \& Charlot tau models predict the integrated spectrum of a stellar population at multiple timepoints though its existence, with a star formation rate that depends exponentially on time.  We use a \citet{sal55} initial mass function (IMF).  These models are obtained at high redshift by blueshifting the filter pass functions of the requested filters as well as blueshifting the input A0 spectrum (as in Melbourne et al 2008).  All model grids are computed over a range of three variables: (1) Age since star formation begins; (2) tau parameter (e-folding time) in years; and (3) the extinction resulting from the dust mixed with the stars.  AGN have been shown to have solar or super-solar metallicities at a range of redshifts \citep{gro06}, so we assume solar metallicity.  The Bruzual \& Charlot models are interpolated for 210 ages, 16 values of dust extinction, and 7 values of the tau age.  The best model is chosen in a least-squares sense, using the measured photometry with errors as derived above.  Since the number of interpolate models is discrete, and uncomfortably low in the case of the tau parameter, a cubic interpolation is used on the final model values to obtain finer accuracy.  

The errors on the stellar population parameters are computed with a further Monte Carlo simulation utilizing knowledge of the photometry errors.  For each ring and source, the least-squares fitting code is repeated on 100 manifestations of the photometry, varied according to the known errors.  The resulting distributions of age, tau, and dust are recorded and used to derive the one-sigma errors for each parameter.  

The fractions of the population (by mass) are computed from the best-fit tau model for three different age ranges: Young (age $<$ 100 Myr), Intermediate (100 Myr $<$ age $<$ 1 Gyr), and Old (age $>$ 1 Gyr).  All references to ``young,'' ``intermediate,'' and ``old'' refer to these age ranges.  These fractions are calculated with integrals over an exponentially declining star formation rate.  These fractions by mass do not refer to the fraction of a population after aging and undergoing total mass loss from the death of young stars; they refer to the fraction of \textit{all stars formed} by mass in a particular time frame since star formation began.  

We stress that since all models are searched for every reasonable value of age, tau parameter, and dust extinction, the error bars we generate with Monte Carlo simulations are a good indication of the degeneracies in the model for a particular SED.  Some SEDs may be fit by vastly different stellar populations models to equivalent $\chi^2$ in the presence of equivalent photometric error, while other SEDs are much less degenerate.  For example, SEDs with extremely red (optical - K$'$) colors may rule out all dust-less old or young populations, as these have a limit to their redness; in this case, the well-known age/dust degeneracy is mitigated.  As such, one can determine if model degeneracy is playing a role by observing the size of the error bars relative to the fraction in each population; if the error bars in each population overlap, little conclusion can be drawn.  We also point out that systematic errors in the models and insufficiencies in the type of model chosen (tau models with fixed solar metallicity) are present for both the AGN and comparison sample.  The principal conclusions of this work are drawn from relative differences between subgroups of this sample.

To illustrate the degeneracies of the models, Figure 11 plots photometry and best-fit spectra for two cases:  (1) the SED from the fourth ring of XID 155, which is representative of non-degenerate cases and (2) the SED from the fifth ring of XID266, which is representative of more degenerate cases.  Each plotted spectrum represents a fit to the photometry when stellar age has been fixed, but dust extinction and the tau parameter are minimized.  The solid lines in both cases represent the best-fit models, fixing no parameters (except metallicity).  Notice that the particular SED of the degenerate case may fit ages between 500 Myr and 4 Gyr, but the SED in the non-degenerate case does not fit the 500 Myr case.  In both SEDs, young stellar ages ($ < 100$ Myr) are ruled out  by the near-IR point.  Note that in Figures 13 and 14, the error bars on the best-fit model parameters reflect the level of model degeneracy apparent in Figure 11.

The stellar populations fractions as described above are plotted for each host galaxy as a function of radius for the AGN sample in Figures 12-19.  Also shown in each figure is a plot of radial profiles for each band.  Appendix C contains the stellar population fraction plots, radial profiles, and short discussions for the normal, comparison sample galaxies.

\subsection{Type II AGN}

\subsubsection{XID 56}
The merging system XID 56 displays predominantly young stellar populations in the core, older stellar populations in the range $0\farcs4 < r < 0\farcs7$, and intermediate age populations in the outer regions.  The outer photometry is likely from the linear feature in the Northeast, so the intermediate ages refer to this structure.  The underlying red spheroidal structure discussed in the previous section is composed of older stellar populations.

\subsubsection{XID 155}
The inclined disk XID 155 possesses a central core composed of dusty younger stellar populations.  The outer disk ($ r > 0\farcs2$) is largely old, with 4-6\% of an intermediate age population by mass and negligible young populations.  

\subsubsection{XID 266}
The large star-forming disk XID 266 displays a prominent negative color gradient, with a red core and blue outer regions.  The central bulge appears to be a mixture of all ages, but is dominated by older stellar populations.  The star forming disk ($r > 0\farcs2$) is a mixture of intermediate aged populations.  The fraction of older population associated with the spheroidal component decreases with radius.  

\subsection{Type I AGN}

\subsubsection{XID 83}
The red spheroid XID 83 with small central spiral arms displays intermediate aged populations in the central $0\farcs6$ and a mixture of old and intermediate aged populations beyond this radius.  This suggests that the central spiral arms are the remnant of a recent star formation event.

\subsubsection{XID 536}
The elliptical merger XID 536 is dominated by old stellar populations at all radii, although intermediate aged populations make a strong contribution between $0\farcs2$ and $0\farcs7.$  Both the two galaxy cores and the streaks that appear in the optical residuals fall in this region and may represent these populations.

\subsubsection{XID 594}
The smooth elliptical host XID 594 is dominated by old stellar populations at all radii greater than $0\farcs15$.  The central core is mostly composed of intermediate aged populations, although the error bars indicate that the models may be degenerate here.  

\subsubsection{XID 15}
The broad-line AGN XID 15 is largely composed of intermediate-aged stellar populations at all radii.  The radial profiles shown in the left panel of Figure 18 and used to reconstruct the stellar populations statistics plotted in the right panel have been measured following a subtraction of the central point source in all bands, which is the reason that the colors of the central core do not resemble the colors elsewhere at higher radii.  The point source contribution is estimated with a simultaneous sersic/PSF fit using GALFIT.  As a result, the residuals are forced to resemble a galaxy-like sersic profile.  There is a peak in the fraction of the youngest stellar population at a radius of $\sim 0\farcs7$, which corresponds to the location of the star forming ring.

\subsubsection{XID 32}
The broad-line AGN host XID 32 is dominated by older stellar populations at all radii.  Similar to XID 15, the radial profiles and reconstructed stellar populations statistics have been measured following a subtraction of the central point source in all bands using GALFIT.  The old stellar populations are consistent with the redness of the spheroidal component underlying the central point source.  The contribution from young stellar populations is less than 0.3\% for $r > 0\farcs2.$

\subsection{Summary of Stellar Populations Analysis}

A summary of the dominant stellar populations in the inner ($r < 0\farcs3$) and outer regions ($r > 0\farcs7$) is shown for all galaxies in Table 4.  A number of observations about the stellar populations of the normal sample inform the comparison to the AGN sample.  First, young and intermediate age populations are rarely seen as dominant in the outer regions of the systems ($r > 3-6$ kpc, depending on redshift).  Those galaxies with this property tend to be spirals, disks, and irregulars (Norm3 and Norm6).  Second, young populations are only seen as dominant in the central regions of spirals, disks, and irregulars (Norm2 and Norm5).  Third, ellipticals are more likely to have old stellar populations at all radii compared to spirals, disks, and irregulars (Norm1, Norm4, and Norm7).  Finally, as these results are expected for normal galaxies, they lend credence to the stellar populations fitting method.

The stellar populations of the AGN hosts are consistent with their morphological classification.  All three observations made in the above paragraph can be made separately for the AGN population, without considering AGN Type (but provided that central point sources resulting from non-stellar continua are subtracted off).  In fact, for every AGN host it is possible to find a rough analogue in the normal sample with respect to broad morphological and stellar population characteristics.  This being stated, the correlations seen in \S 5 between morphology and AGN Type are manifest as correlations between the presence of young and intermediate stellar populations and AGN Type.  A majority of Type I AGN (4/5) have old stellar populations in the outer regions.  All of these 4 galaxies are ellipticals.  A majority of Type II AGN (2/3) have intermediate stellar populations in the outer regions, although this is not significant due to small number statistics.  

The strongest summarizing observation that can be drawn from this small sample is that the Type II AGN representatives are more likely to be spirals, disks, or irregulars and the Type I AGN representatives are more likely to be ellipticals according to corroborating evidence from both morphology and stellar populations.  

\section{CONCLUSIONS AND DISCUSSION}

\subsection{Stellar Populations of AGN in Local Universe}

The stellar populations of this $z\sim1$ population of AGN may be compared to those at lower redshift.  A study of SDSS AGN in \citet{kau03}, selected via optical spectroscopic line ratios find that narrow-line AGN (Type II, $0.02 < z < 0.3$) of all luminosities reside almost exclusively in massive galaxies and have distributions of sizes, stellar surface mass densities and concentrations that are similar to those of ordinary early-type galaxies in our sample.  Locally, the host galaxies of low-activity AGN have stellar populations similar to normal early types, AGN activity/strength being measured  with the line luminosity of the [OIII] line.\footnote{AGN strength is independent of AGN Type.  Strong AGN are defined as log $L_{[OIII]} > 7$ in units of bolometric solar luminosities ($L_{\sun} = 4\times10^{33}$ ergs s$^{-1}$).}  The hosts of high-activity AGN have much younger mean stellar ages. The young stars are not preferentially located near the nucleus of the galaxy, but are spread out over scales of at least several kiloparsecs.  \citet{kau03} also examines the stellar populations of the host galaxies of a sample of broad-line AGN and conclude that there is no significant difference in stellar content between Type 2 \citet{sey43} hosts and quasars (QSOs) with the same [OIII] luminosity and redshift. This establishes that a young stellar population is a general property of AGN with high [OIII] luminosities.  Strong AGN have stellar populations similar to late type galaxies (young ages). 

Although the correlation between AGN strength measured with $L_{[OIII]}$ and the hard X-ray luminosity is good at low and high redshift \citep{net06}, the correlation between AGN Type determined with optical spectroscopic methods and X-ray hardness ratio is less clear.  The AGN in \citet{kau03} are predominantly Type II AGN, determined by narrow-line ratios in optical spectra.  The traditional typing method used for decades \citep[e.g.,][]{ost89} requires the presence of optical broad lines to flag a Type I AGN.  The X-ray Type classification relies primarily on the X-ray hardness ratio (HR) and the neutral hydrogen column density that can be computed from the HR through models.  This typing confusion has been addressed with deep hard X-ray surveys, like the \textit{Swift} BAT survey \citep{tue08}.   This survey of $\sim100$ local AGN finds that the correspondence between hard X-ray typing and optical typing is good (94\% correspondence) when archival optical spectra of X-ray AGN are reexamined for broad-lines.   We continue with the knowledge that X-ray AGN Types and strengths are comparable to Types and strengths derived primarily from optical spectra.  

\subsection{Comparison to Local Results}

As \citet{kau03} divides the local Type II sample into two categories (strong and weak), we divide the high redshift sample into four categories (including Type I AGN):  Strong Type I AGN, Weak Type I AGN, Strong Type II AGN, and Weak Type II AGN.  The distinction between Type I and Type II is made at a neutral hydrogen column density of $10^{22}$ cm$^{-2}.$  The division between weak AGN and strong AGN is made at log $L_{[OIII]} = 7.5.$ \citet{kau03} uses a division of log $L_{[OIII]} = 7.0$, but we require a higher division to include 3 galaxies in the weak sample rather than 2 (XID 32 is in the Strong Type I sample if the sample is divided here).   Our sample of 8 AGN hosts with $0.4 < z < 1.76$ has morphologies, stellar population distributions, and color distributions that are consistent with findings of \citet{kau03} at lower redshift. 

Type I AGN in this sample are not dominated by young (age less than 100 Myr) populations at any radii.  Type I AGN may be dominated by intermediate aged stellar populations, but these are restricted to the central ($r < 3-6$ kpc) regions.  Overall, 5 of 5 strong AGN (both Type I and II) hosts are dominated by young or intermediate aged populations at some radius, while 1 of 3 weak AGN (both Type I) are dominated by young or intermediate AGN at some radius.  To enable comparison across the AGN Type dimension as well as the AGN strength dimension, we have plotted the average stellar population statistics for each of three groups (strong Type I, strong Type II, and weak Type I AGN) in Figure 20.   Notice that the overall level of young and intermediate stellar populations is clearly different in the strong and weak populations, regardless of AGN Type.

It is difficult to directly compare these observations with Kauffmann's finding that young stellar populations are correlated with AGN strength, as this result was obtained for both strong and weak Type II AGN.  We do not sample weak Type II AGN in this paper.  However, combining Type I and Type II populations, there is a correlation between [OIII] line luminosity and the presence of dominating young and/or intermediate aged populations at any radii.  This is consistent with Kauffmann's low-redshift finding that (1) AGN strength is correlated with the presence of young stellar populations provided that (2) the stellar populations of Type I AGN are indistinguishable from Type II AGN.   Figure 21 plots [OIII] line luminosities for all AGN sources against the radially averaged fraction of young stellar populations (age $< 100$ Myr).  This plot mixes Type I and Type II sources.

The morphologies of the strong Type II AGN in this study follow trends outlined in Kauffmann's low-redshift sample.  The morphologies of the galaxies in Kauffman's sample with considerable [OIII] (log $L_{[OIII]} > 7.8$), a group of which this study's XID 155, 266, and 56 are members, fall into three broad categories:  (1)  Single blue spheroids/amorphous galaxies, with no close companions and no significant structure.  These look like E/S0 galaxies with anomalous blue colors.  (2) Single disk galaxies.  In these, structure is seen outside the bright core, like arms, bars, and dust lanes.  No close companions are visible in this category.  (3) Disturbed, interacting galaxies in which the core system possesses one or more close companions and obvious tidal debris.   Of this study's strong Type II AGN, XID 56 falls into category (3), XID 266 falls into (2), and XID 155 falls into either (2) or (3), depending on whether the extended blue disk is a disk or tidal debris.  As Kauffmann says, this shows that there is a variety of ways to feed the central black hole, but common to all of them are large patches of gas and extra star formation. 

We find that 3 of our 3 Type II AGN are LIRGs, as compared to 1 of 3 weak Type I AGN.  2 of 2 strong Type I AGN are LIRGS.  In the Type I AGN sample (both strong and weak), the 24 micron to 8 micron flux ratios indicate that the source of the mid-IR emission is due to hot dust in close proximity to the central accretion disk itself.  However, the strong Type I AGN are distinguished from the weak Type I AGN by the level of 24 micron flux, which elevates them to LIRG-status.  The powerful Type II AGN, in sharp contrast, display 24/8 micron flux ratios indicative of a strong starburst contribution to the mid-IR emission, which is more than enough to classify them as LIRGs.  It is unclear if the Type II AGN would remain LIRGs if the starbursting component were removed.  These observations are consistent with the notion that starbursting in Type II AGN is somehow connected with their status as obscured AGN.  All of the strong AGN (log $L_{[OIII]} > 7.5$ in bolometric solar luminosities) are classified as LIRGS regardless of obscuration; however, certainly the unobscured AGN are LIRGs because of mid-IR emission associated with the AGN.

There is additional information in the presence of point sources and their SEDs in the sample.  The Type II population does not display point sources in any band, according to the criteria that the $\chi^2$ with a fitted point source be significantly ($5\sigma$) improved compared to that with a wider ($100$ mas) Gaussian.  2 of 5 Type I AGN display clear point sources in all bands.  These 2 AGN additionally display broad-lined emission.  The SEDs of the point sources are consistent with those of unobscured QSOs.  No other Type I AGN (XID 83) display point sources in any bands.  

Overall, the morphologies, stellar population distributions, and color distributions of the high-redshift sample of 8 AGN are similar to the low-redshift samples, except for various small differences arguing for larger gas reservoirs in the strong Type II population.  The evidence for this is drawn from two critical observations:  (1)  High-z strong Type II AGN are more likely to be LIRGS via starbursting events.  This compares to all Type I AGN, which, if presenting LIRG activity, are excited to that level via hot dust surrounding the AGN.   (2)  Type II AGN hosts at high-z tend to be Sc/d/Irr galaxies, while Type I AGN tend to be earlier types (verified in Pierce et al 2007).   With the findings that Type II AGN uniquely possess extended star formation, LIRG starbursting, and are morphologically distinct from Type I AGN, this raises the question:  What is truly causing the obscuration that fundamentally distinguishes these two categories of AGN?  If the obscuring medium is associated with these large scale starbursts, then this leads to the supposition that high-z Type II AGN are obscured via kpc-scale dusty features that are byproducts of starbursting in addition to small ($r < 300$ pc) dust tori, as in the case of Type I AGN.

\subsection{Significance of Stellar Populations Distinctions}
The discussion above must be treated with caution, as the number of galaxies in each bin (Strong Type I, Strong Type II, and Weak Type I) is either two or three.  A student's t-test can be used to estimate the probability that the hosts in these groups have been selected from fundamentally different populations.  We use the radially averaged fraction of young and intermediate aged stars (age $< 1$ Gyr), without weighting by luminosity, to compare the different populations.  Comparing the Strong Type I population to the Weak Type I population gives a probability of 83\% that the hosts are drawn from a different sample.  Using just the radially averaged fraction of young stars (age $ < 100$ Myr) gives a probability of 91\% that these two populations are distinct.  However, when comparing the Strong Type I population and the Strong Type II population for stars less than 1 Gyr in age, the probability is 61\%.  This places higher weight on the conclusion that AGN strength ($L_{[OIII]}$) is correlated with the presence of YSPs, but less weight on the observation that the stellar populations of Type I and Type II hosts at equivalent AGN strength are distinct.  A larger, carefully-controlled sample of Type I and Type II AGN hosts at high redshift will be necessary to examine this question in more depth.

\section{SUMMARY}
We have completed a pilot study of a small sample of X-ray selected AGN in the GOODS field, combining K$'$ band imagery from the Keck LGSAO system with deep ISAAC Ks imagery and HST ACS imagery in four optical bands.  We have combined evidence from multiple independent AGN selection methods to verify that the eight galaxies in this sample are AGN.  A morphological analysis of the galaxies compared to a sample of nonactive systems has revealed that the Type I AGN hosts are more likely to be ellipticals and the Type II AGN hosts are more likely to be spirals, disks, or irregular types.  A stellar populations analysis of the five band photometry at a resolution better than 100 milliarcseconds in all bands has demonstrated that the stellar populations of the AGN hosts are consistent with their morphological identities, i.e., early types have older ages and later types have younger ages.  We combine Type I and Type II populations and find that there is a correlation between AGN strength ($L_{[OIII]}$) and the presence of dominating young and/or intermediate aged populations at any radii, similar to what is seen at low-redshift \citep{kau03}.  The mid-infrared properties of the AGN hosts identify all of the strong Type II AGN hosts as LIRGs via starbursting events, the strong Type I AGN hosts as LIRGS via hot, localized dust surrounding the central engine, and the weak AGN as non-LIRGs.  The strong Type II AGN hosts are also more likely to possess extended star formation than the Type I AGN.  Although this study cannot come to rigorous conclusions about the nature of Type II AGN obscuration due to the small sample size, these last two pieces of evidence taken together with the previously seen correlation between morphology and AGN Type at $z\sim1$ suggest that Type II obscuration is due to kpc-scale dusty features that are associated with LIRG-level starbursting.

\section{ACKNOWLEDGEMENTS}	

This work has been supported in part by the NSF Science and Technology Center for Adaptive Optics, managed by the University of California (UC) at Santa Cruz under the cooperative agreement No. AST-9876783.

Some of the data presented in this paper were obtained from the Multimission Archive at the Space Telescope Science Institute (MAST). STScI is operated by the Association of Universities for Research in Astronomy, Inc., under NASA contract NAS5-26555. Support for MAST for non-HST data is provided by the NASA Office of Space Science via grant NAG5-7584 and by other grants and contracts.

The laser guide star adaptive optics system was funded by the W.M. Keck Foundation.  The artificial laser guide star system was developed and integrated in a partnership between the Lawrence Livermore National Labs (LLNL) and the W.M. Keck Observatory.  The laser was integrated at Keck with the help of Curtis Brown and Pamela Danforth.  The NIRC2 near-infrared camera was developed by CalTech, UCLA, and Keck (PI Keith Matthews).  The data presented herein were obtained at the Keck Observatory, which is operated as a scientific partnership among the CalTech, UC, and NASA.  This work is supported in part under the auspices of the US Department of Energy, National Nuclear Security Administration and by LLNL under contract W-7405-Eng-48.  S.M.A acknowledges fellowship support by the Allen family through UC Observatories/Lick Observatory.  This work is based in part on observations made with the European Southern Observatory telescopes obtained from the ESO/ST-ECF Science Archive Facility.  Observations have been carried out using the Very Large Telescope at the ESO Paranal Observatory under Program ID(s): LP168.A-0485.

The authors wish to recognize and acknowledge the very significant cultural role and reverence that the summit of Mauna Kea has always had within the indigenous Hawaiian community.  We are most fortunate to have the opportunity to conduct observations from this superb mountain.

\facility{CXO, HST, Keck:II, Spitzer, VLT:Antu}

{}

\appendix

\section{APPENDIX A:  LGSAO PSF ESTIMATION}

In Laser Guide Star AO systems, the spatially and temporally variant PSF is more difficult to model than in an NGS system, due to additional error terms and the shifting atmospheric geometries (i.e., the cone effect). In a typical CATS field with LGS at the center and a tip/tilt star near the edge of the field, the PSF distribution is approximately given by the sum of three error components:  (1) Anisoplanaticism and cone effect from the LGS itself, (2) Tilt anisoplanaticism (or anisokineticism), and (3) non-spatially dependent error sources such as shot noise from the tip/tilt star.  LGS anisoplanaticism is assumed to be a radially symmetric distribution centered on the LGS.  Anisokineticism is also radially symmetric, but is centered on the tip/tilt star.  Error due to source (3) are assumed to not depend on geometry.

We use the Steinbring et al. (2005) formulation to estimate the PSF on a target-by-target basis using an exposure of a star cluster and a single stellar source in the target field.  This semiempirical method assumes that the spatial and temporal components of PSF variation are largely separable.  It utilizes an LGS image of a cluster taken once throughout the night to capture the spatial variance of the PSF (i.e., LGS anisoplanaticism and anisokineticism).  Ignoring LGS dimming due to intervening clouds or technical problems with the laser, the remaining temporal variance throughout the night can be captured with an additional convolved spatially-variant Gaussian kernel to model image degradation due to anisokineticism from the tip/tilt.  In the images presented, the tip/tilt star is imaged in the corner of the field to serve as the reference for the Gaussian.  The tip/tilt star and laser guide star are not necessarily oriented in the same geometry for cluster calibration images and science frames.  Practically, this requires an estimation of the anisokineticism in the cluster image.  

In the \citet{ste05} method, PSFs are computed for each science target.  A star is first chosen in the cluster image for which the orientation relative to the laser guide star is the same as the tip/tilt star in the science frame of interest.  A second star is chosen in the cluster image close to the location of the particuliar science target relative to the laser guide star.  The second star is deconvolved from the first star using the Max\_Likelihood deconvolution function in IDL, which provides very good agreement upon reconvolution ($< 5\%$ at all pixels in a radial profile).  The resulting kernel, which to first order translates PSFs at the tip/tilt star in the science frame to PSFs at arbitrary locations, includes the effects of LGS anisoplanatism and the difference in anisokineticism that these two cluster stars have seen.  The tip/tilt star in the science frame is convolved with this kernel and further convolved with a Gaussian of full-width at half-max:

$$FWHM = R \sqrt{(\theta_s / \theta_a)^2 - (\theta_2 / \theta_a)^2 + (\theta_1 / \theta_a)^2}$$

Here $\theta_s$ is the angle between the science target and the tip/tilt star.  $\theta_1$ is the angle between the first cluster star (chosen to correspond to the science tip/tilt star) and the cluster tip/tilt star.  $\theta_2$ is the corresponding angle between the second cluster star (chosen to correspond to the science target) and the cluster tip/tilt star. $\theta_a$ is the anisokinetic angle and R is a constant.  The ratio $(R/\theta_a)$ is calibrated by a second star in the science frame.  Note that the anisokinetic angle is assumed to be the same between the cluster frame and the science frame.  This is certainly not guaranteed to be true.  \citet{ste05} recommends that a second calibration image of the cluster be taken with tip/tilt correction off (to obtain an independent measure of the anisokinetic angle), but these calibration images were not taken for the frames presented here.  

This uncertainty in the anisokinetic angle, which may vary by a factor of two, causes an uncertainty in the final Gaussian FWHM.  We translate this into a quantitative PSF error via Monte Carlo simulation of multiple PSFs.  Targets near the tip/tilt star will have less of this error because their error is dominated by LGS anisoplanatism instead.  Additionally, this error is concentrated in the central 300 mas aperture because the Gaussian error functions due to anisokineticism are not widely extended.  

\citet{ste05} reports that the full method (with a both cluster calibration frames, with and without tip/tilt correction) gives 20\% one-sigma error on the Strehl ratio on stars $8\farcs$ from the LGS and $28''$ from the tip/tilt star.  This is likely dominated by the uncertainty in the anisokineticism that we see.  

\section{APPENDIX B:  MORPHOLOGICAL PROPERTIES FOR COMPARISON SAMPLE}

Image tiles, GALFIT fits, model residuals, and morphological analyses are displayed for the normal, comparison sample in Figures 22-30.

\subsection{Normal 1}
Normal 1 is an elliptical galaxy, with smooth profiles in all bands.  The only residuals remaining after subtraction of a best-fit sersic + Gaussian model are associated with the improperly-fit width of the central core.  The axis ratio of the sersic component is greater than 0.9 in all bands.  There are elongated features at $r = 1\farcs0$ in the optical residuals that resemble shells and rings.
    
\subsection{Normal 2}
Normal 2 is an inclined disk with a central bluish, elongated feature.  As seen in the GALFIT residuals, the central structure possesses knot-like signatures of star formation.  The morphology outside of the central region ($r > 0\farcs2$) is smooth and disk-like in all bands, with no knot-like star formation signatures.  This suggests that Normal 2 is an Sa or S0 type galaxy.  

\subsection{Normal 3}
Normal 3 is a nearly edge-on disk with a redder outer disk and indications of central star formation.  The outer disk ($r > 0\farcs5$ deprojected) is more prominent in bands redder than V.  There are two major central features, including a star forming ring of deprojected radius $r = 0\farcs3$ and a central point source.  The point source is visible in bands redder than V and the star forming ring is visible in bands bluer than z.  

\subsection{Normal 4}
Normal 4 is an elliptical galaxy with smooth profiles in all bands.  There is a strong central blue component, while the redder bands possess more extended profiles.  The axis ratios of the primary sersic component in the best-fit sersic + Gaussian models are greater than 0.9 in all bands.  There are few features in the residuals after subtracting galaxy models.  

\subsection{Normal 5}
Normal 5 is a nearly face-on spiral galaxy (Sc) with clear evidence of star formation.  Multiple blue star-forming knots are pinned to several spiral arms and a central star forming ring of radius $0\farcs4$.  The central core is apparent in all bands, but all other knots and features disappear in the observed K$'$, which only consists of the core embedded in an extended, spheroidal component.  This spheroidal component is not apparent in the bluer bands.  The two major spiral arms appear to emerge from the central core, but are not symmetric. 

\subsection{Normal 6}
Normal 6 is a slightly inclined disk with a star forming ring and a central core.  The central bulge is not visible in the B band, but is seen in all redder bands.  The ring is composed of several pinned star-forming knots that are prominent in the bluer bands.  Underlying the blue ring is a red, disk-like smooth structure seen primarily in the z and K$'$ bands.  The observed K$'$ image is smooth and free of knots and rings, except for the central core.

\subsection{Normal 7}
Normal 7 is a flattened elliptical (axis ratio $\sim 0.7$ in all bands) with a smooth morphology.  

\subsection{Normal 8}
Normal 8 is a barred, grand design spiral galaxy with a central red bulge.  The K$'$ band (rest-frame J-band) image displays a spheroidal bulge on top of an elongated bar seen in the bluer bands as well, but does not show the spiral arms.  The spiral arms are prominent in the bluer bands.  The morphology is largely smooth, with an absence of star-forming knots and rings.

\subsection{Normal 9}
Normal 9 is a compact, inclined disk with several star-forming knots in the central region ($r < 0\farcs5$).  The observed K$'$ morphology is different from the optical bands, with two weak spiral arms embedded within an ellipsoidal structure.  

\section{APPENDIX C:  STELLAR POPULATIONS PROPERTIES FOR THE COMPARISON SAMPLE}

Stellar population fraction plots, radial profiles, and short discussions for the normal, comparison sample galaxies, as in Figures 12-19 for the AGN sample, are displayed in Figures 31-39.

\subsection{Normal 1}
This elliptical nonactive galaxy is dominated by older stellar populations at all radii.  The shells and features apparent in the model-subtracted residuals, which were pointed out in \S 6, appear also to be composed of older stellar populations, meaning that they remnants of some long ago event.

\subsection{Normal 2}
Morphologically, this disk-like galaxy is an Sa/S0 with a central blue core.  The central core is composed of young stellar populations primarily.  The region between $r = 0\farcs3$ and $r = 0\farcs9$ is a mixture of intermediate-aged and older stellar populations, and the region beyond this radius is dominated by an older population.  This system has a positive age gradient.
    
\subsection{Normal 3}
This inclined disk is a mixture of populations of various ages, but intermediate ages dominate at small and large radii.  Although this disk only possesses knot-like structure and a ring in the central $0\farcs5$, it is clear that some star formation had occurred more recent than 1 Gyr to construct the outer disk.
    
\subsection{Normal 4}
This elliptical galaxy with a blue core and smooth features is composed of intermediate-aged populations in the central regions ($r < 0\farcs4$) and older stellar populations beyond this radius.
    
\subsection{Normal 5}
Normal 5 is a clear face-on spiral with star-forming knots across the length of two spiral arms.  Young stellar populations are dominant in the regions filled with the central star-forming ring, or $r  < 0\farcs6$.  The spiral arms themselves, which emerge from the star forming ring at this radius, are described as an intermediate-aged component at this point, although the underlying older stellar population dominates at this and higher radii.  
    
\subsection{Normal 6}
This irregular galaxy appears to have a red, bulge-like component in the center and a blue ring at $r = 0\farcs3-0\farcs4$.  The red core is largely an older stellar population, and the star forming ring clearly shows as a younger contribution.  At higher radii, the system is a combination of equal fractions of populations, but the overlap of these points compared to the errors bars suggest that the photometric errors are too large to draw conclusions.  
    
\subsection{Normal 7}
This smooth elliptical with very little blue contribution is dominated by older stellar populations at low and high radii ($r < 0\farcs3$ and $r > 0\farcs7$).  Intermediate-aged populations dominate at radii between these two regions, but since there are no morphological features at these radii that might correspond to these populations, this might be a result of photometry error in the B-band.
    
\subsection{Normal 8}
Normal 8 is a clearly barred spiral with no knot-like structure and a core bluer than the surrounding system.  This system appears to be the sum of three morphologically orthogonal components that are composed of different populations:  A blue, star-forming component in the central core, an old bar, and intermediate aged spiral arms.  The central core is an equal mixture of these different components.  At higher radii ($r > 0\farcs2$), the red bar dominates the photometry with an older stellar population.  The spiral arms appear at the highest radius with an intermediate-aged contribution of 10\% by mass.  
    
\subsection{Normal 9}
This inclined disk with the appearances of central star formation turns out to be dominated by a young stellar population for $r < 0\farcs6$.  There is a significant old component in the core, possibly due to the peak of some red spheroidal, bulge-like component.  Beyond $0\farcs6$, the population is dominated by a weak, underlying old stellar population.  

\clearpage
\begin{figure}
  \caption{SEDs of point sources subtracted from the center of XID 32 (left) and XID 15 (right), with wavelength plotted in rest-frame units, plotted on top of Figures 1 and 2 from \citet{fra00}.  The solid lines in the left plot represent a power-law model of quasar emission plus different amplitudes of power-law synchrotron infrared emission.  The solid lines in the right plot represent power-law quasar emission with a varying absorption due to intervening dust.  The SED of XID 15 is overlaid on the model with $E(B-V) = 0.3.$}
  \begin{center}
    \includegraphics[width=4.2in,height=3.0in]{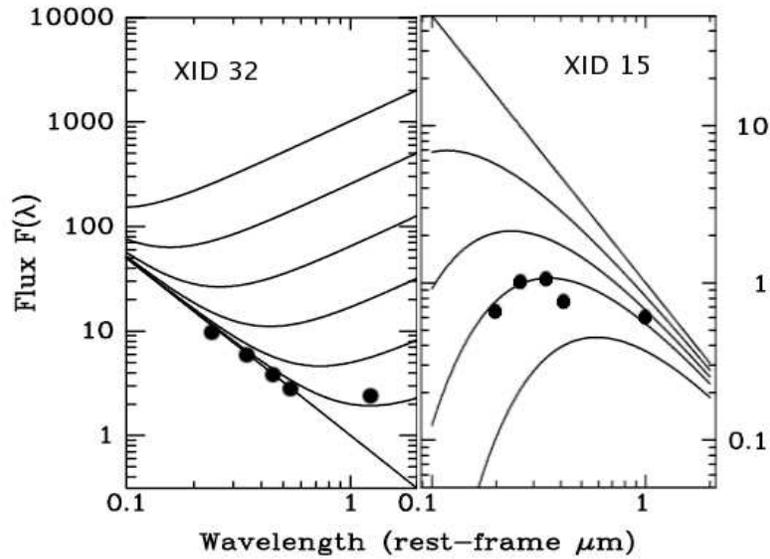}
    \\[.5cm]
  \end{center}
\end{figure}

\begin{figure}
  \caption{Log-log radial profiles for XID 56 (K$'$) images at various intermediate steps during the calculation of aperture corrections.  The solid line is the measured radial profile for the raw K$'$ image.  The dotted-dashed line is the radial profile of the best GALFIT fit to this galaxy, including a convolution to the known PSF.  The diamond points represent the annulus photometry directly measured on the raw data.  The dashed line is the radial profile of the GALFIT model without convolution with the PSF, denoting the best estimate of the intrinsic light distribution.   Aperture corrections are computed by comparing the GALFIT model with and without convolution by the PSF.  The triangle points plot the aperture-corrected photometry.}
  \begin{center}
    \includegraphics[width=4.2in,height=3.0in]{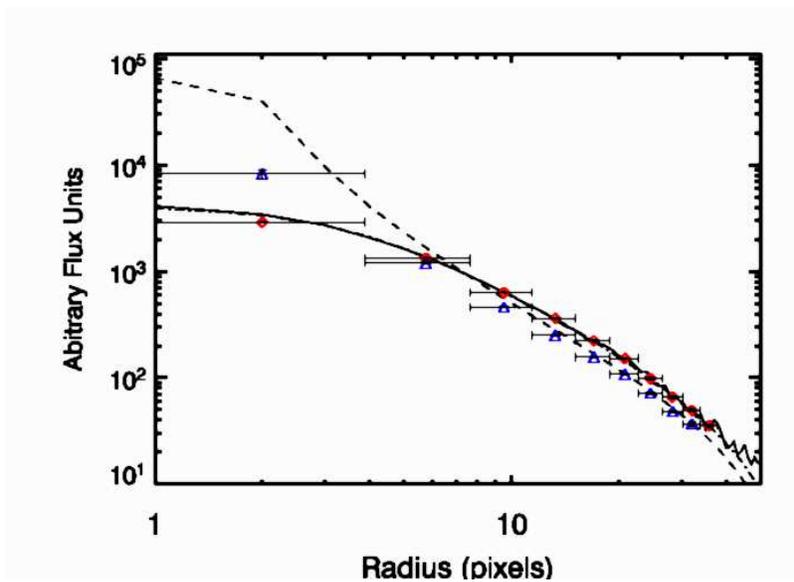}
    \\[.5cm]
  \end{center}
\end{figure}
\clearpage
\begin{figure}
  \caption{Tiled magery, GALFIT models, and colormaps for XID 56.  The log-scale tiled images are $3\farcs$ across.  The five bands are split column-wise in the order B, V, i, z, and K$'$. The bottom five panels are the raw, inverted images.  The middle five are the corresponding GALFIT models with bulge and disk sersic components.  Residuals are displayed in the top row.  The B-i colormap for XID 56 is shown in the bottom left and the V-K$'$ colormap is shown in the bottom right.  All magnitudes are in the AB system.  K$'$ imaging from adaptive optics is used within the gray square of 2.0'' size;  deep ISAAC imaging in Ks is used outside of the square.  The correction from Ks to K$'$ is small ($\sim0.02$ mag) and ignored for the colormaps.}
  \begin{center}
    \includegraphics[width=5in,height=3.0in]{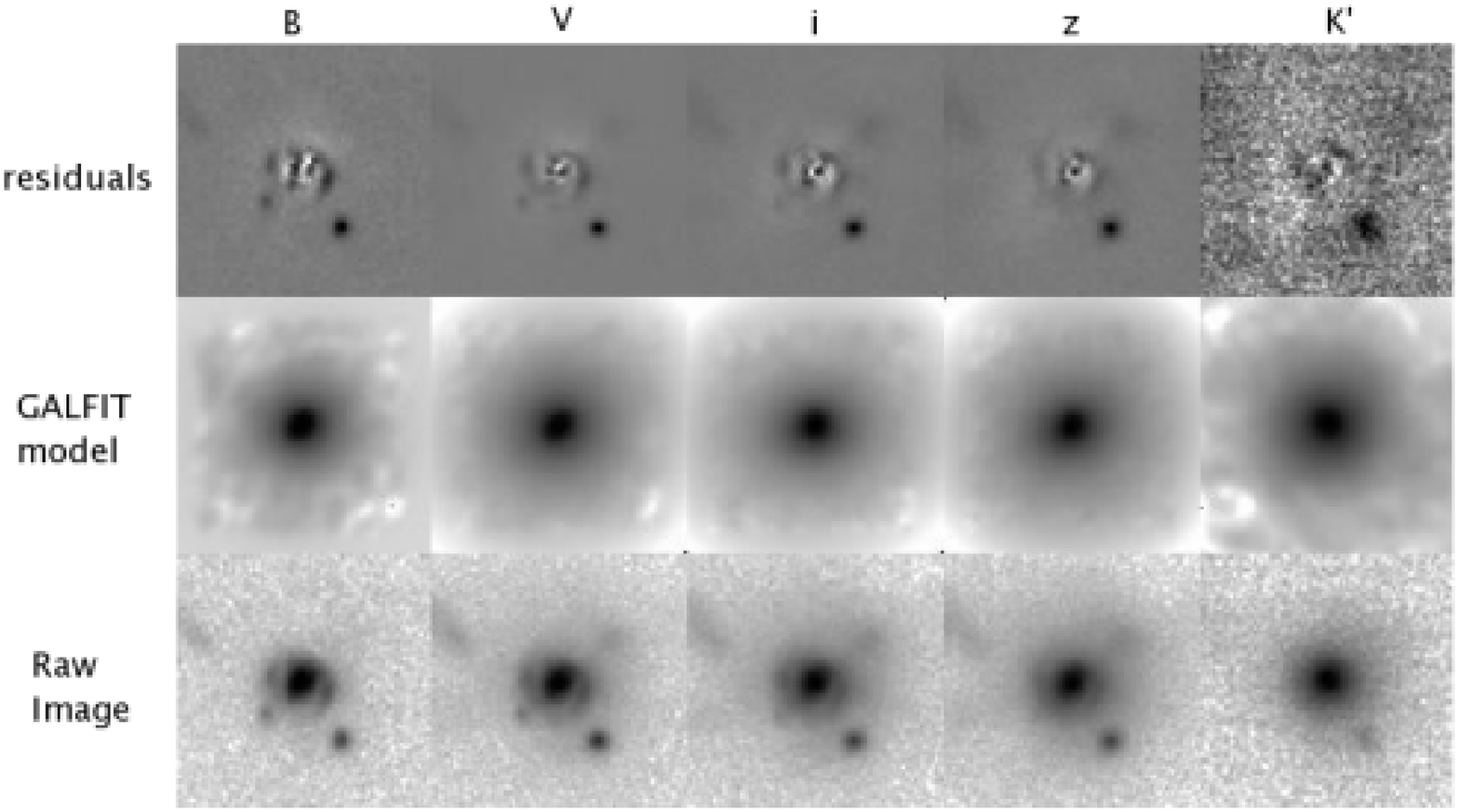}
    \\[.5cm]
    \includegraphics[width = 3in, height = 2.0in]{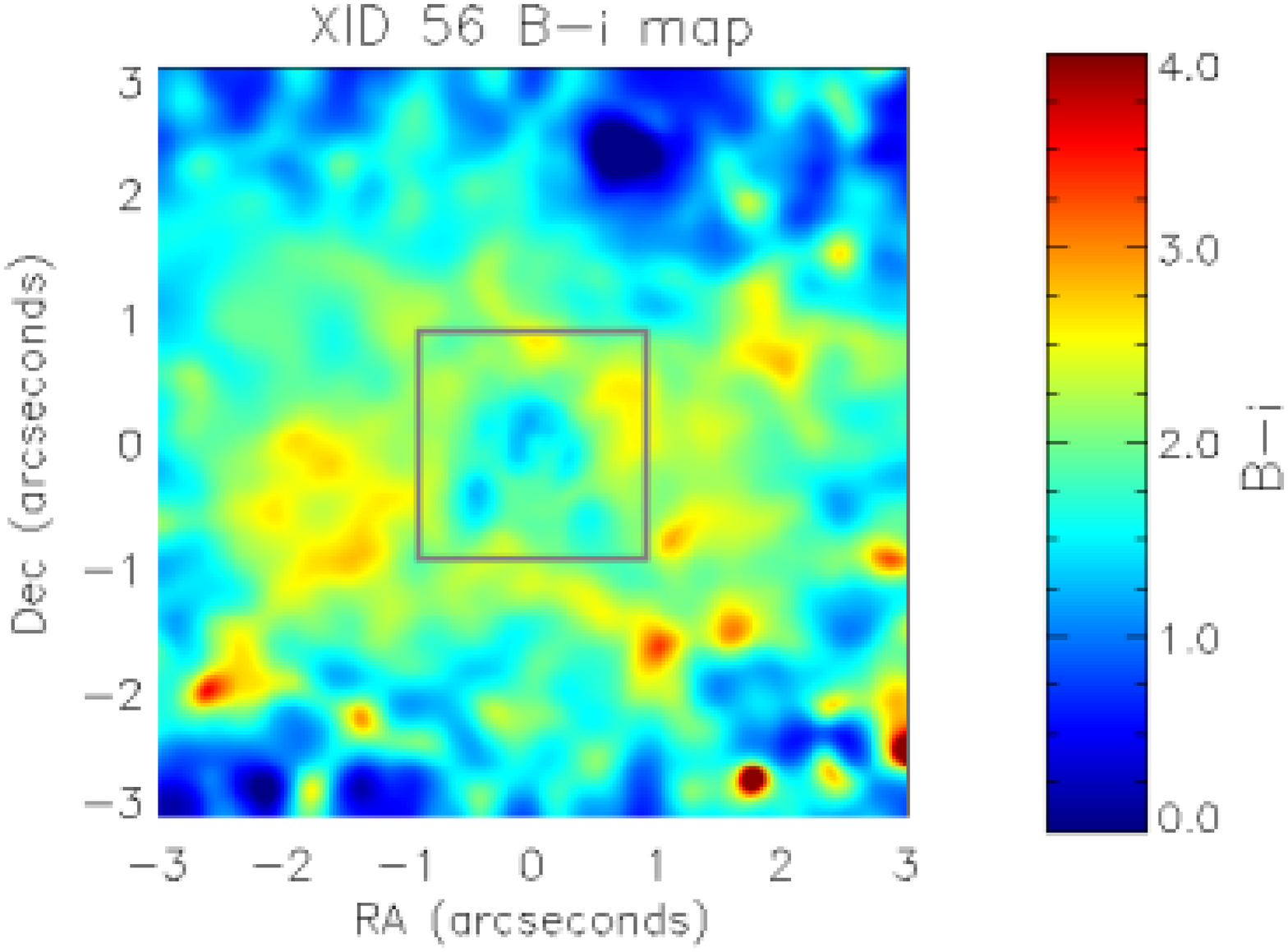}
    \includegraphics[width = 3in, height = 2.0in]{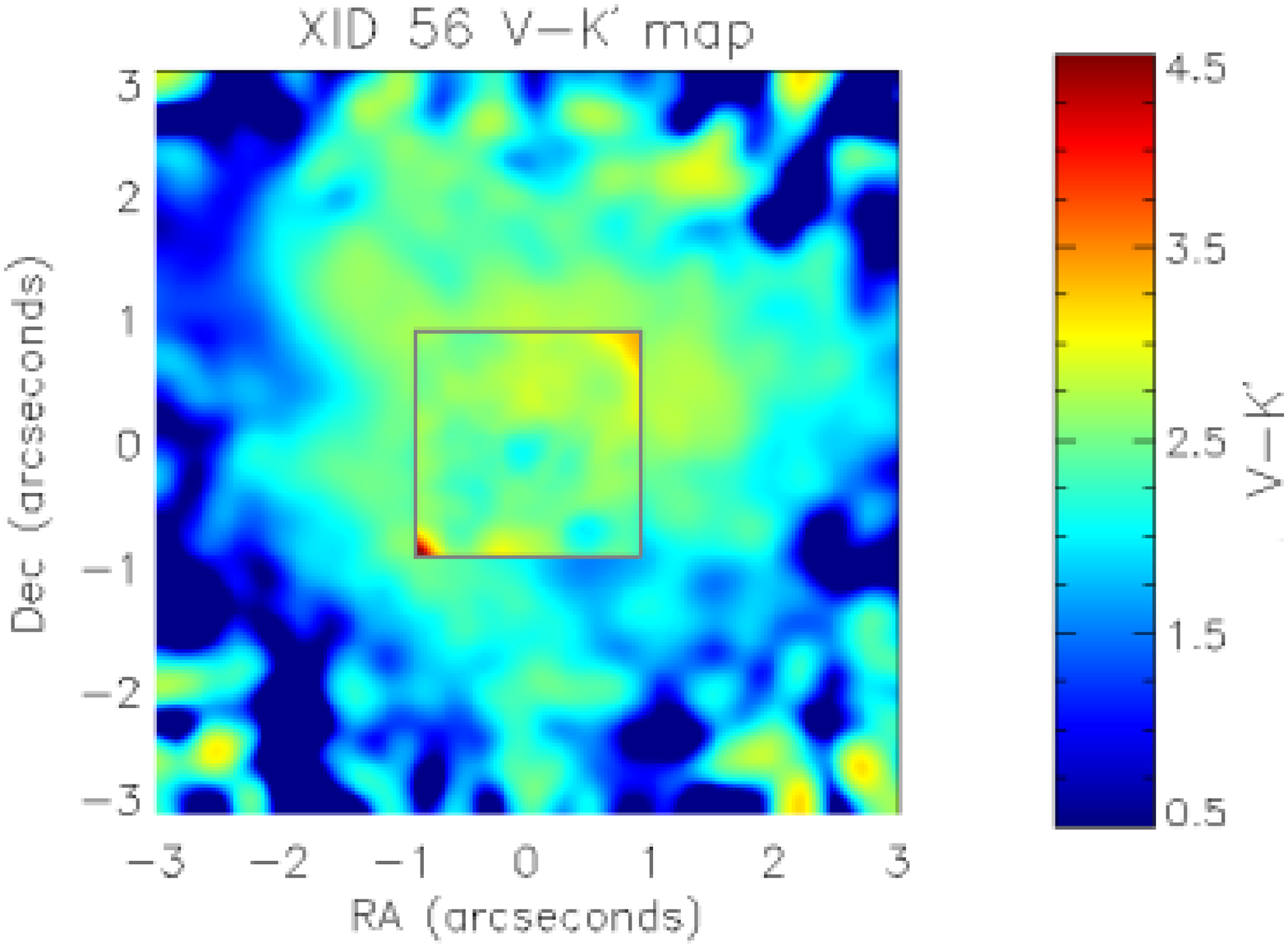}
  \end{center}
\end{figure}

\begin{figure}
  \caption{Tiled magery, GALFIT models, and colormaps for XID 155, as in Figure 3.}
  \begin{center}
    \includegraphics[width=5in,height=3.0in]{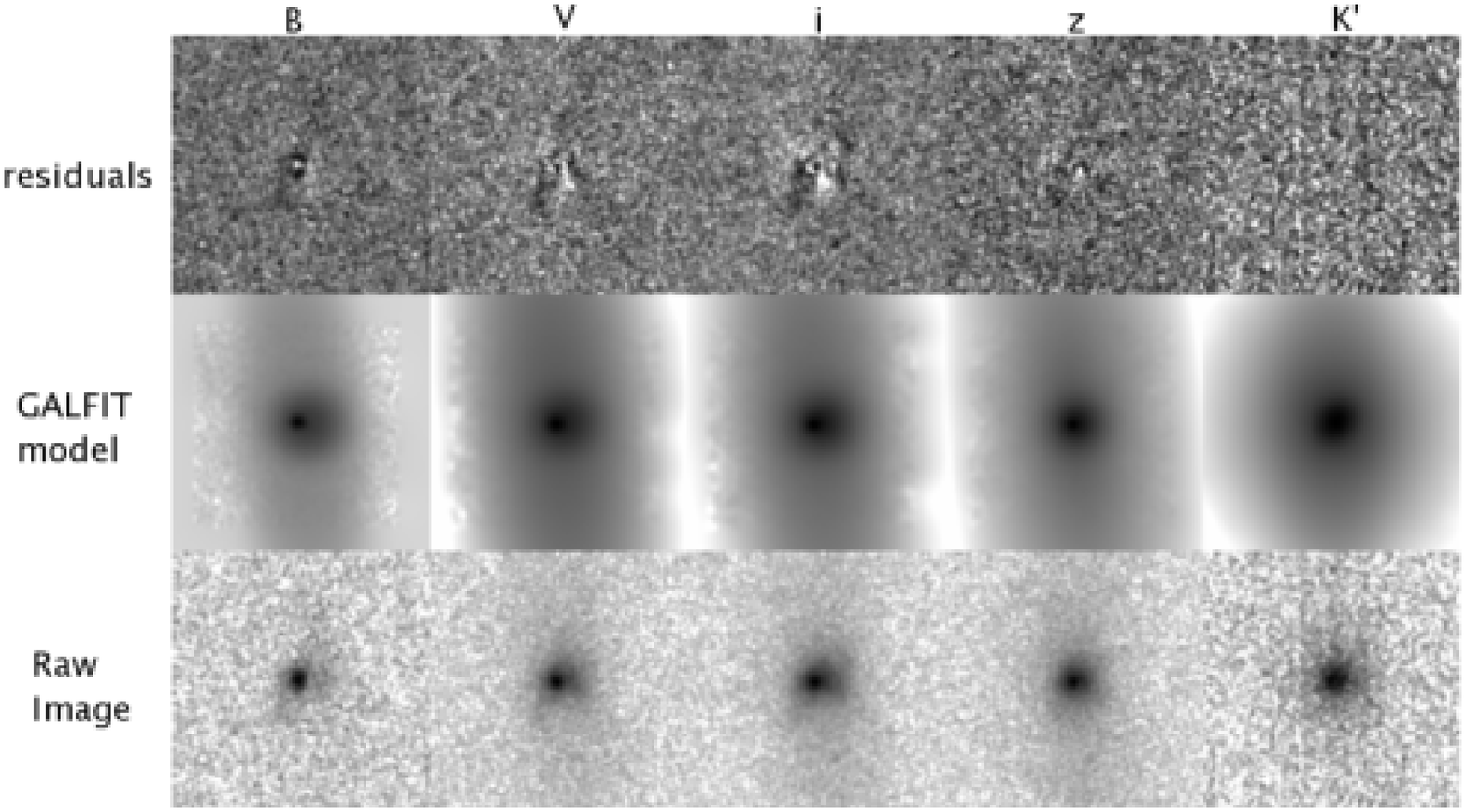}
    \\[.5cm]
    \includegraphics[width = 3in, height = 2.0in]{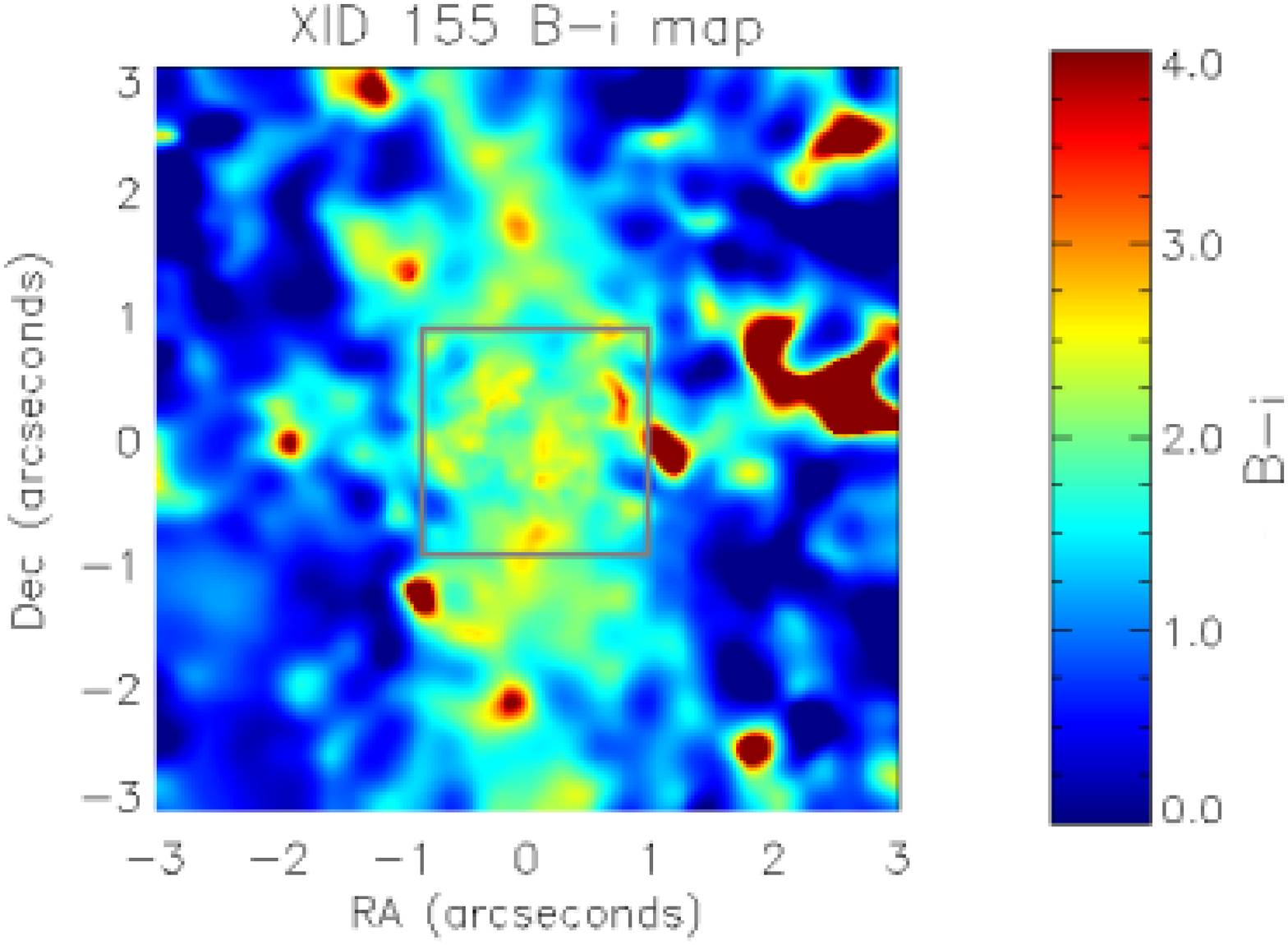}
    \includegraphics[width = 3in, height = 2.0in]{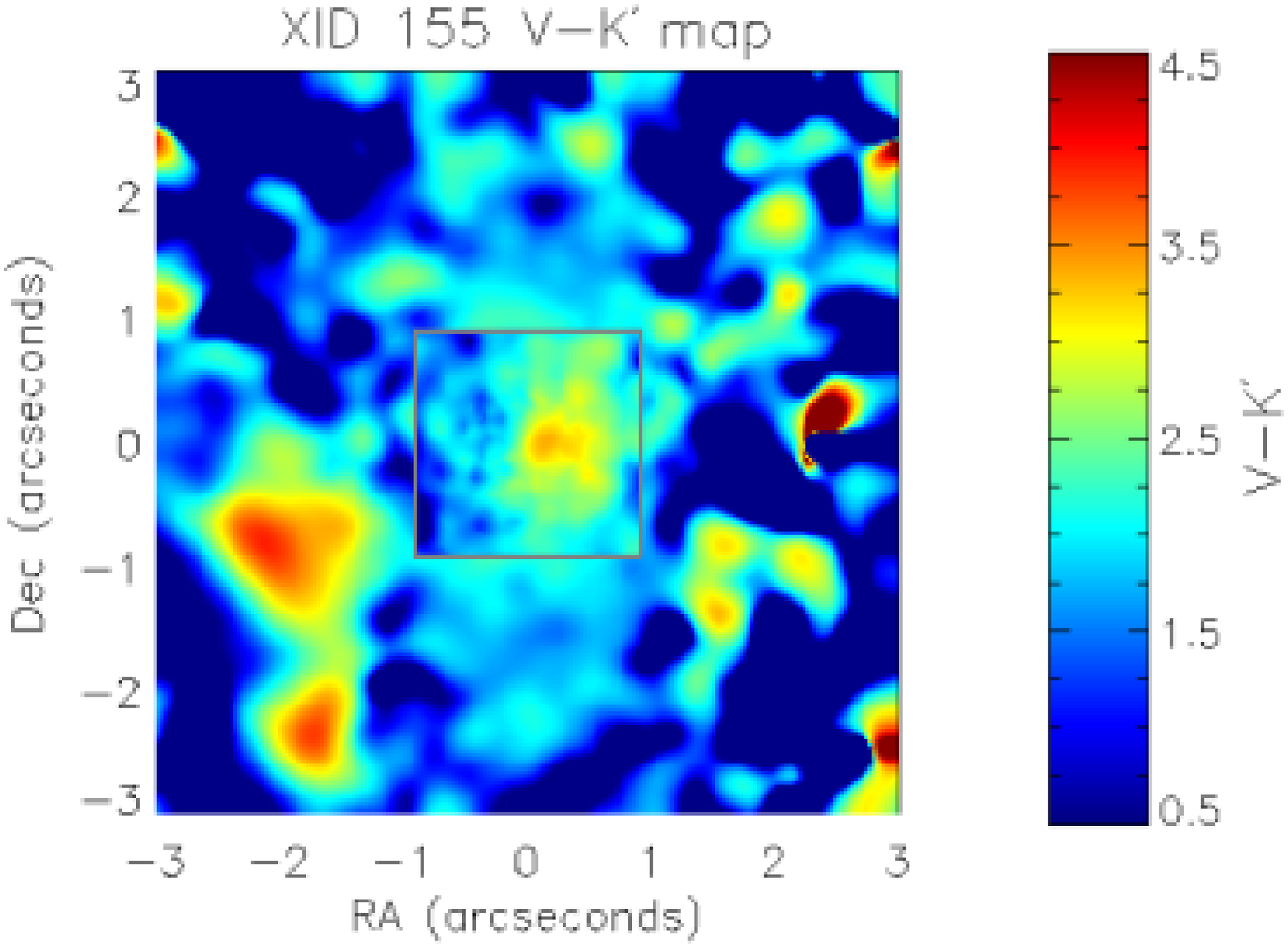}
  \end{center}
\end{figure}
\clearpage
\begin{figure}
  \caption{Tiled imagery, GALFIT models, and colormaps for XID 266, as in Figure 3.}
  \begin{center}
    \includegraphics[width=5in,height=3.0in]{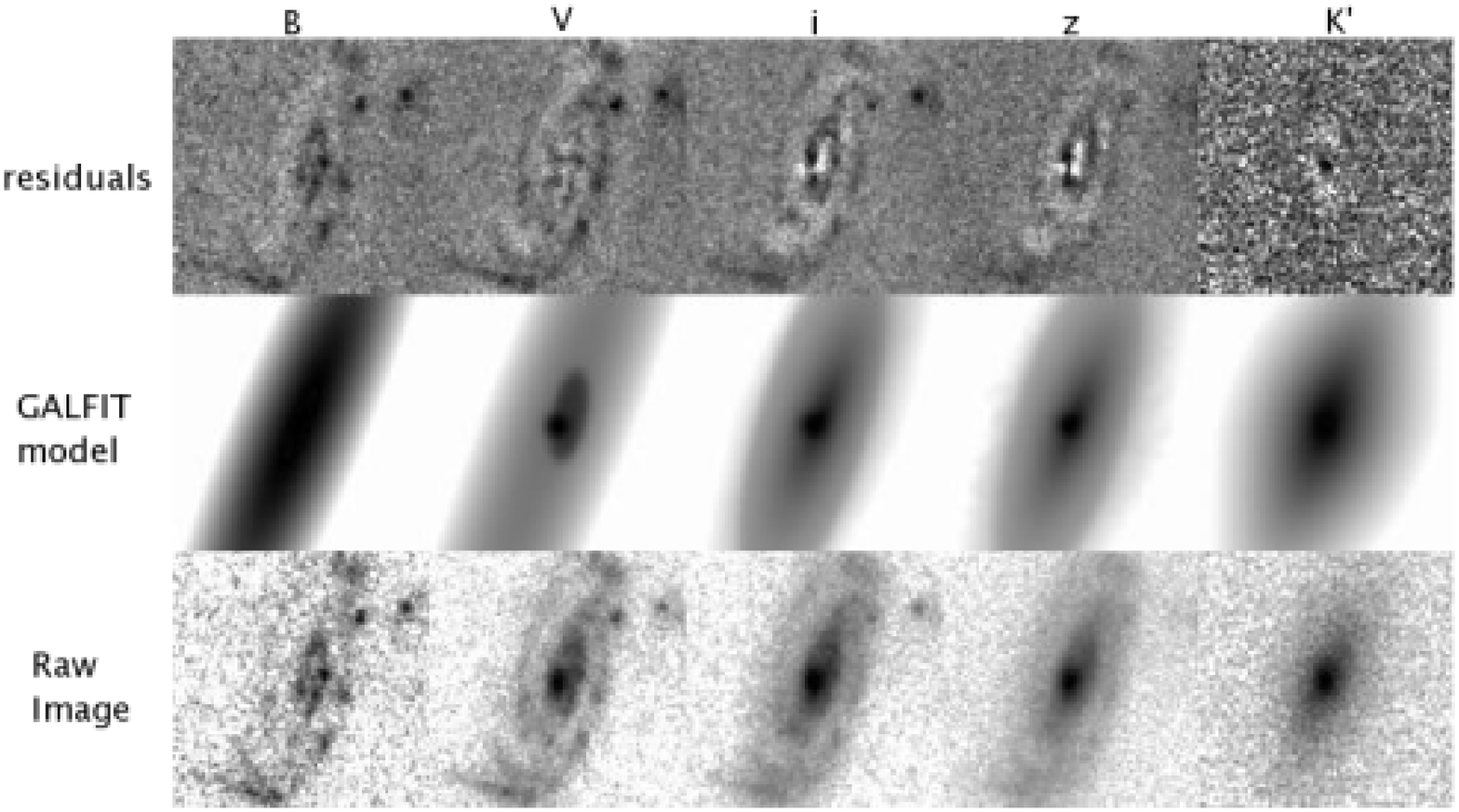}
    \\[.5cm]
    \includegraphics[width = 3in, height = 2.0in]{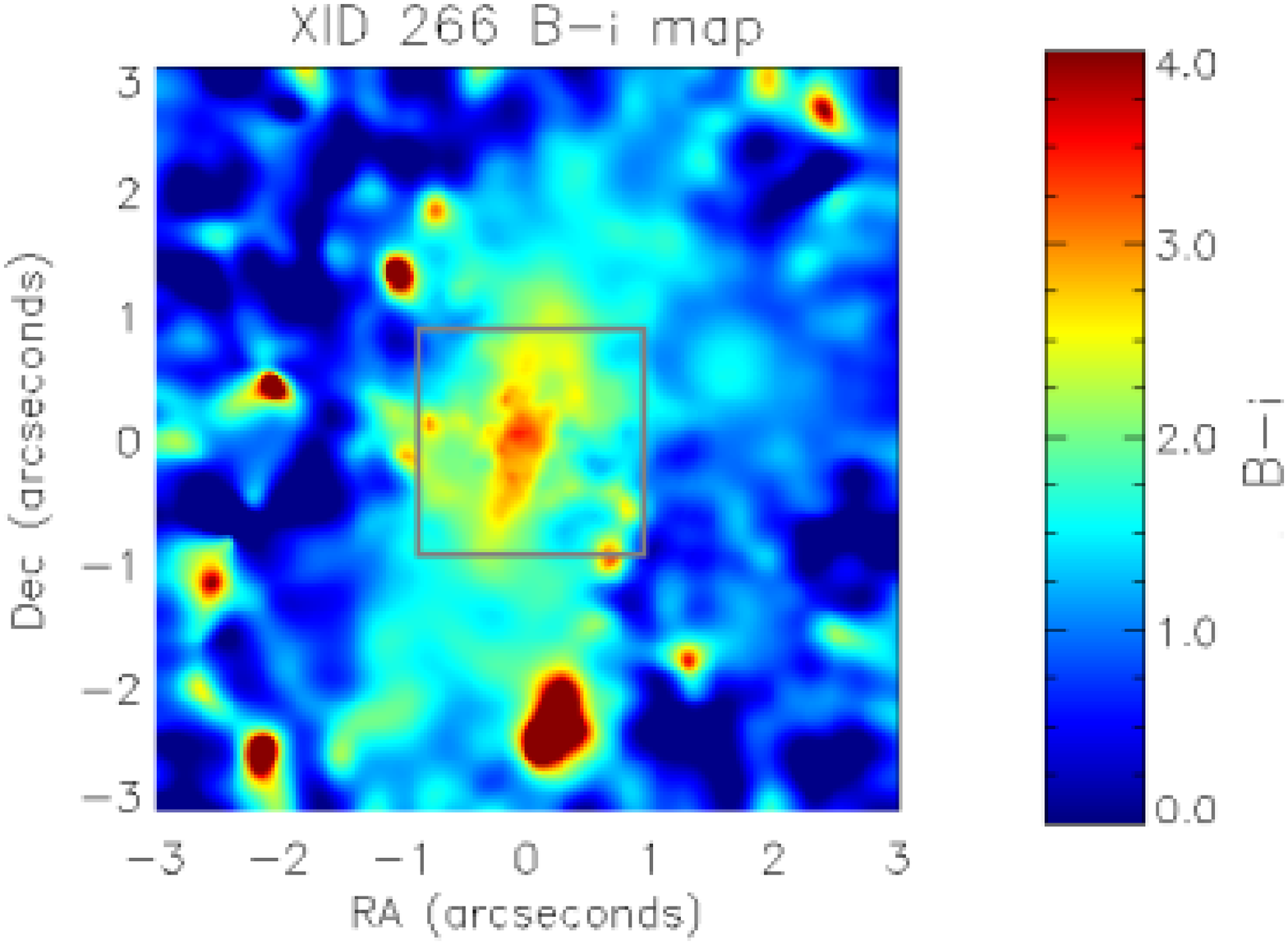}
    \includegraphics[width = 3in, height = 2.0in]{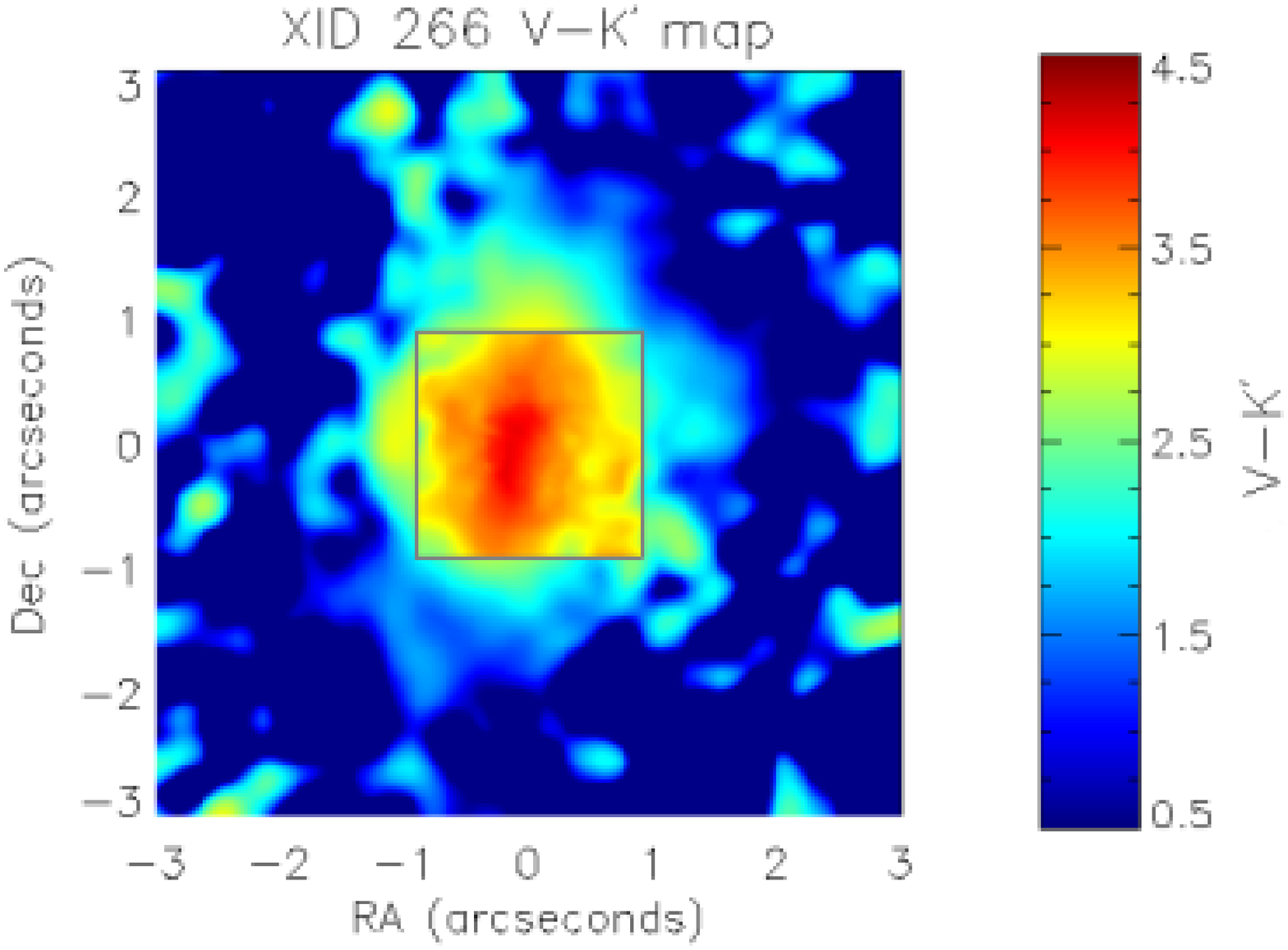}
  \end{center}
\end{figure}

\begin{figure}
  \caption{Tiled imagery, GALFIT models, and colormaps for XID 83, as in Figure 3.}
  \begin{center}
    \includegraphics[width=5in,height=3.0in]{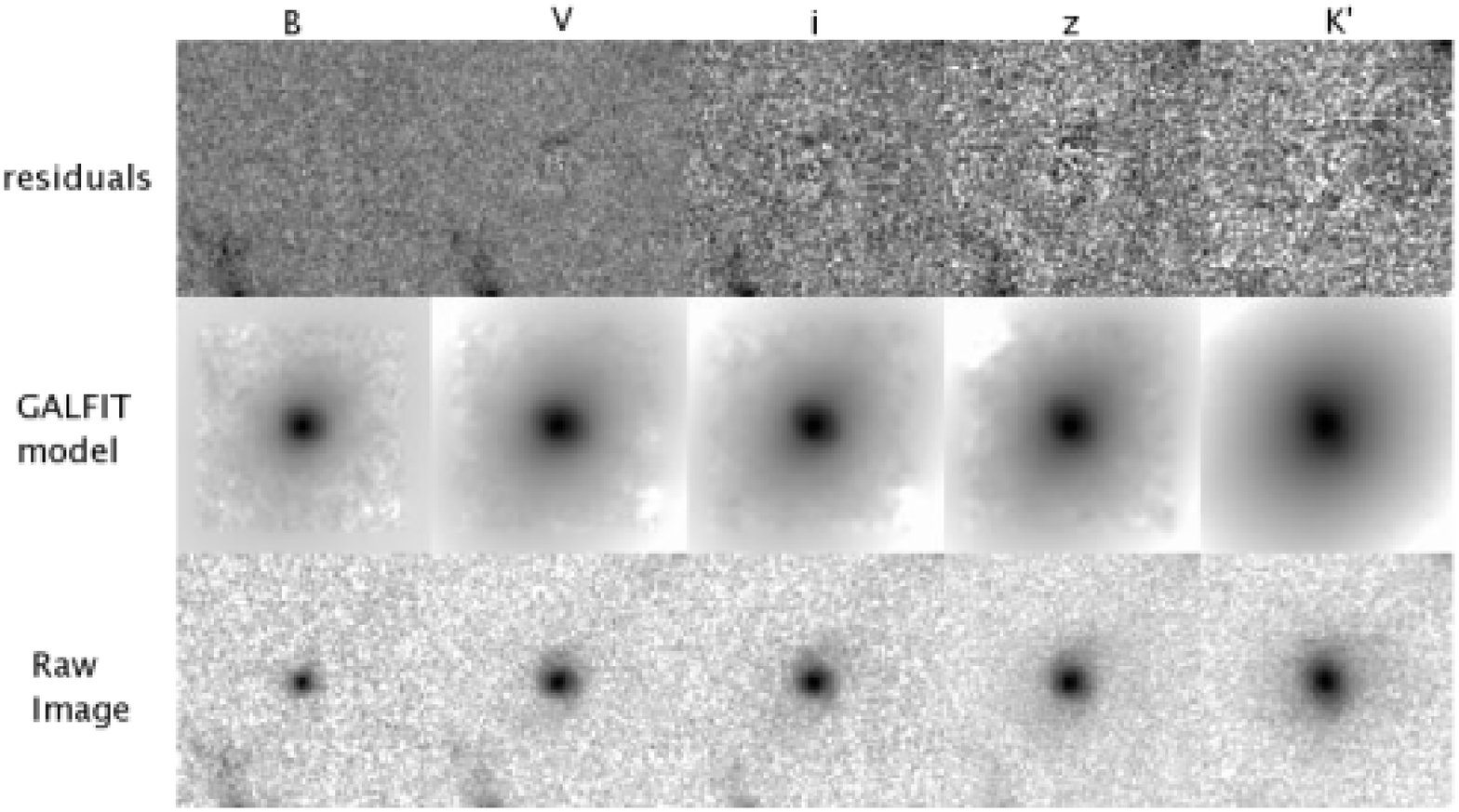}
    \\[.5cm]
    \includegraphics[width = 3in, height = 2.0in]{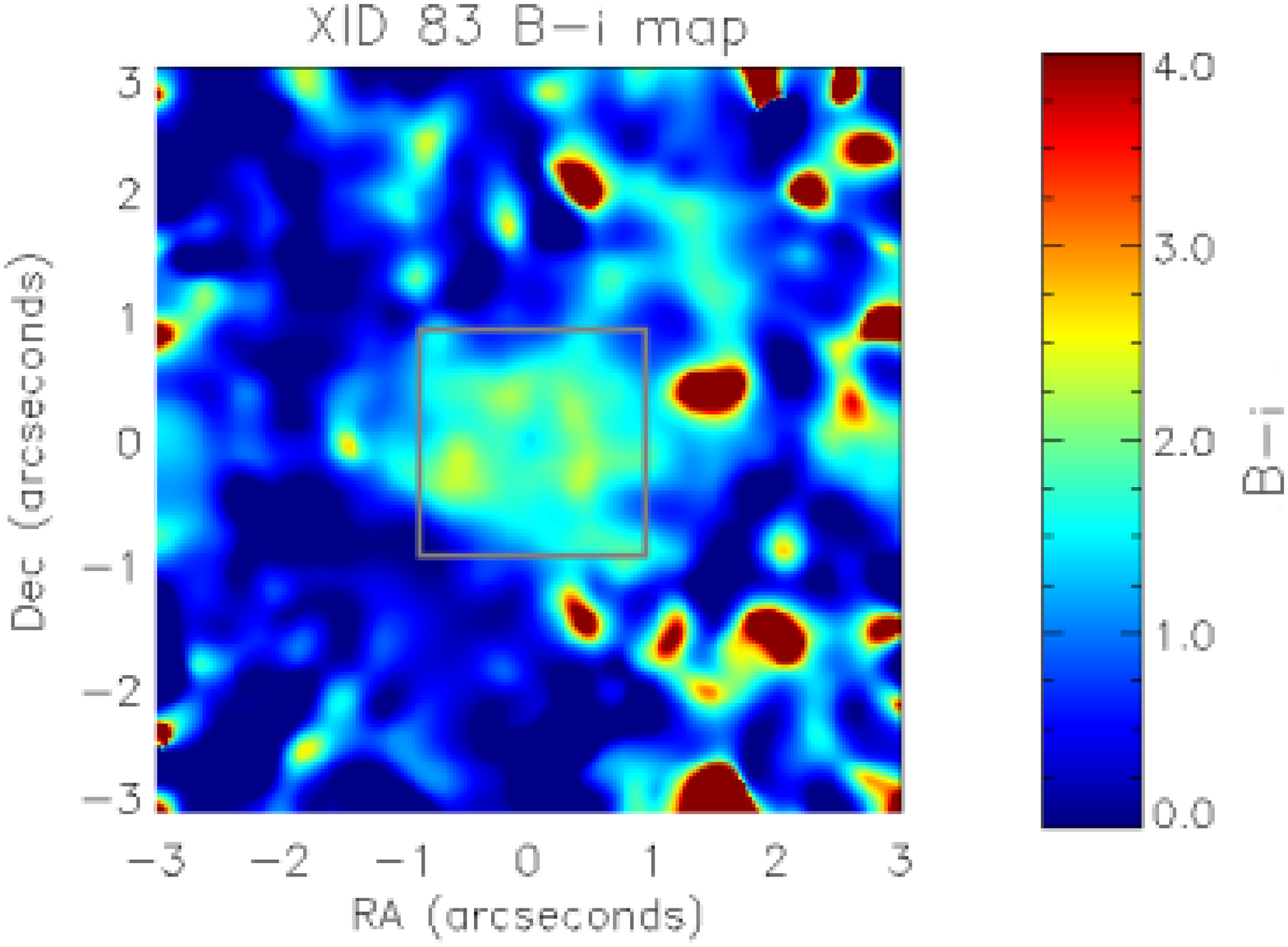}
    \includegraphics[width = 3in, height = 2.0in]{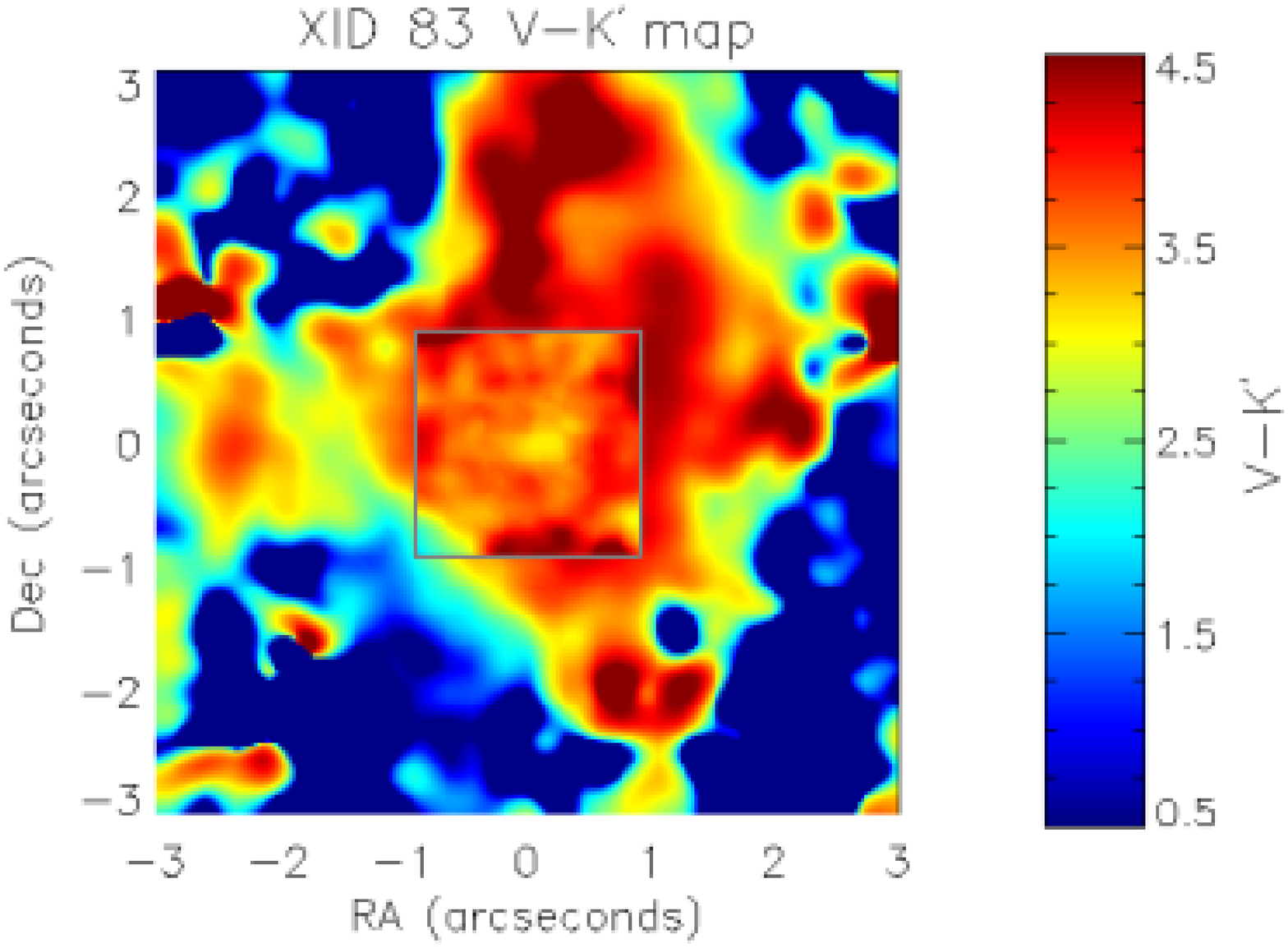}
  \end{center}
\end{figure}
\clearpage
\begin{figure}
  \caption{Tiled imagery, GALFIT models, and colormaps for XID 536, as in Figure 3.}
  \begin{center}
    \includegraphics[width=5in,height=3.0in]{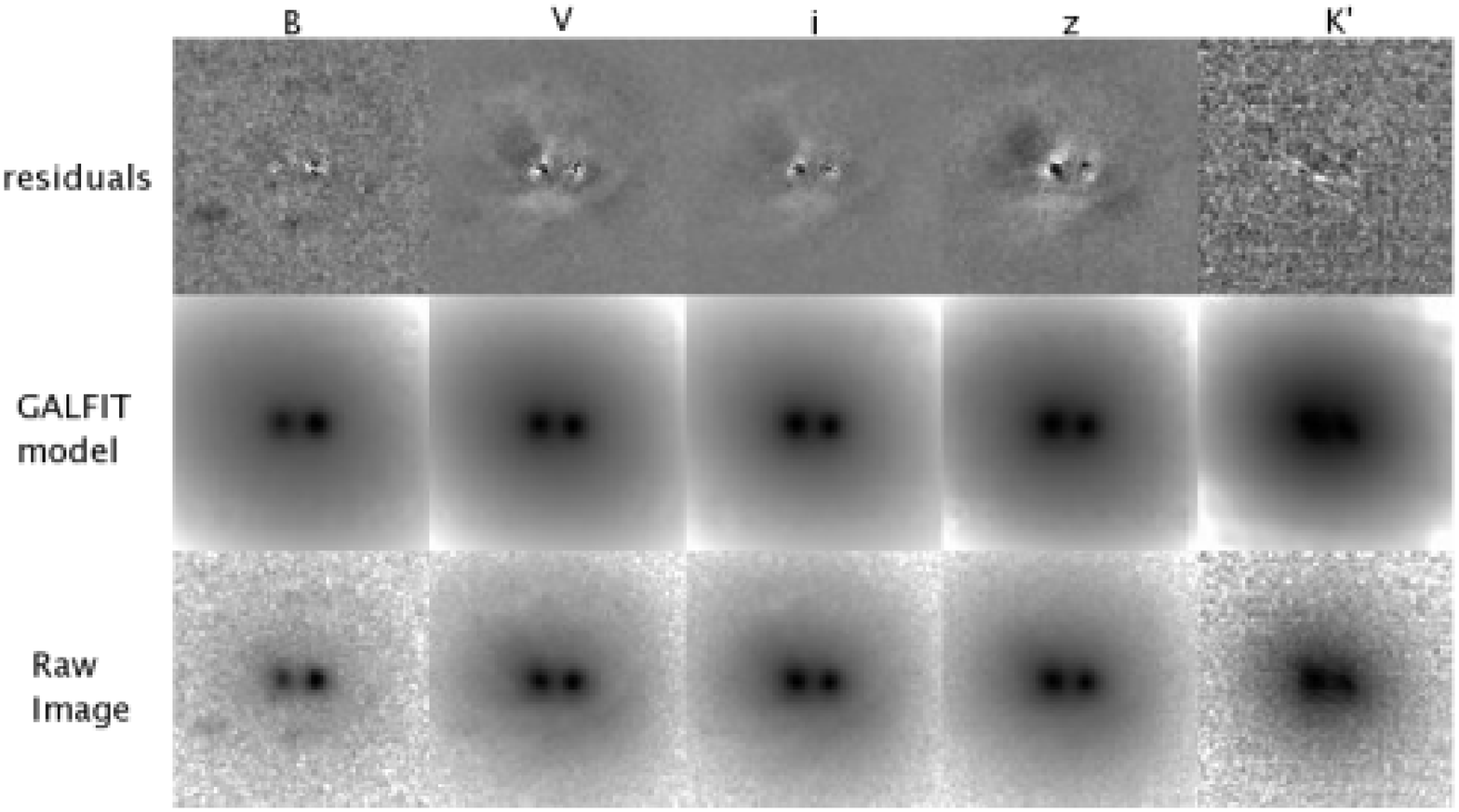}
    \\[.5cm]
    \includegraphics[width = 3in, height = 2.0in]{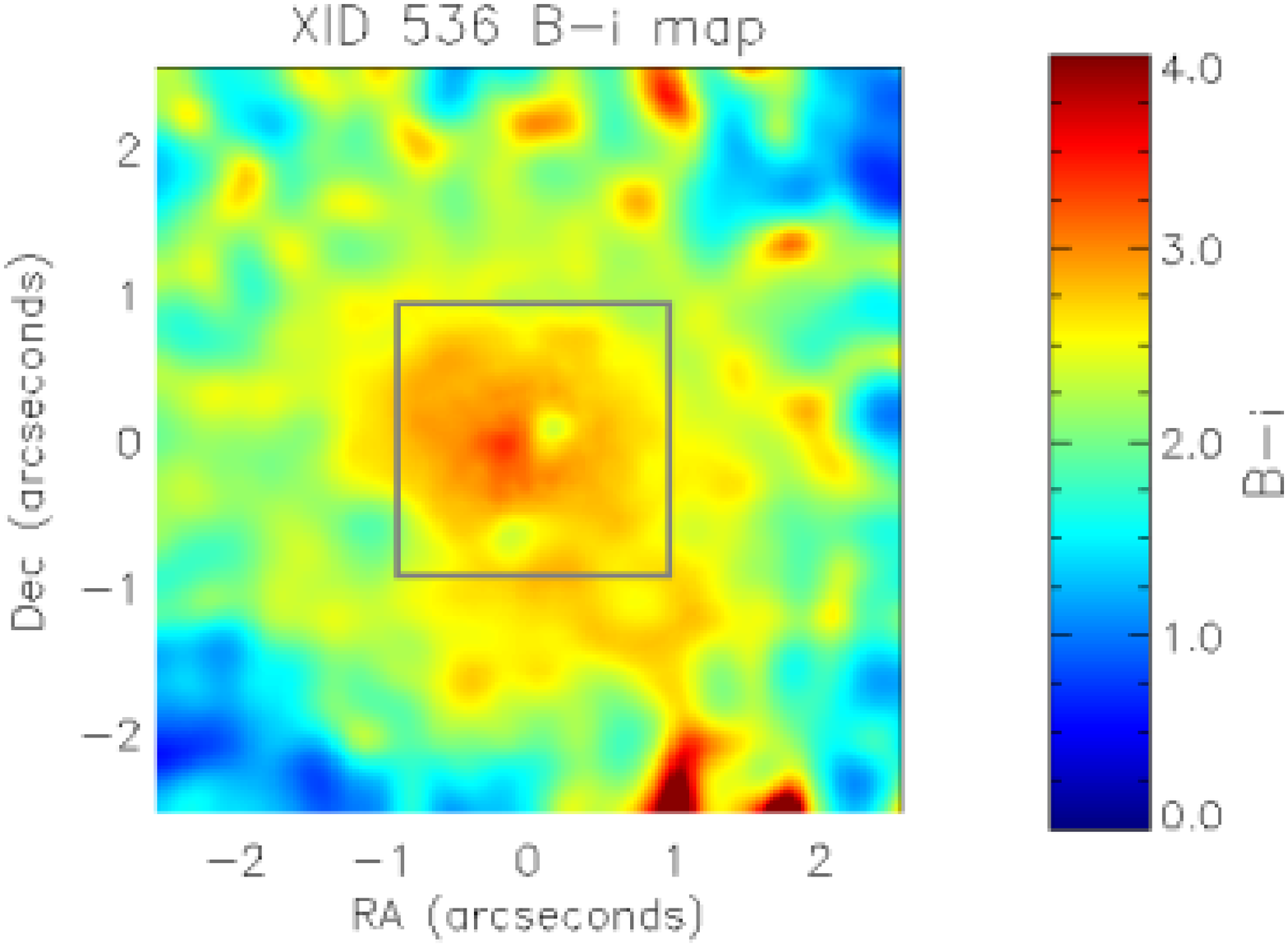}
    \includegraphics[width = 3in, height = 2.0in]{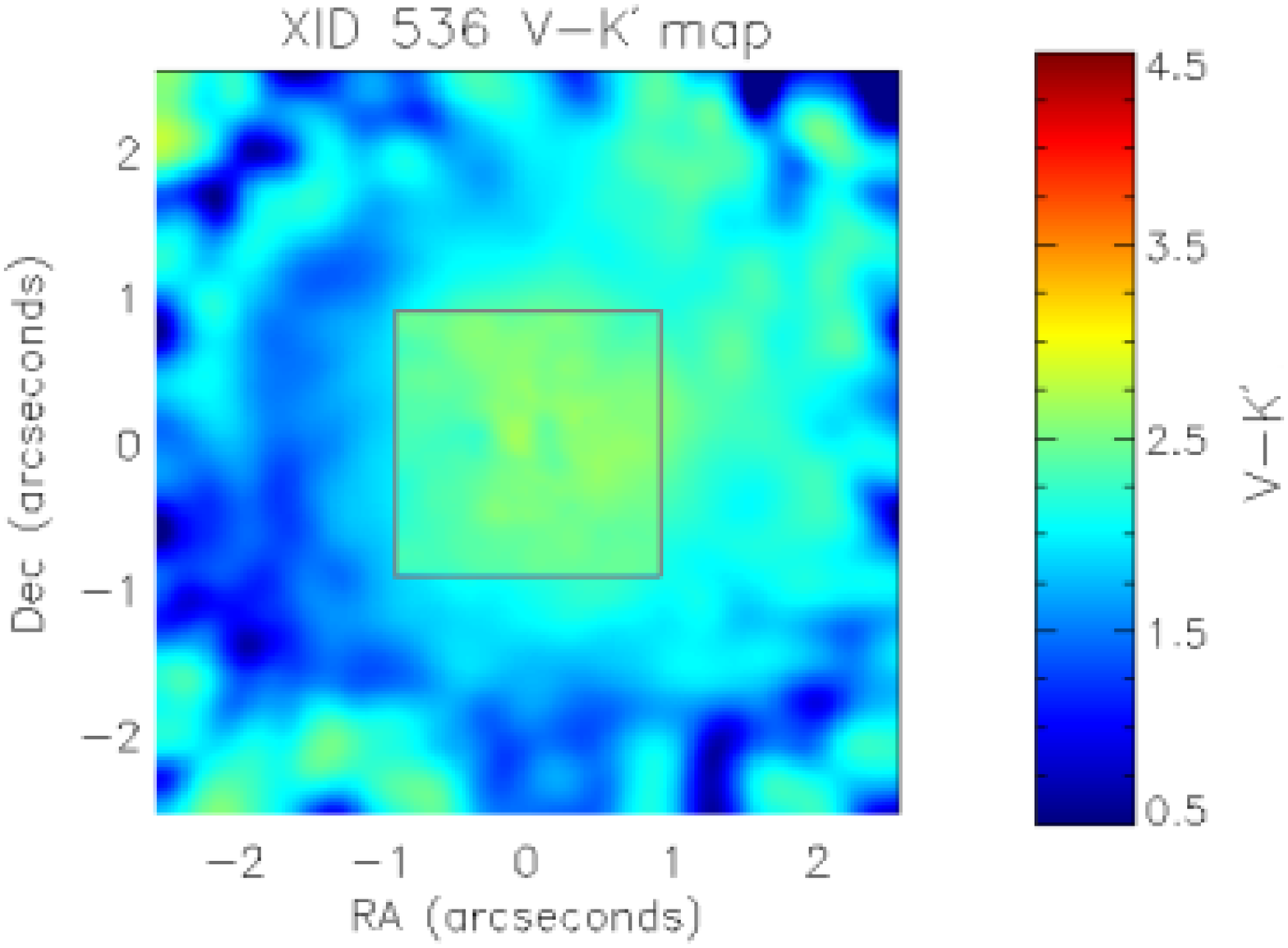}
  \end{center}
\end{figure}

\begin{figure}
  \caption{Tiled imagery, GALFIT models, and colormaps for XID 594, as in Figure 3.}
  \begin{center}
    \includegraphics[width=5in,height=3.0in]{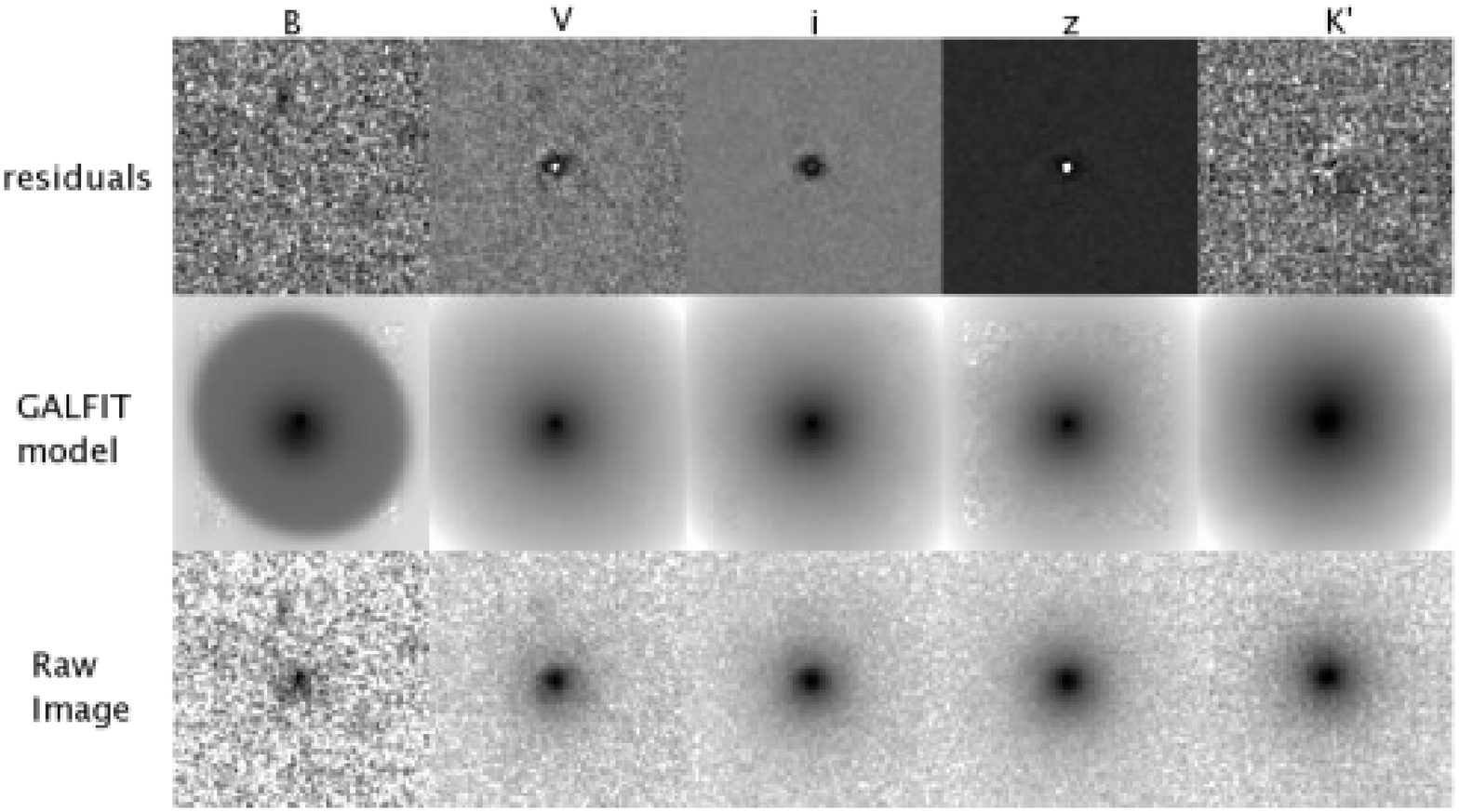}
    \\[.5cm]
    \includegraphics[width = 3in, height = 2.0in]{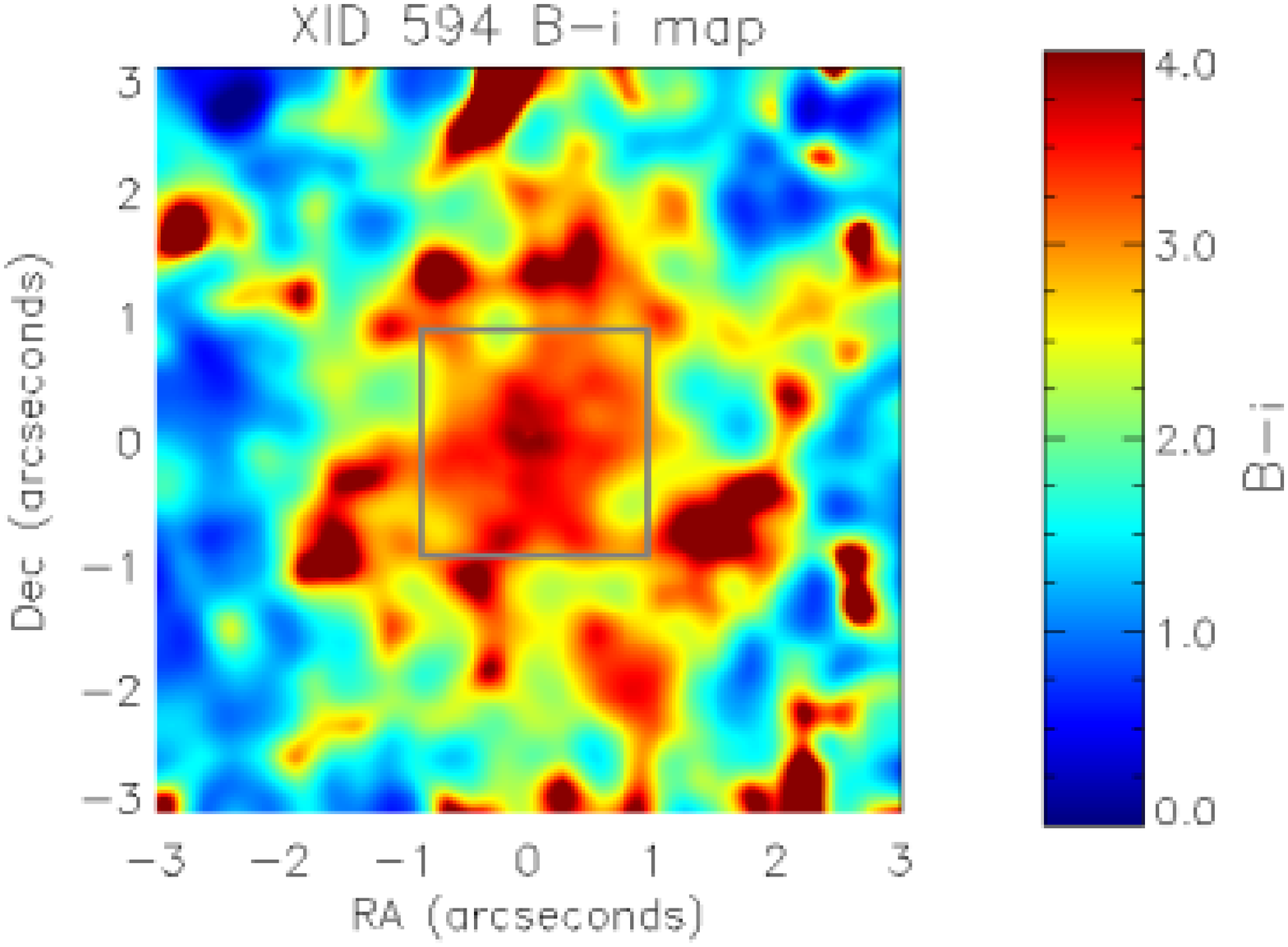}
    \includegraphics[width = 3in, height = 2.0in]{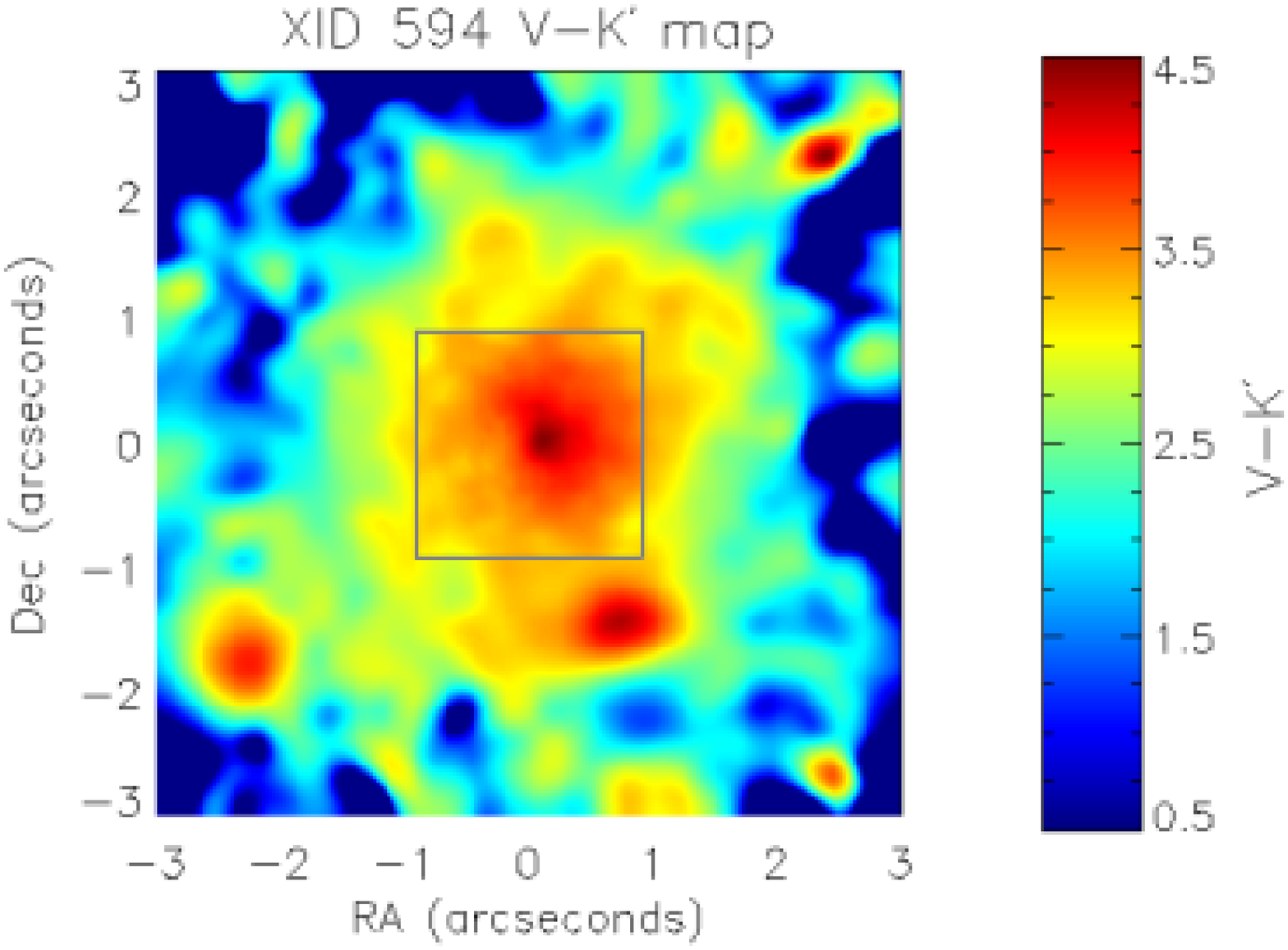}
  \end{center}
\end{figure}
\clearpage
\begin{figure}
  \caption{Tiled imagery, GALFIT models, and colormaps for XID 15, as in Figure 3.}
  \begin{center}
    \includegraphics[width=5in,height=3.0in]{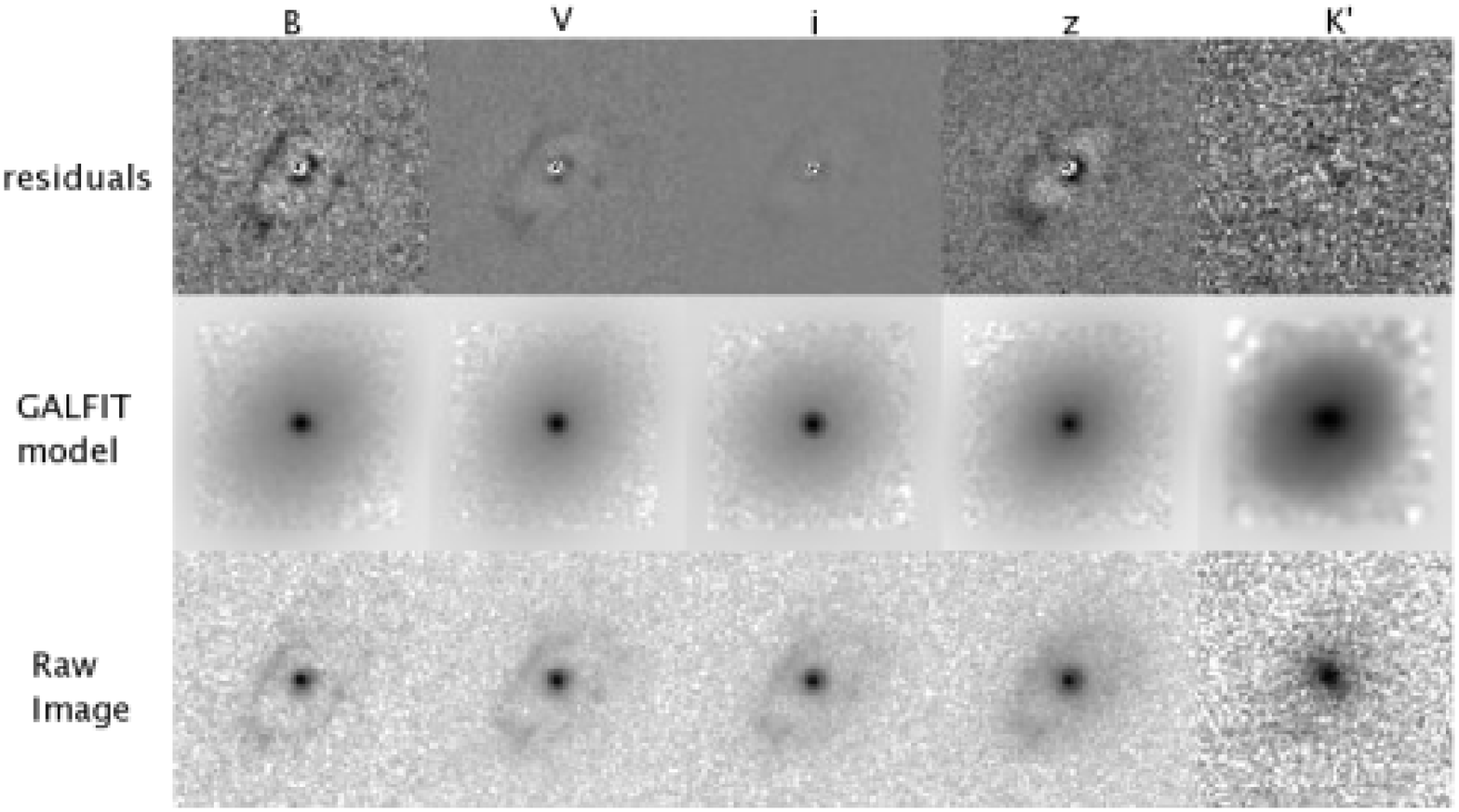}
    \\[.5cm]
    \includegraphics[width = 3in, height = 2.0in]{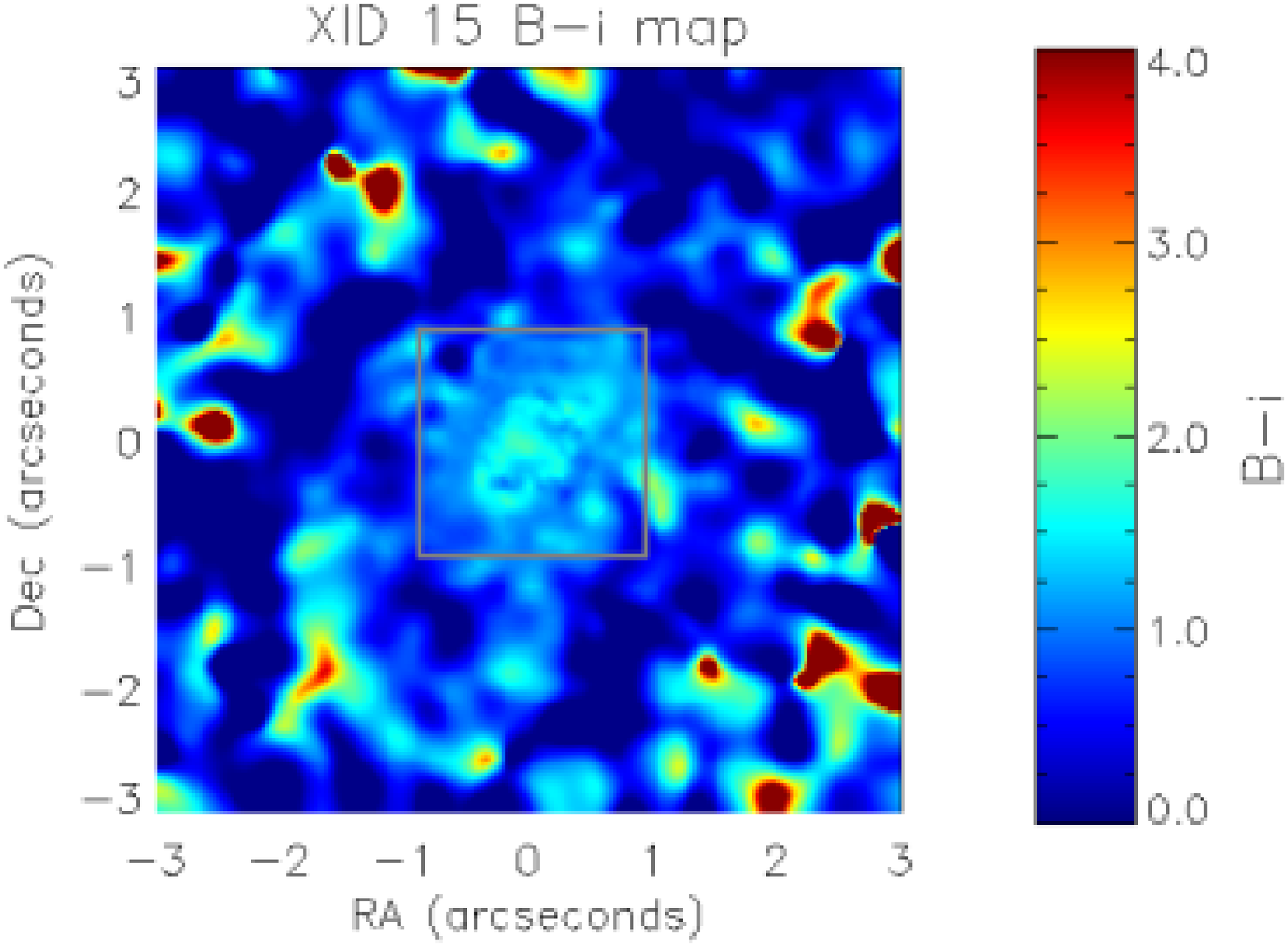}
    \includegraphics[width = 3in, height = 2.0in]{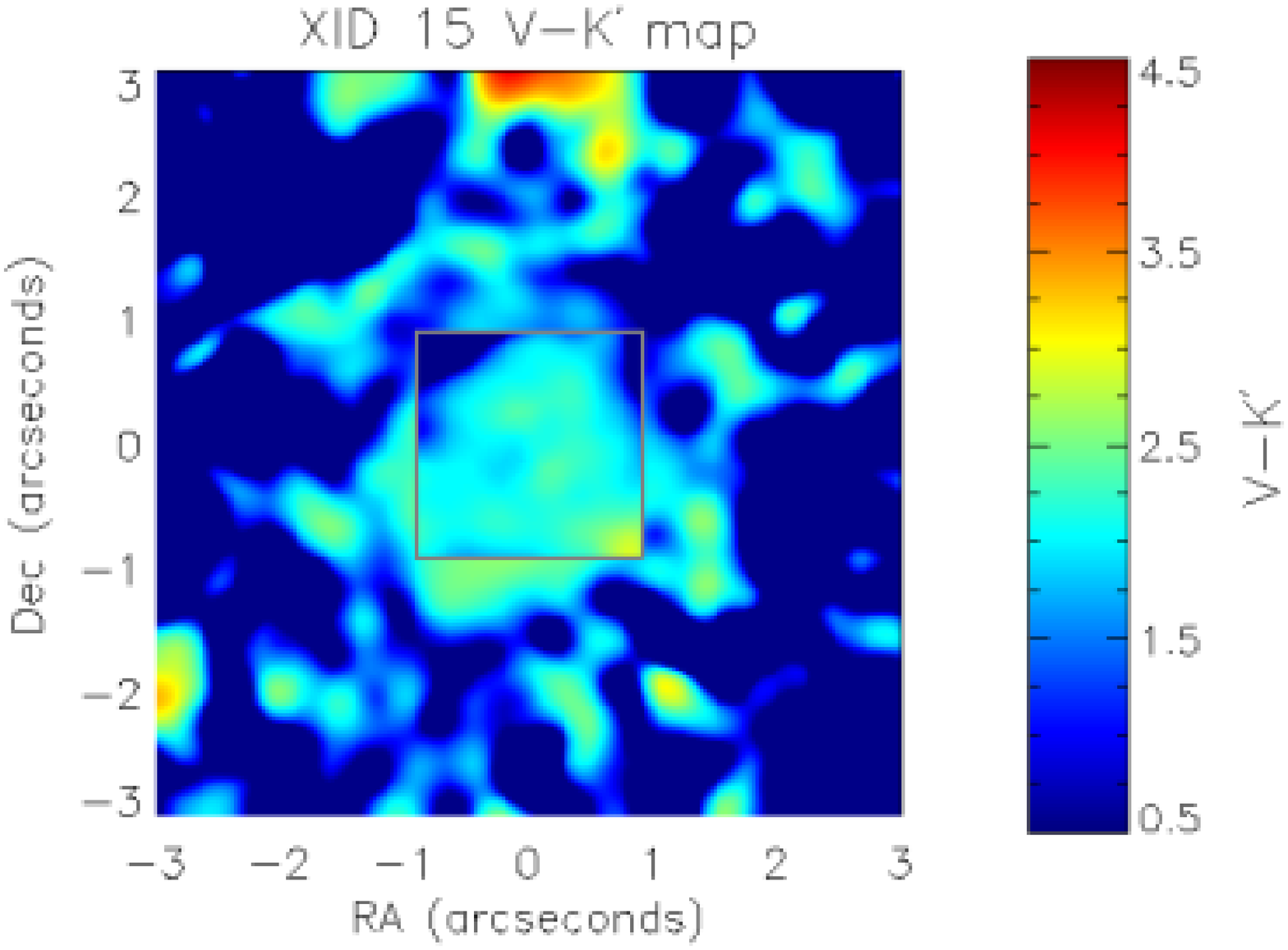}
  \end{center}
\end{figure}

\begin{figure}
  \caption{Tiled imagery, GALFIT models, and colormaps for XID 32, as in Figure 3.}
  \begin{center}
    \includegraphics[width=5in,height=3.0in]{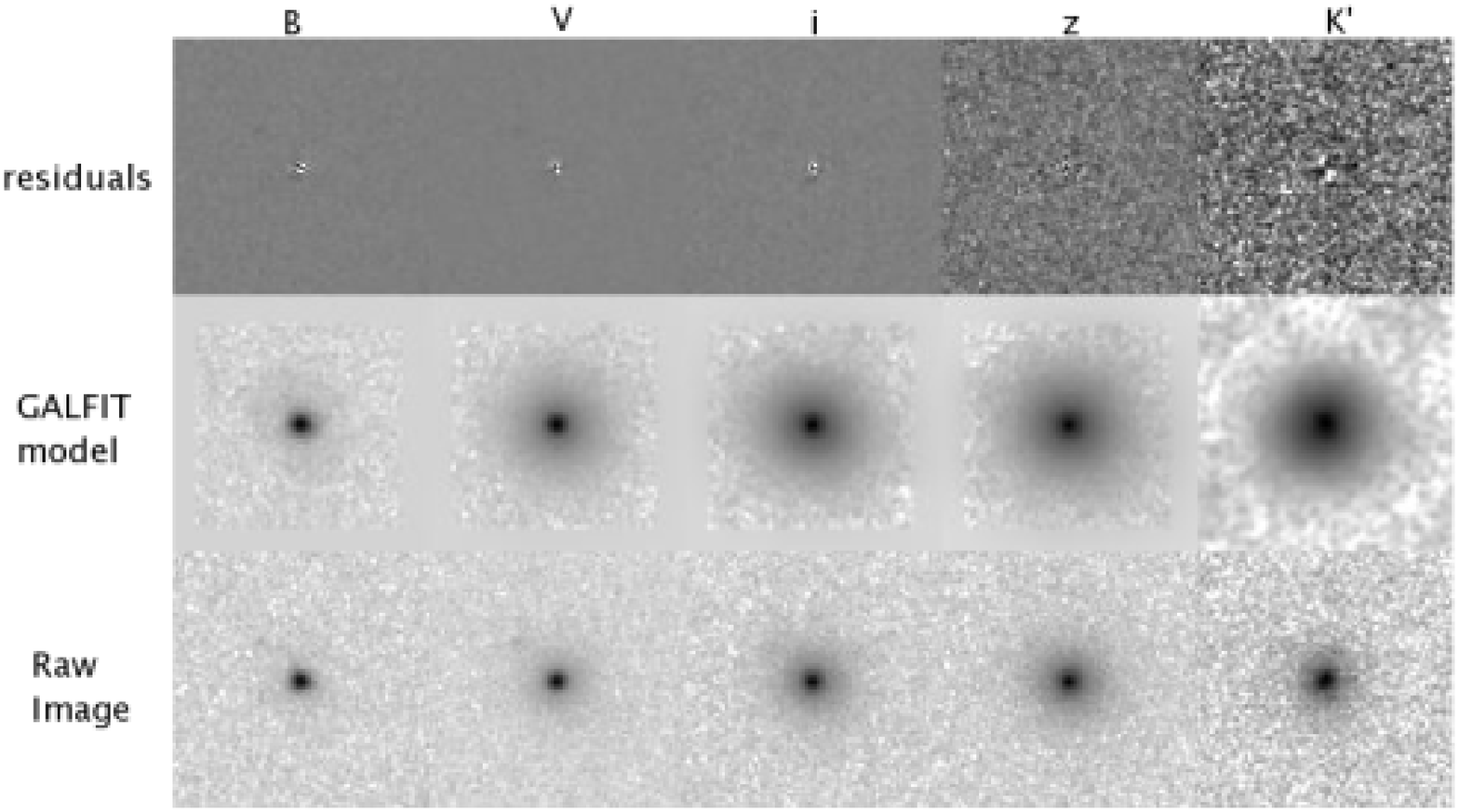}
    \\[.5cm]
    \includegraphics[width = 3in, height = 2.0in]{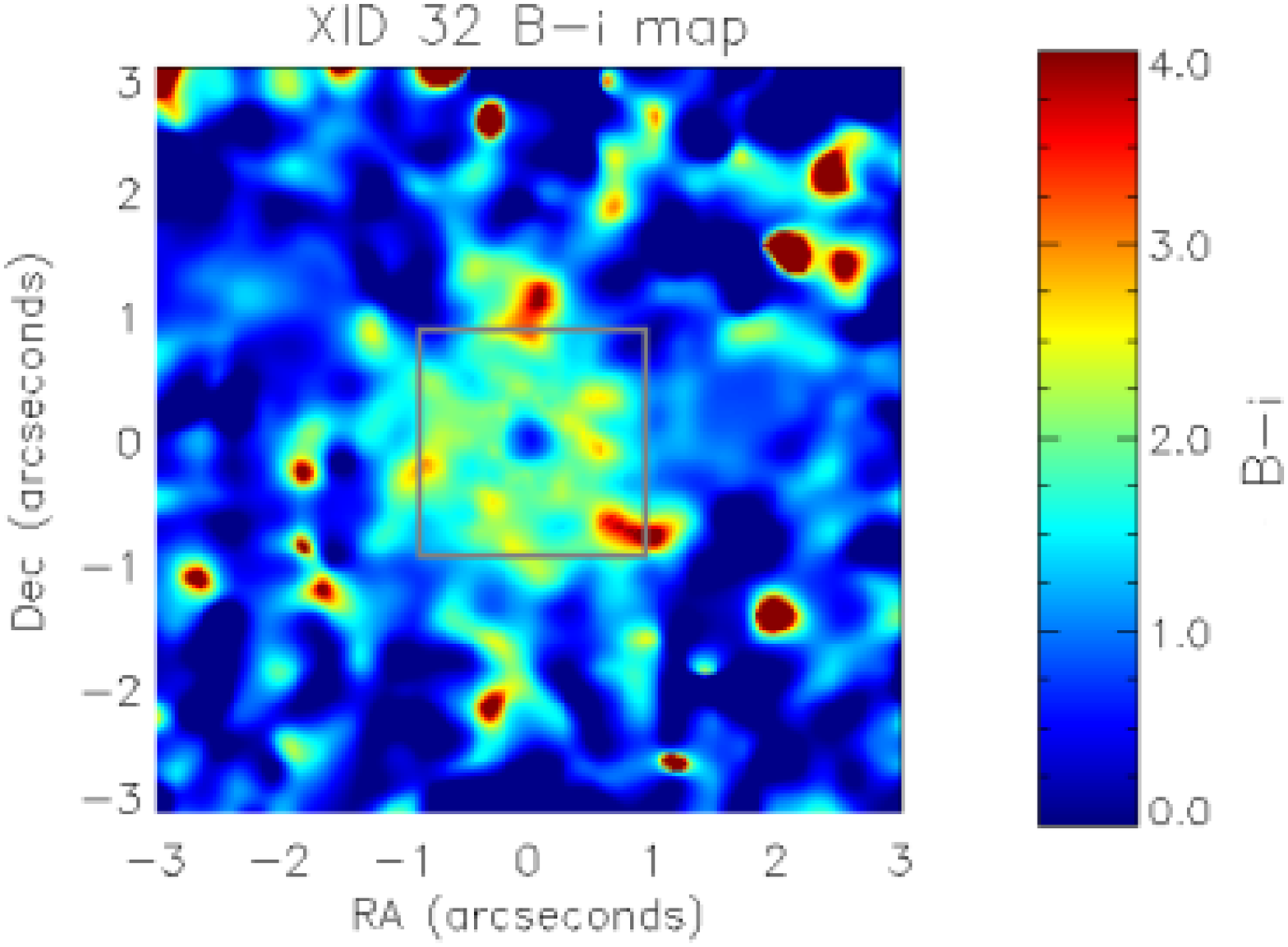}
    \includegraphics[width = 3in, height = 2.0in]{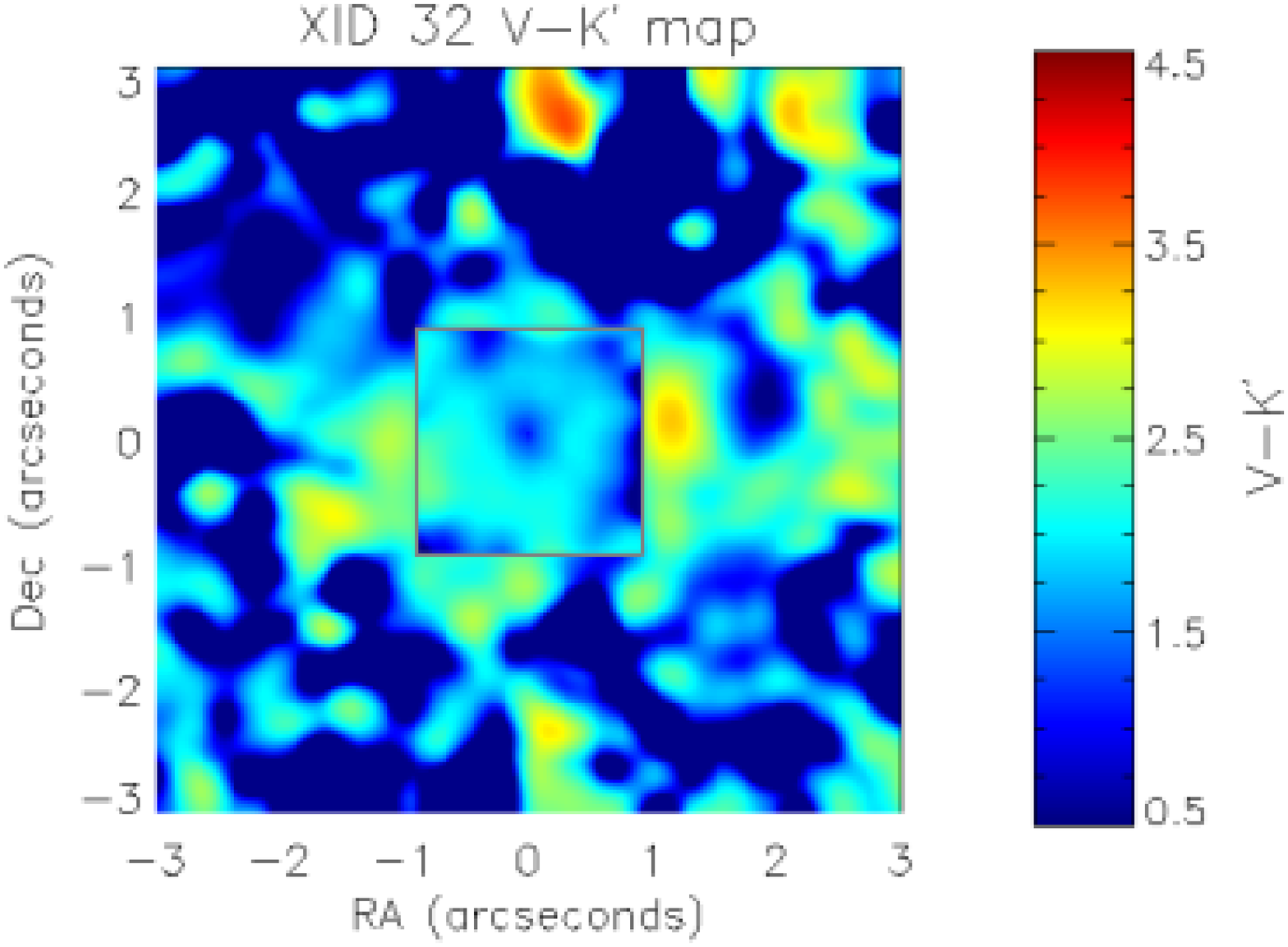}
  \end{center}
\end{figure}
\clearpage
\begin{figure}
  \caption{Measured SEDs and best-fit Bruzual \& Charlot spectra for two cases:  (left) the SED from the fourth ring of XID 155 and (right) the SED from the fifth ring of XID 266.  The plotted spectra represent fits to the SED when the age is fixed at different values and the dust extinction and tau parameter are allowed to float.  The solid line in both cases is the best-fit model to the SED when all parameters are allowed to float.  Note that ages below 1 Gyr are largely ruled out in the left panel, but not so in the right panel, where models between 500 Myr and 4 Gyr are within the photometric error.  The SED in the left panel is representative of non-degenerate cases and the SED in the right panel is representative of degenerate cases.}
  \begin{center}
    \includegraphics[width = 3in, height = 2.0in]{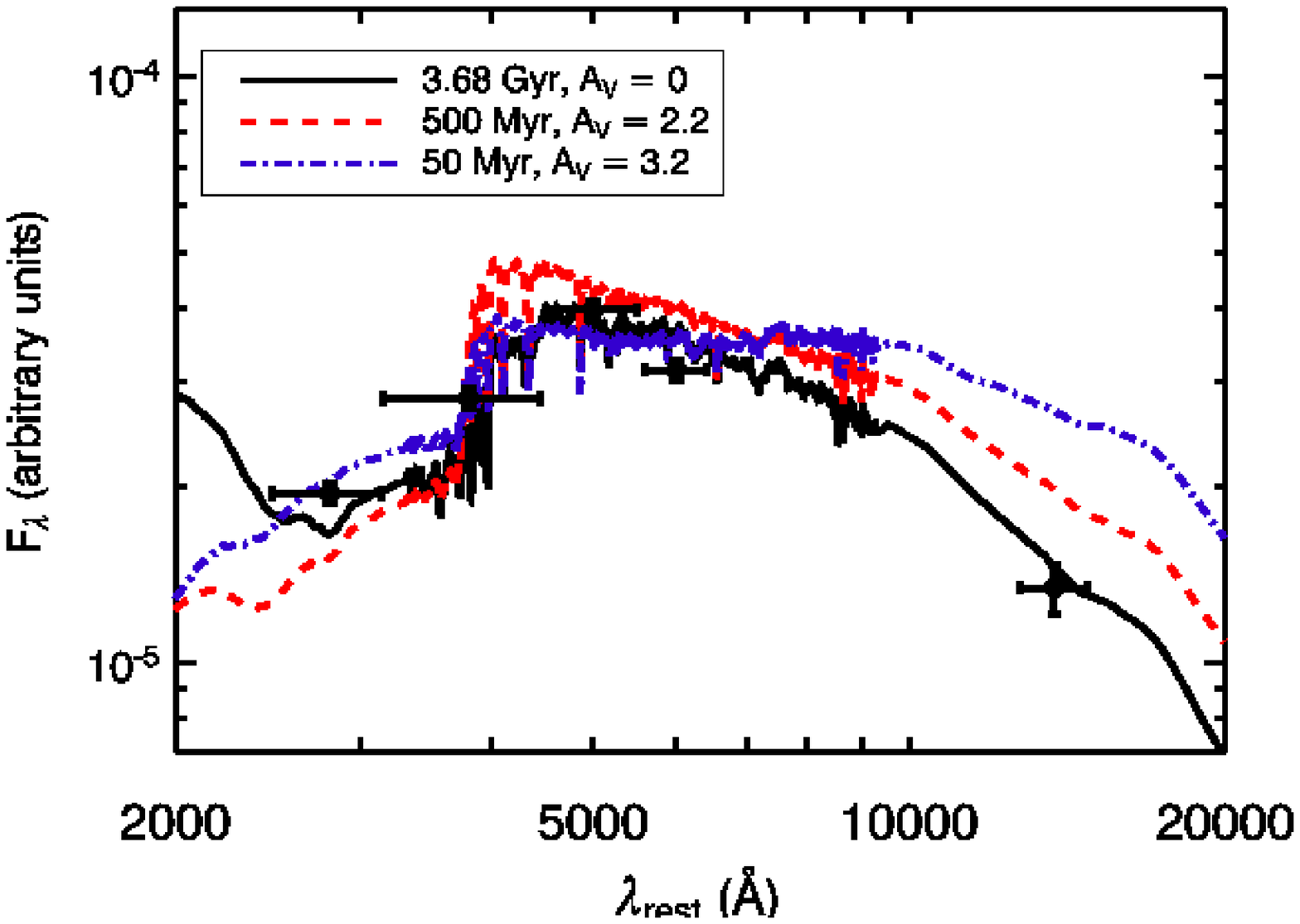}
    \includegraphics[width = 3in, height = 2.0in]{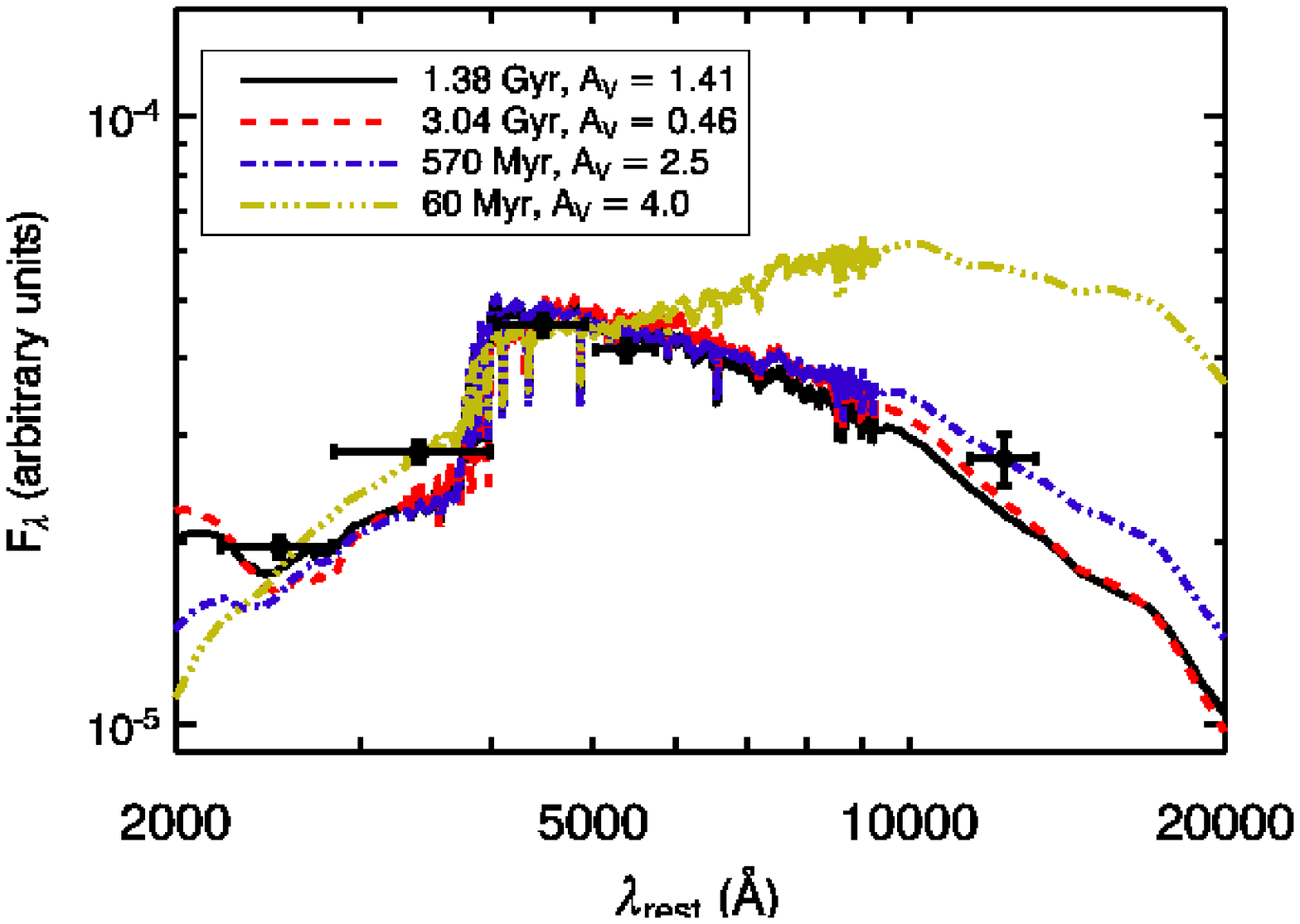}
  \end{center}
\end{figure}

\begin{figure}
  \caption{Radial profiles and fractions of different stellar populations plotted vs. radius for XID 56.  The radial profile plots the surface brightness for each band (AB magnitudes per square arcsecond) as a function of radius.  The fractions of all stars formed by mass is plotted in a log scale versus the radius for three populations:   Young (age $<$ 100 Myr), Intermediate (100 Myr $<$ age $<$ 1 Gyr), and Old (age $>$ 1 Gyr).  The young population is denoted by blue asterisks, the intermediate age population is denoted by green triangles, and the old population is denoted by red diamonds.}
  \begin{center}
    \includegraphics[width = 3in, height = 2.0in]{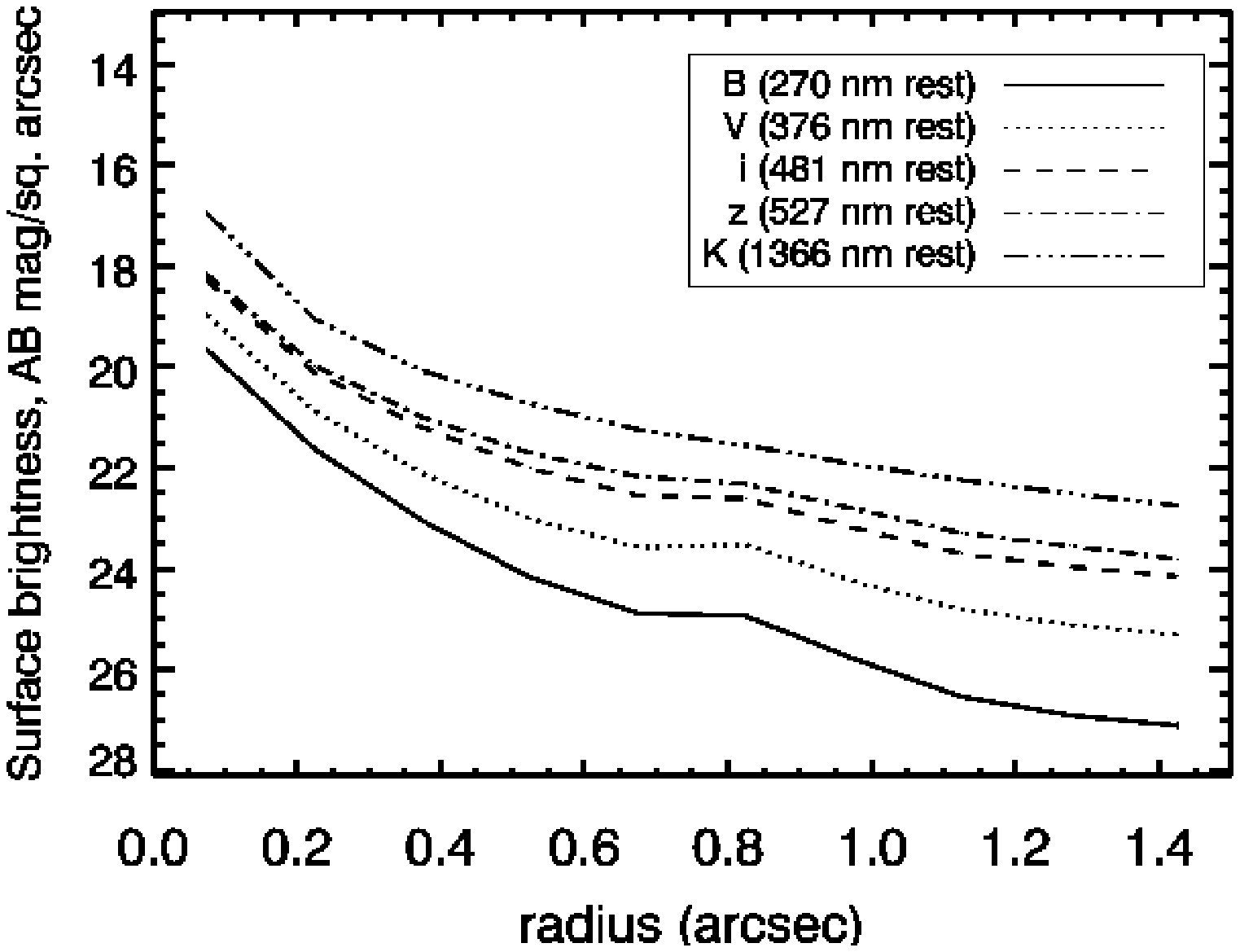}
    \includegraphics[width = 3in, height = 2.0in]{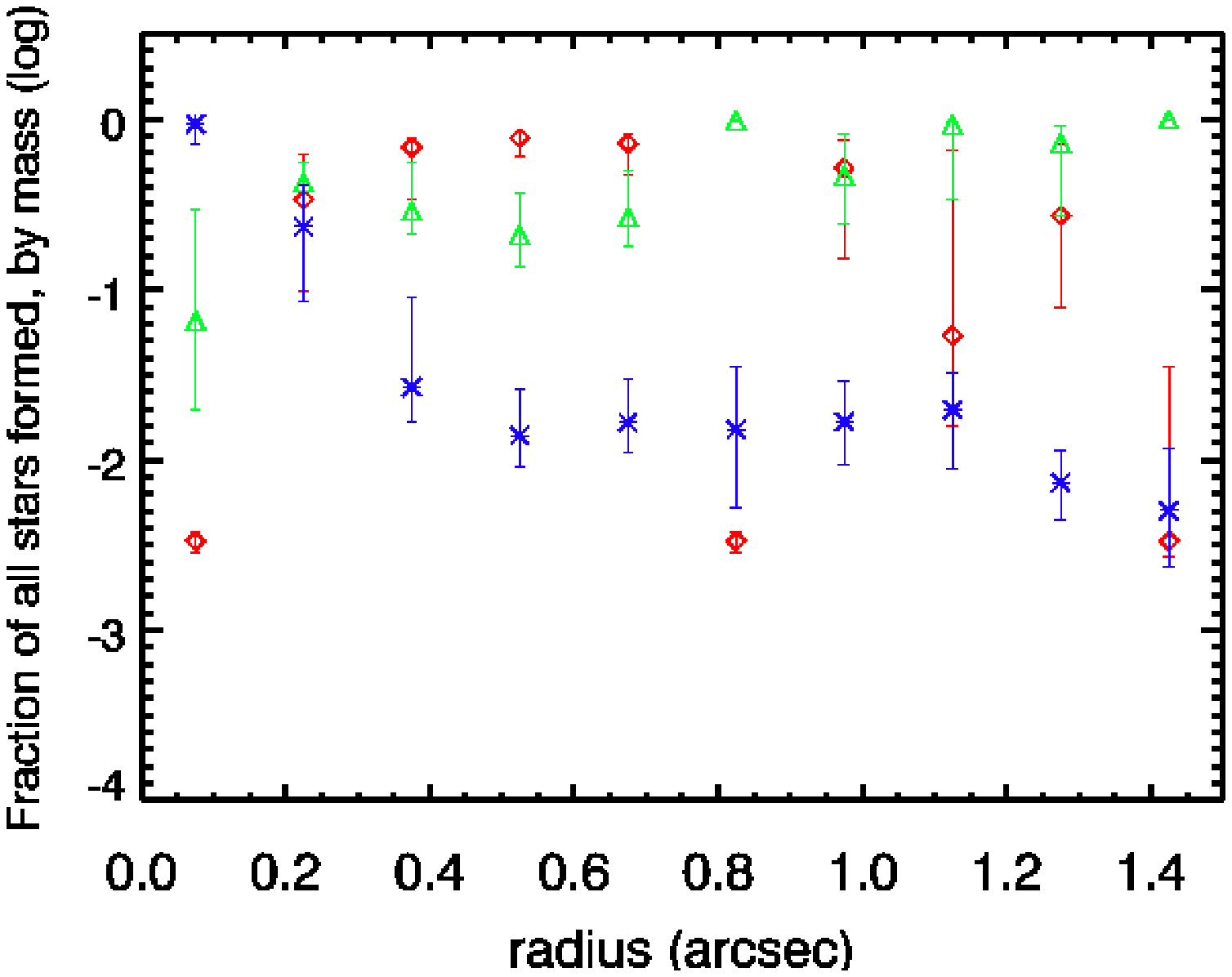}
  \end{center}
\end{figure}
\clearpage
\begin{figure}
  \caption{Radial profiles and fractions of different stellar populations plotted vs. radius for XID 155, with plots and symbols as in Figure 12.}
  \begin{center}
    \includegraphics[width = 3in, height = 2.0in]{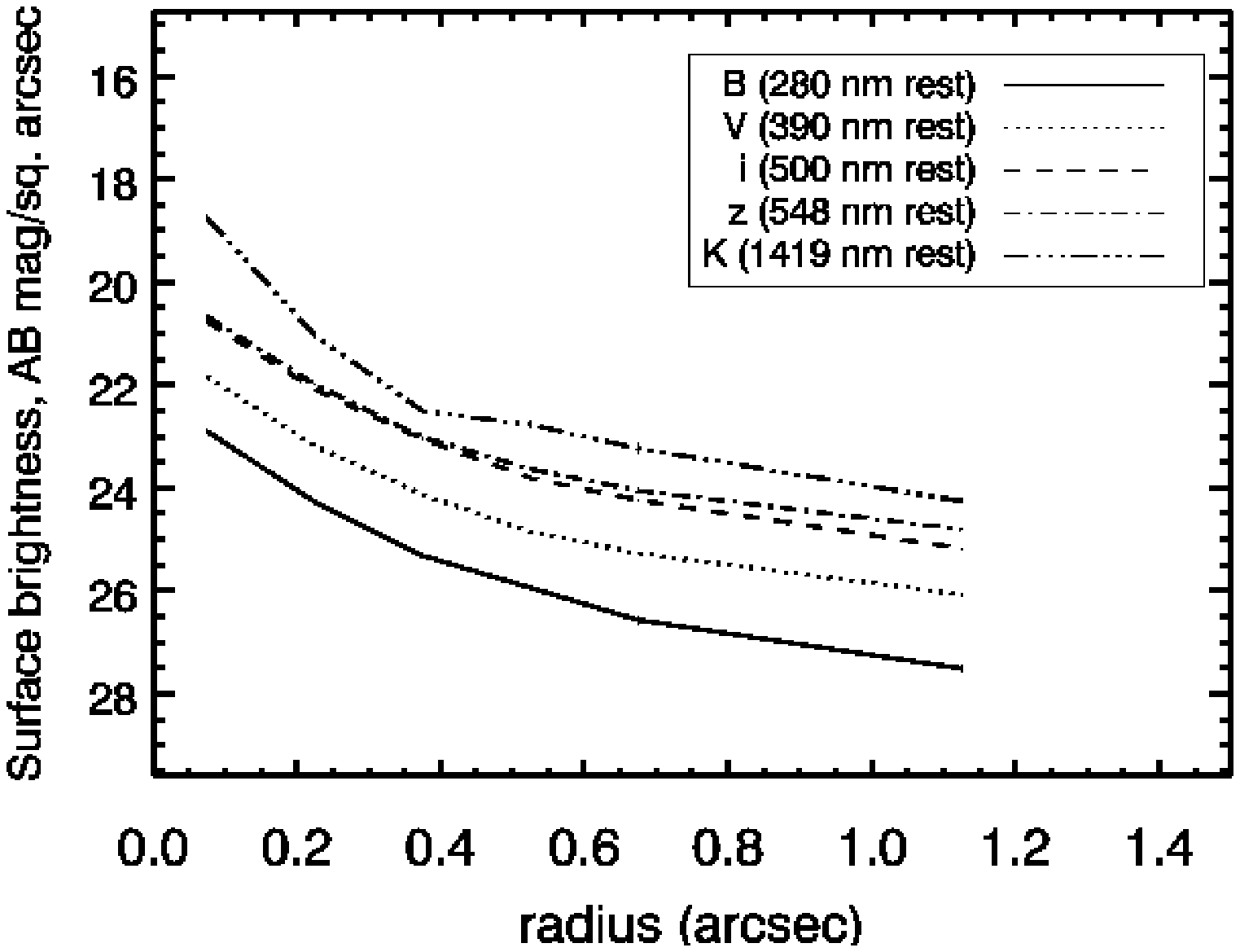}
    \includegraphics[width = 3in, height = 2.0in]{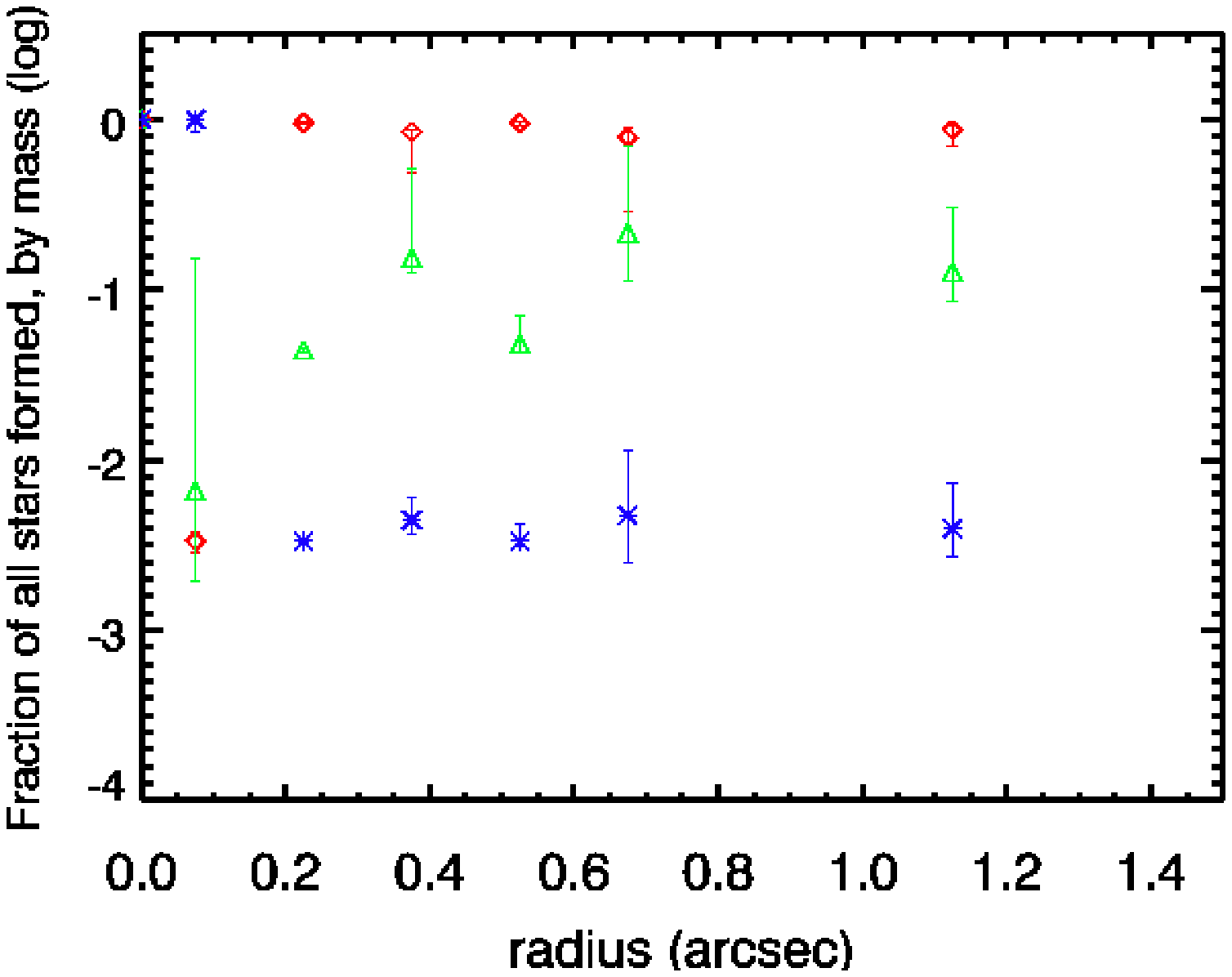}
  \end{center}
\end{figure}

\begin{figure}
  \caption{Radial profiles and fractions of different stellar populations plotted vs. radius for XID 266, with plots and symbols as in Figure 12.}
  \begin{center}
    \includegraphics[width = 3in, height = 2.0in]{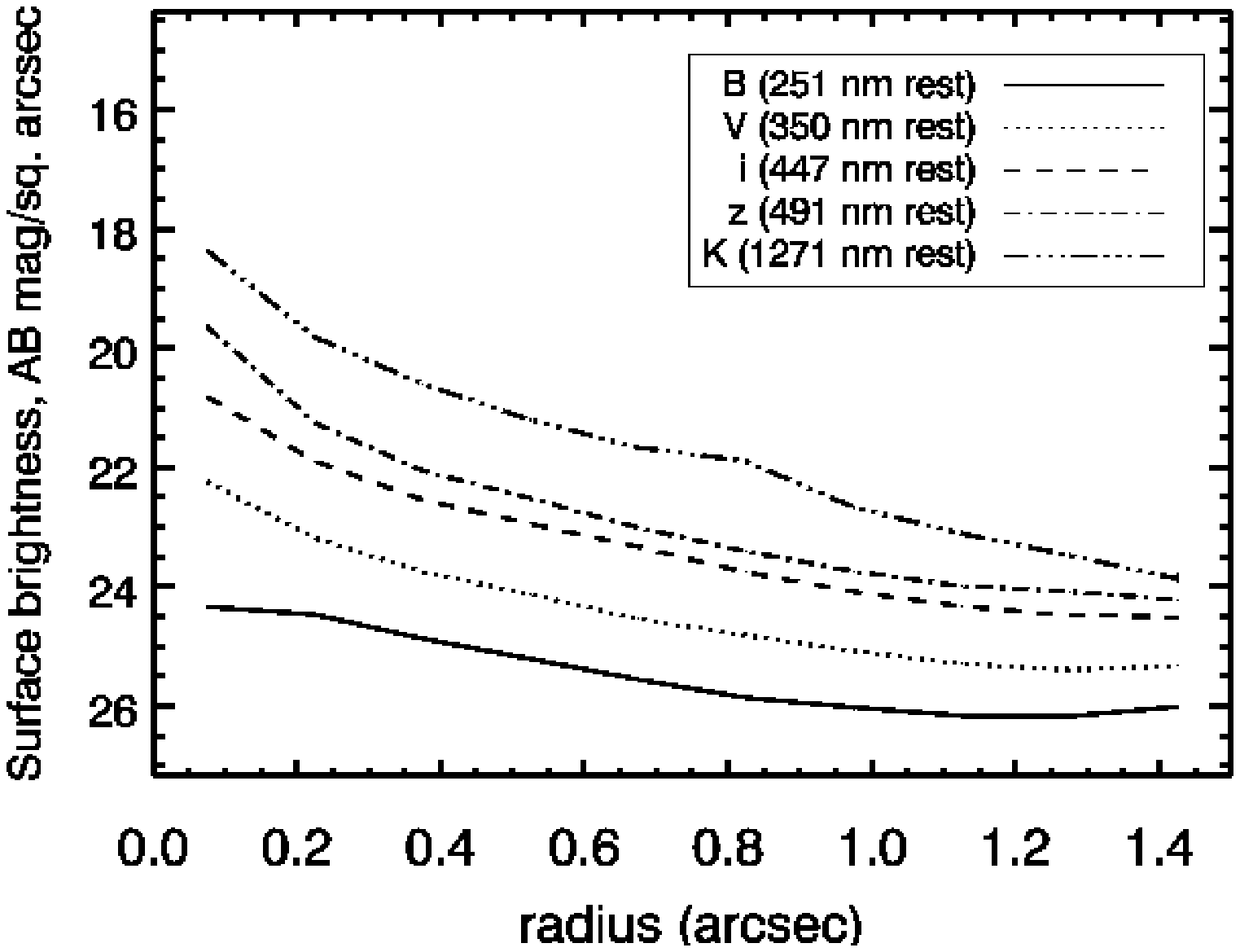}
    \includegraphics[width = 3in, height = 2.0in]{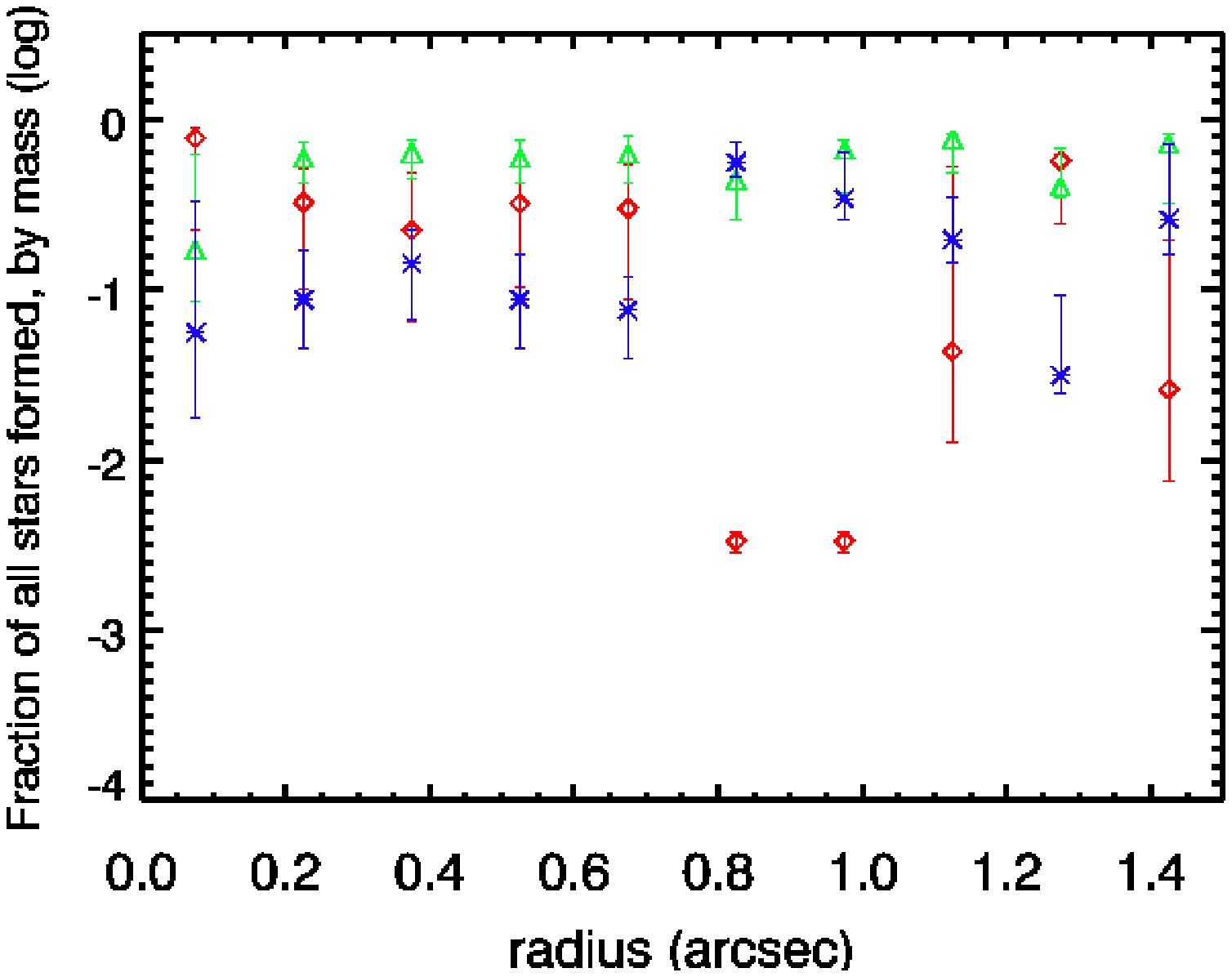}
  \end{center}
\end{figure}
\clearpage
\begin{figure}
  \caption{Radial profiles and fractions of different stellar populations plotted vs. radius for XID 83, with plots and symbols as in Figure 12.}
  \begin{center}
    \includegraphics[width = 3in, height = 2.0in]{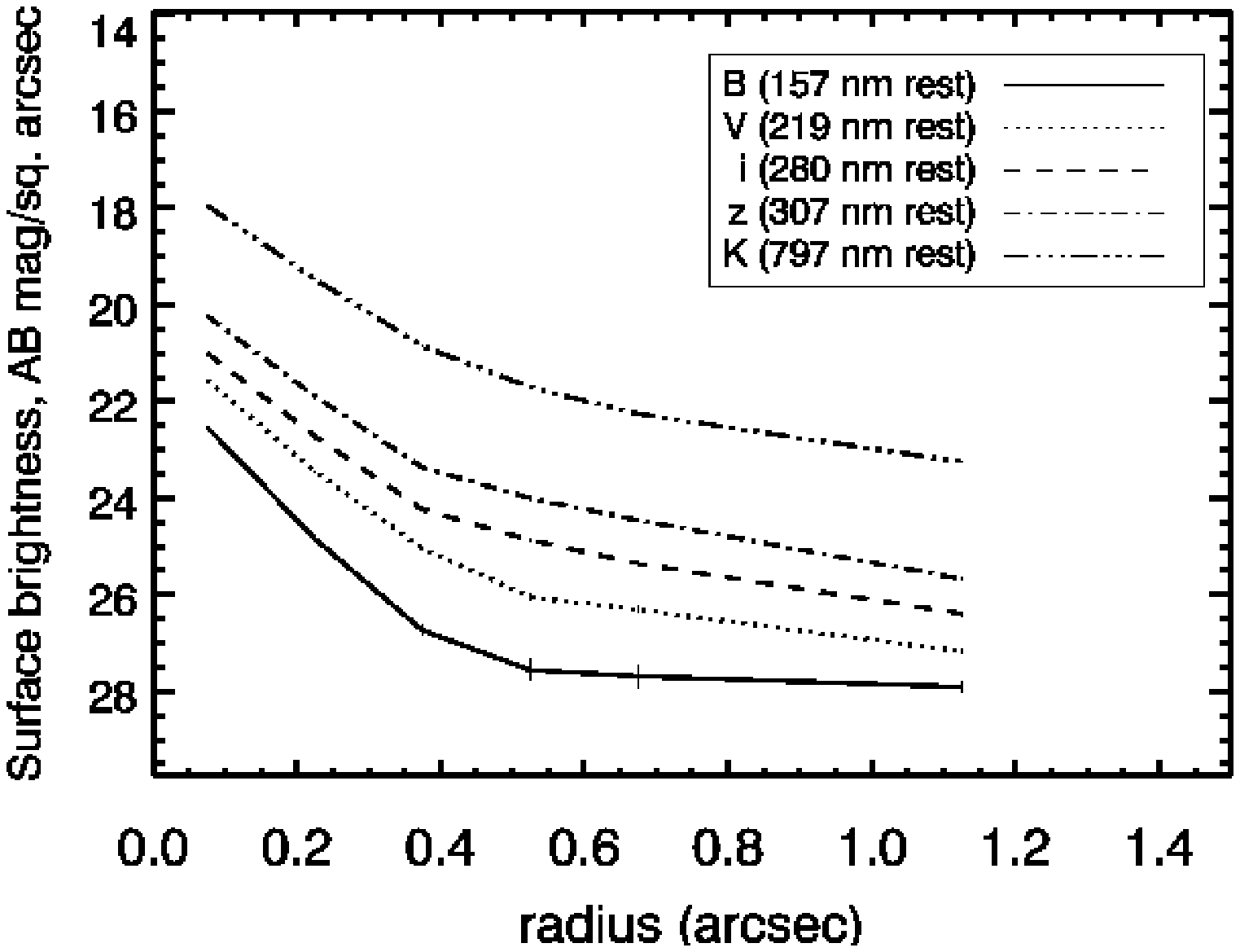}
    \includegraphics[width = 3in, height = 2.0in]{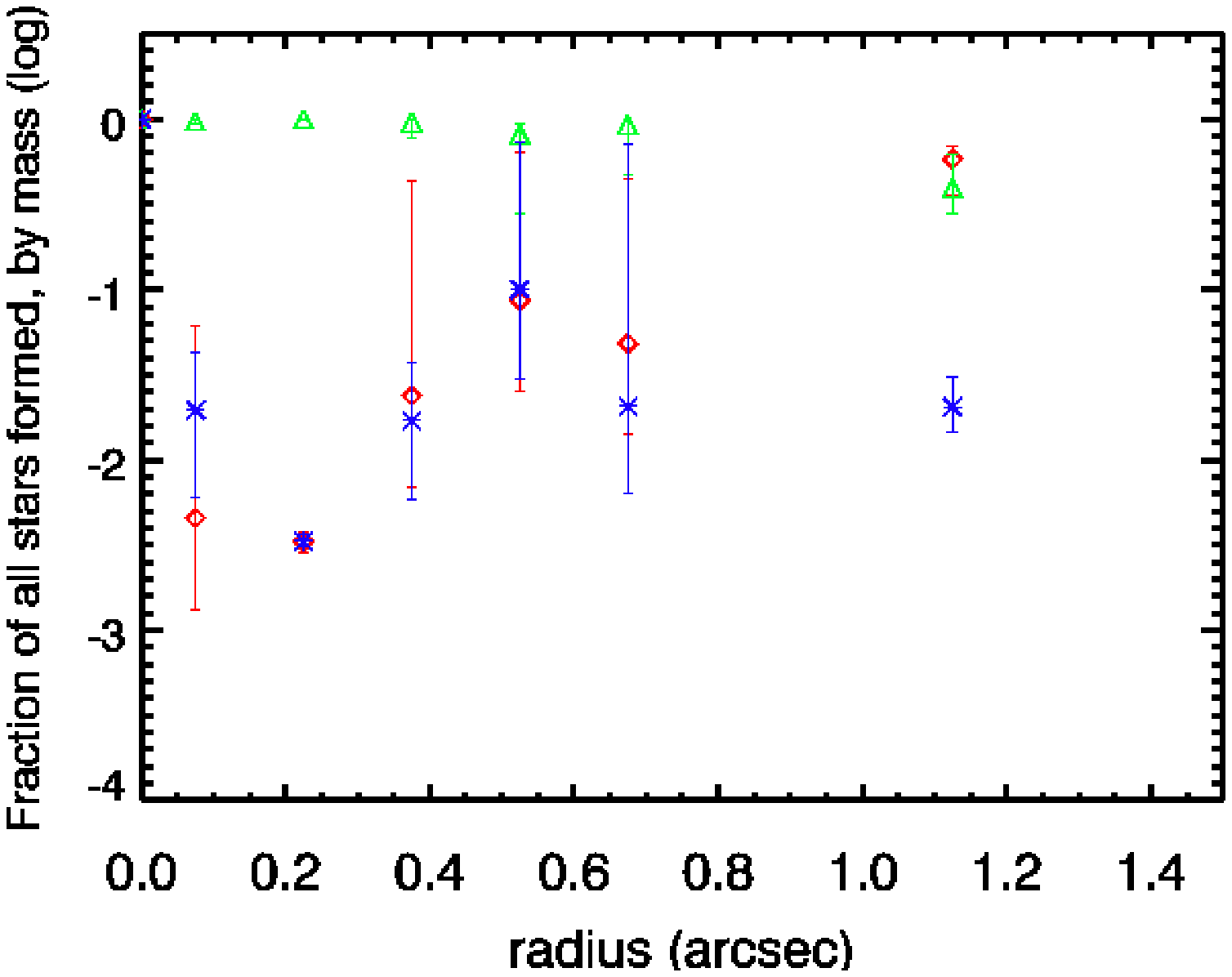}
  \end{center}
\end{figure}

\begin{figure}
  \caption{Radial profiles and fractions of different stellar populations plotted vs. radius for XID 536, with plots and symbols as in Figure 12.}
  \begin{center}
    \includegraphics[width = 3in, height = 2.0in]{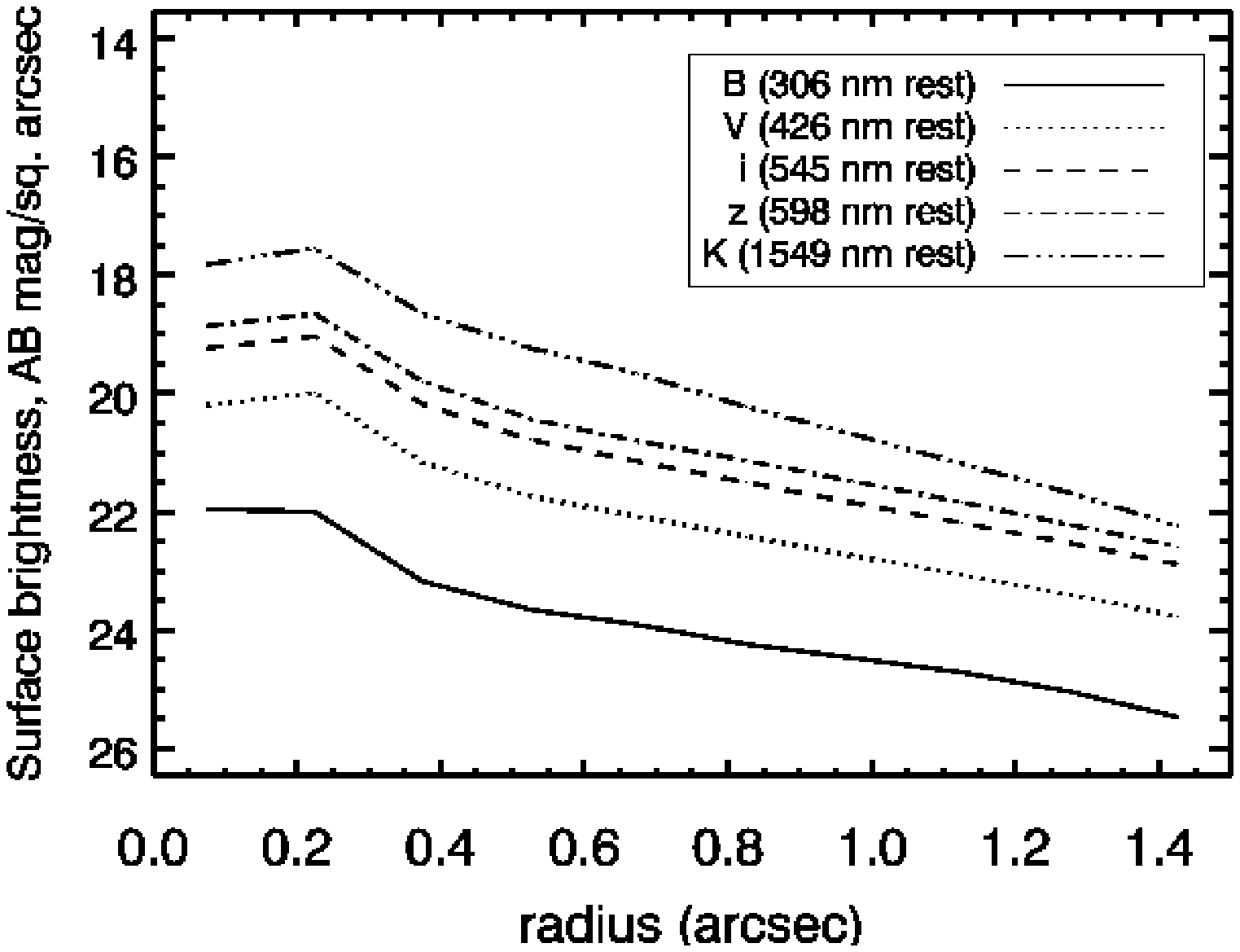}
    \includegraphics[width = 3in, height = 2.0in]{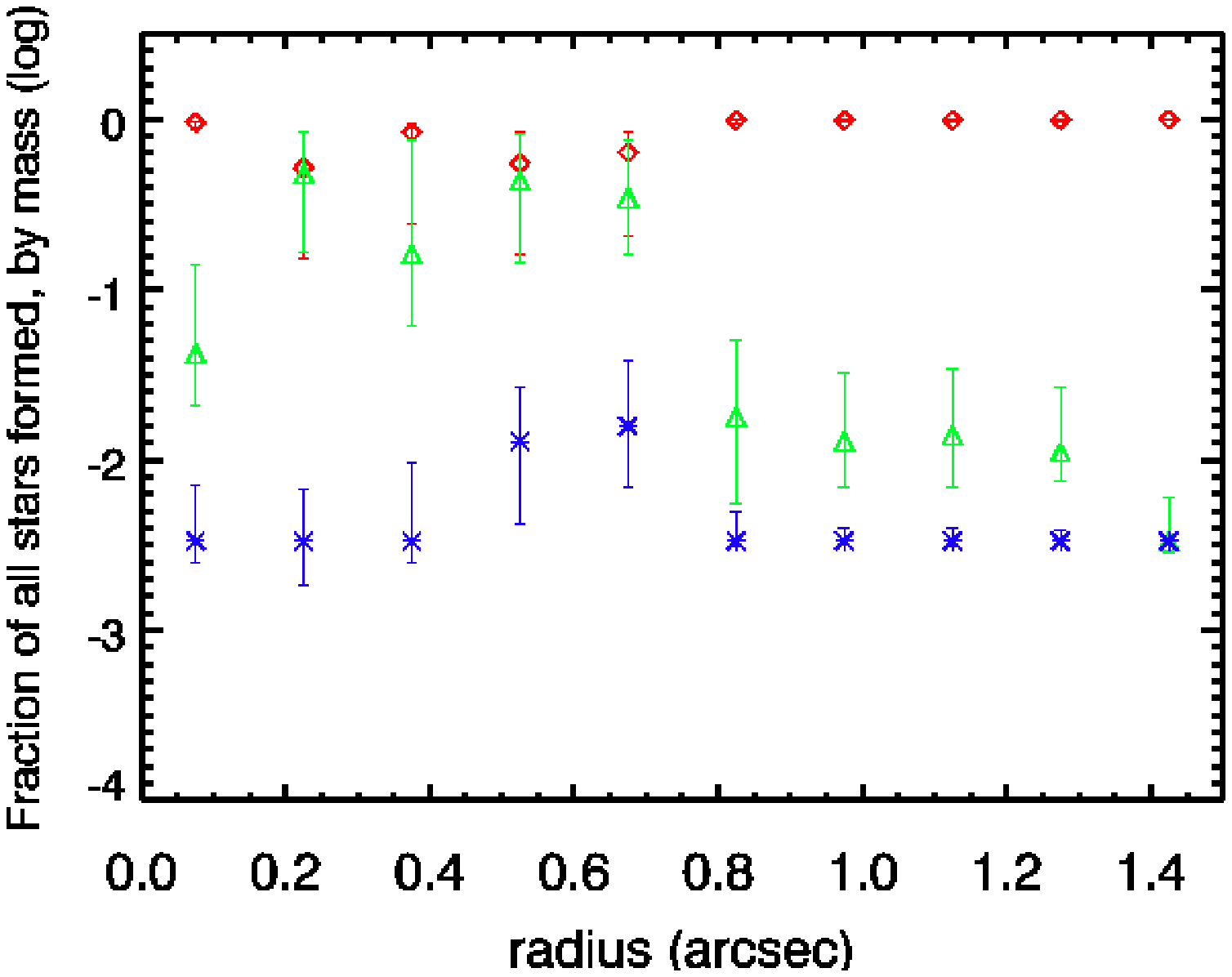}
  \end{center}
\end{figure}
\clearpage
\begin{figure}
  \caption{Radial profiles and fractions of different stellar populations plotted vs. radius for XID 594, with plots and symbols as in Figure 12.}
  \begin{center}
    \includegraphics[width = 3in, height = 2.0in]{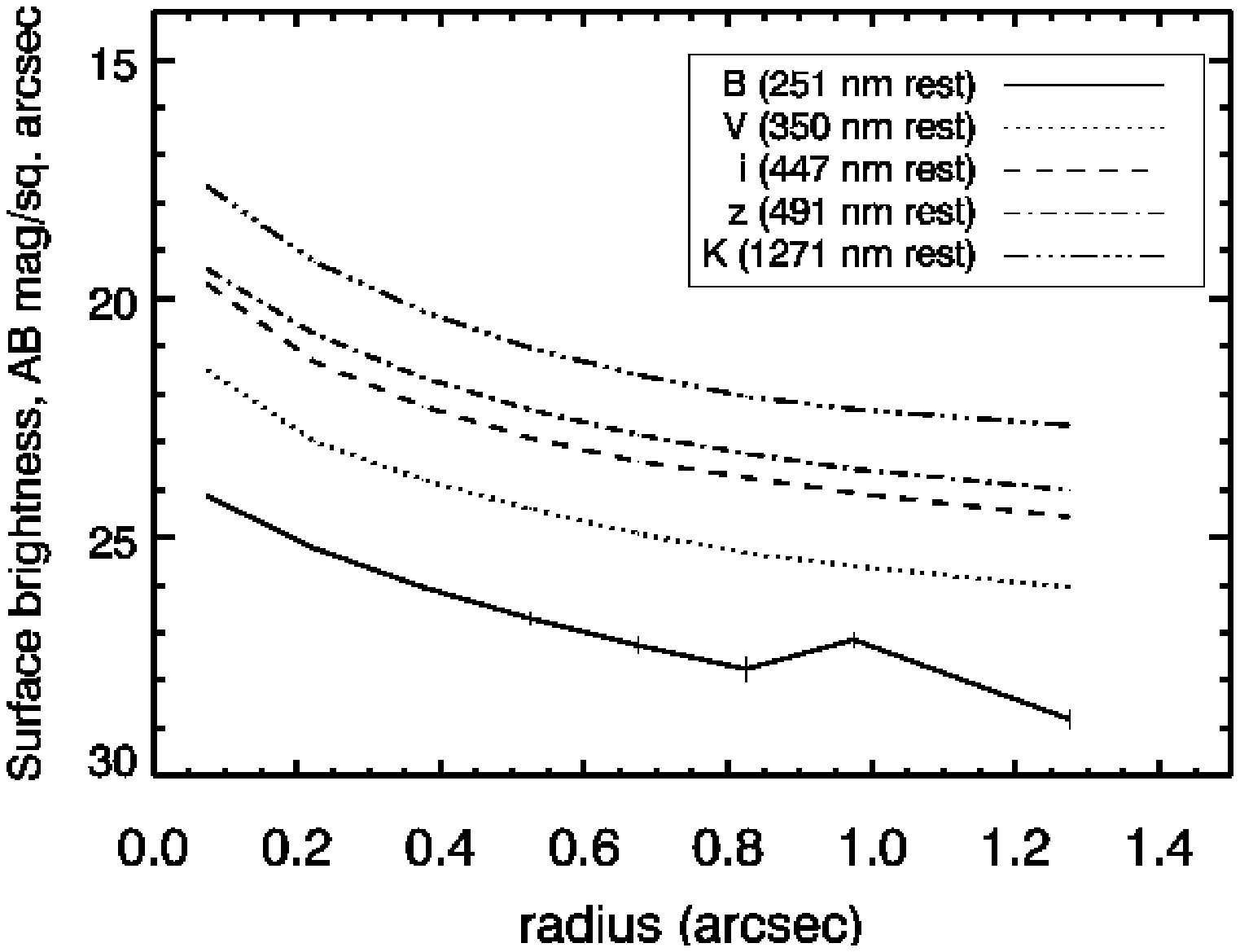}
    \includegraphics[width = 3in, height = 2.0in]{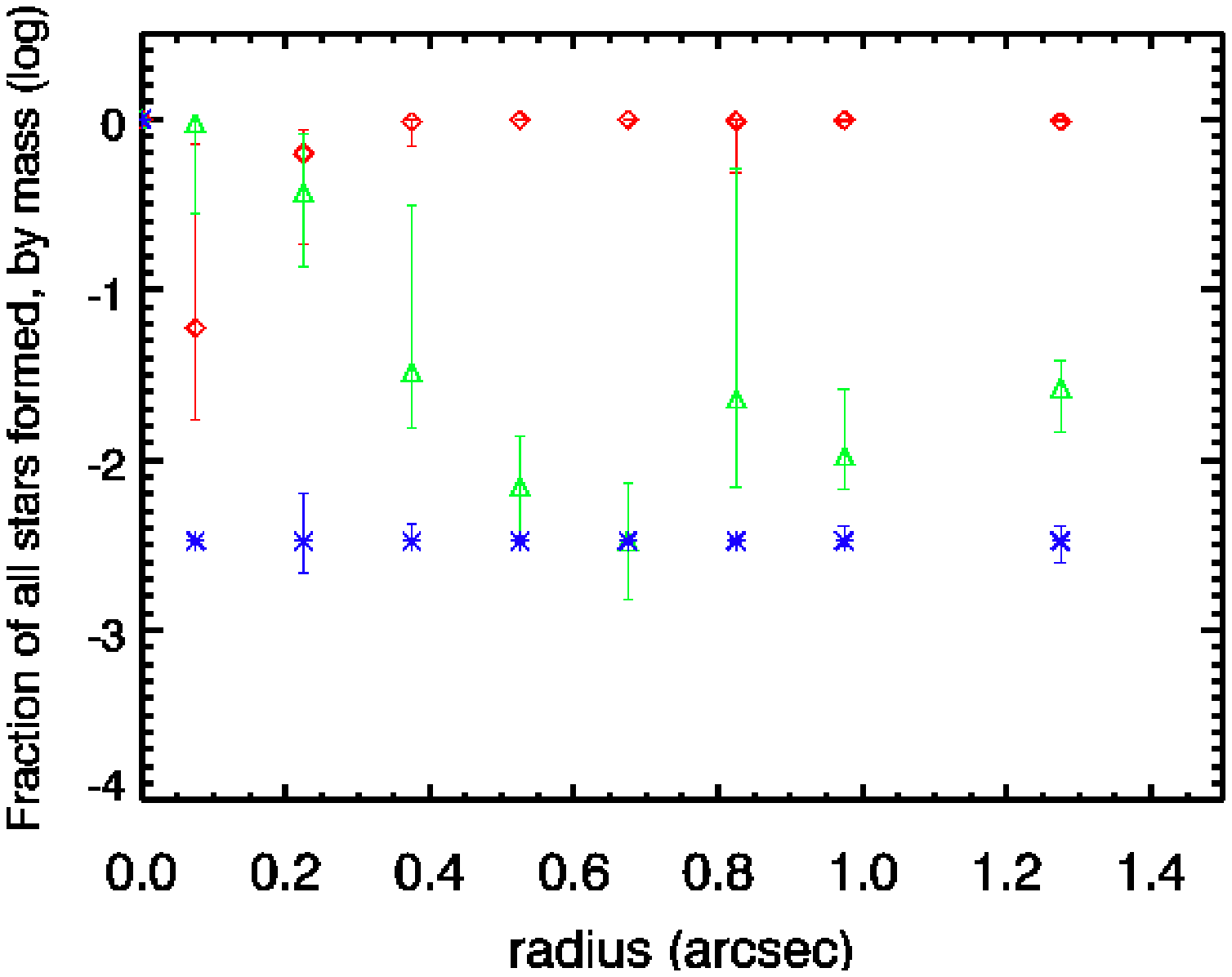}
  \end{center}
\end{figure}

\begin{figure}
  \caption{Radial profiles and fractions of different stellar populations plotted vs. radius for XID 15, with plots and symbols as in Figure 12.}
  \begin{center}
    \includegraphics[width = 3in, height = 2.0in]{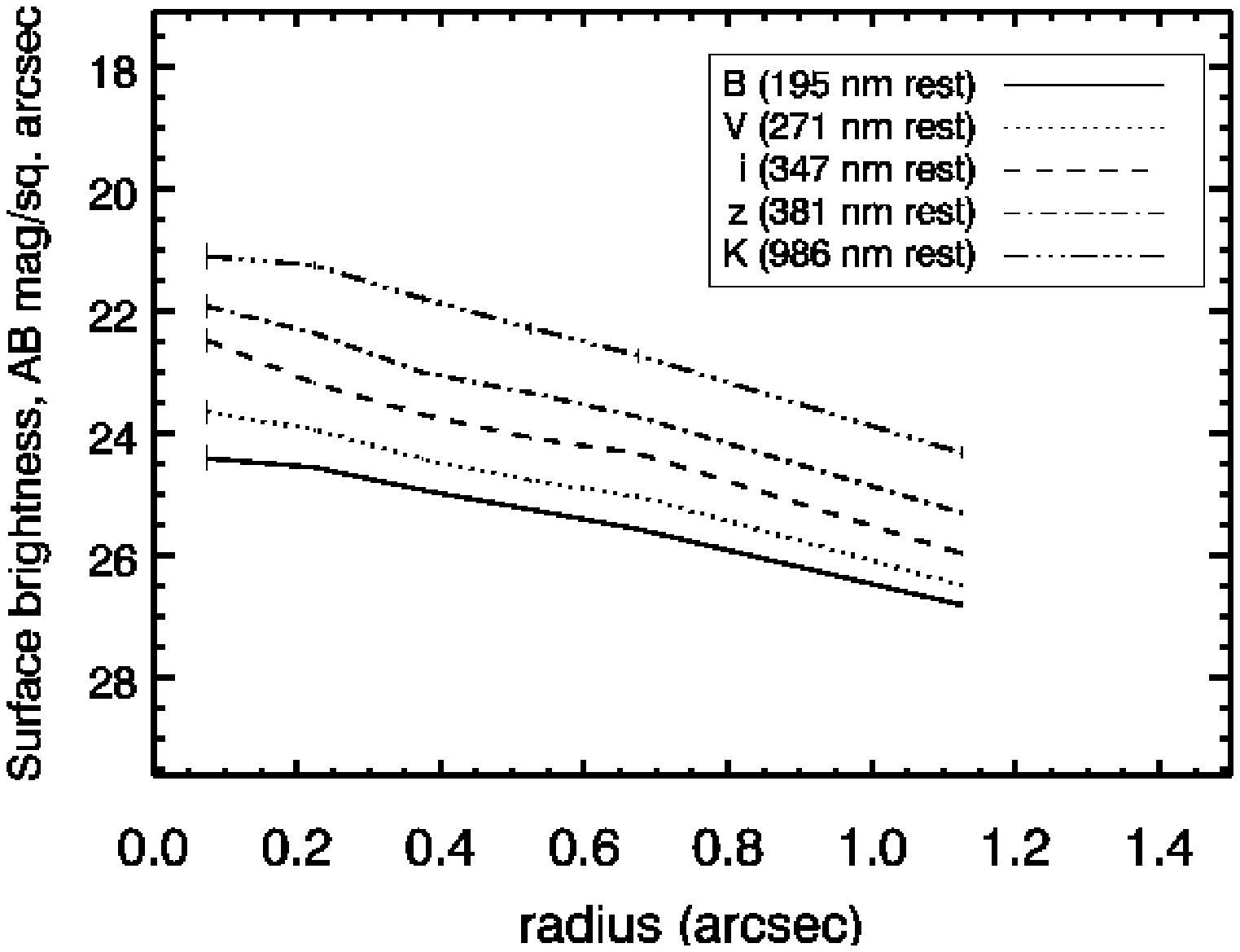}
    \includegraphics[width = 3in, height = 2.0in]{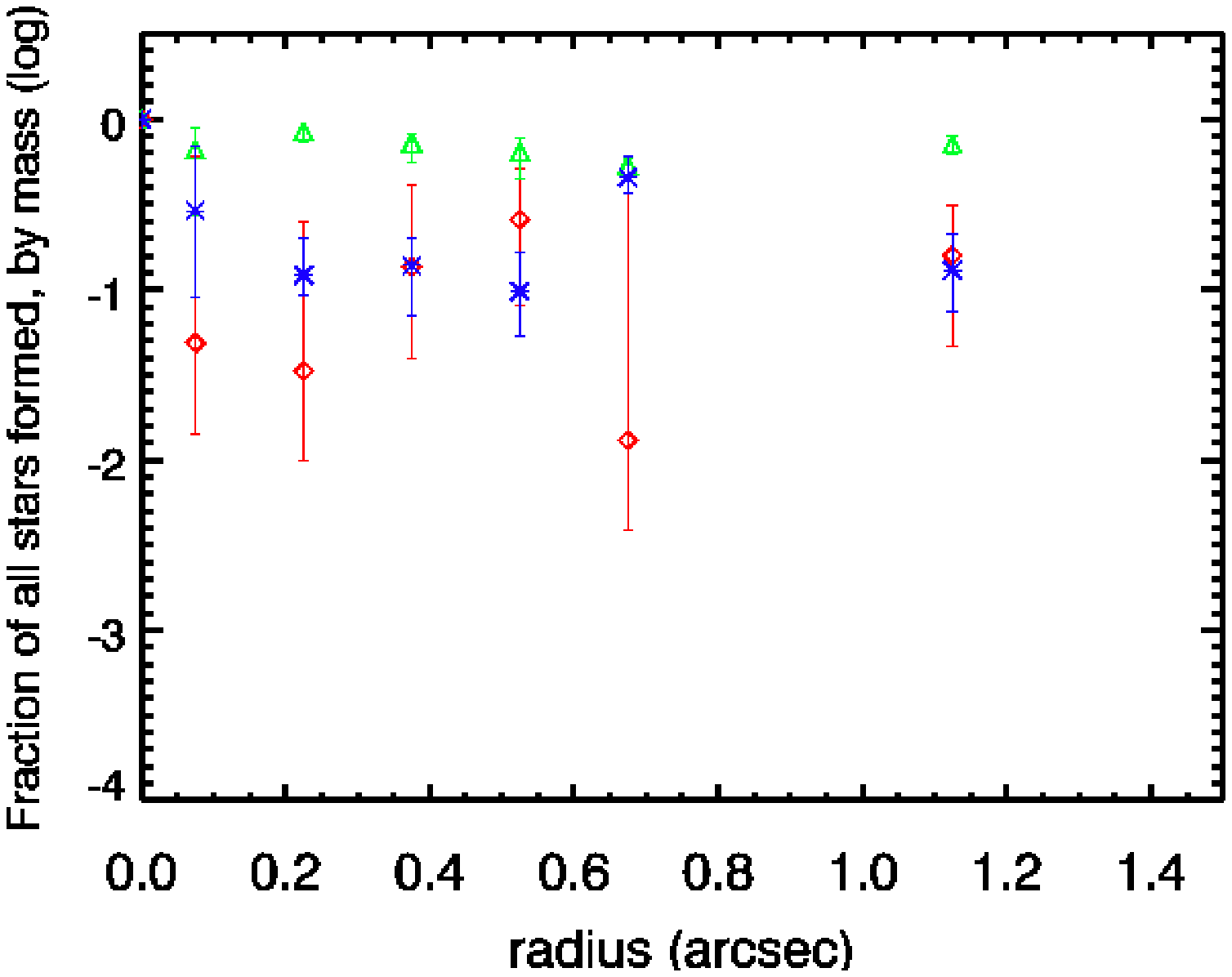}
  \end{center}
\end{figure}
\clearpage
\begin{figure}
  \caption{Radial profiles and fractions of different stellar populations plotted vs. radius for XID 32, with plots and symbols as in Figure 12.}
  \begin{center}
    \includegraphics[width = 3in, height = 2.0in]{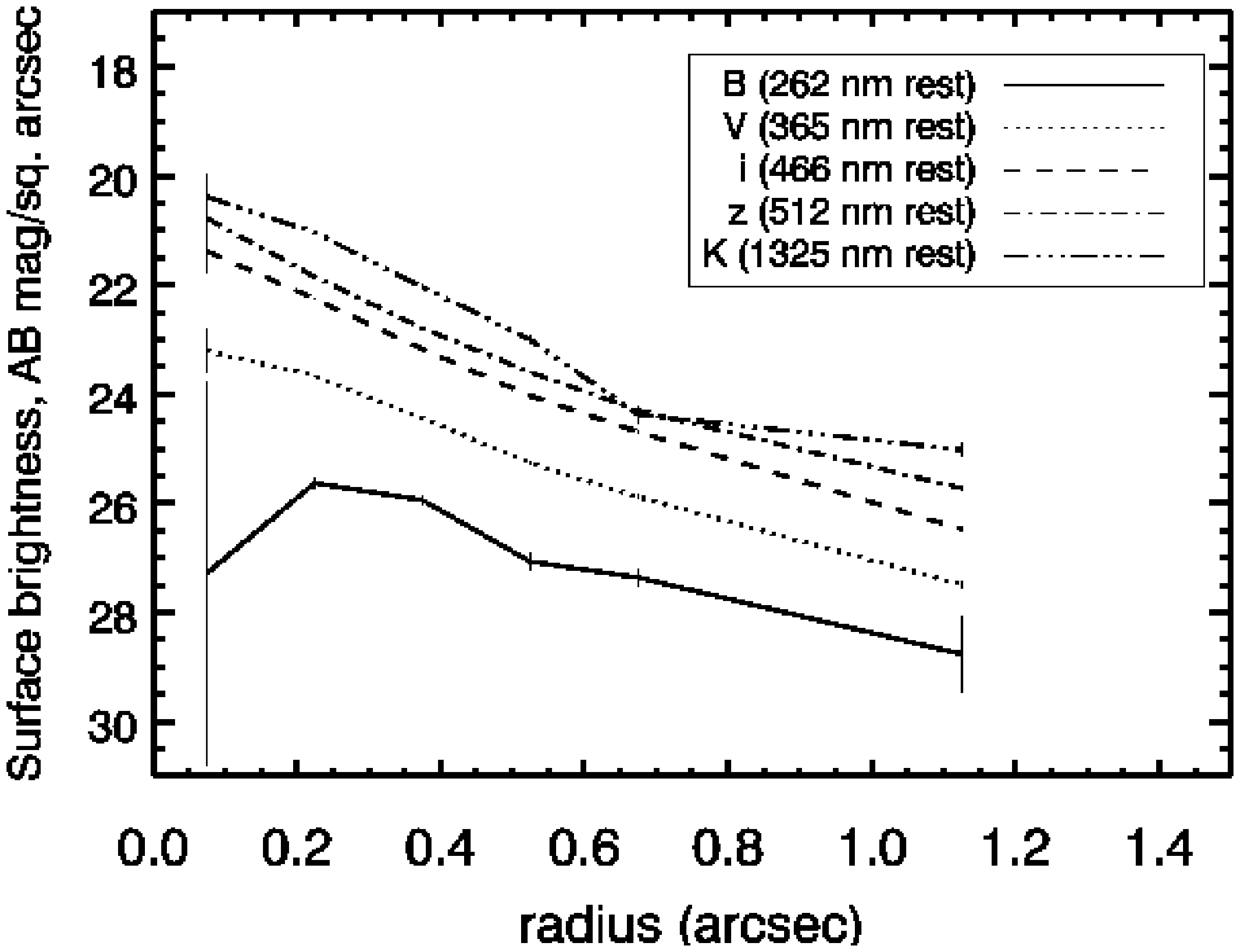}
    \includegraphics[width = 3in, height = 2.0in]{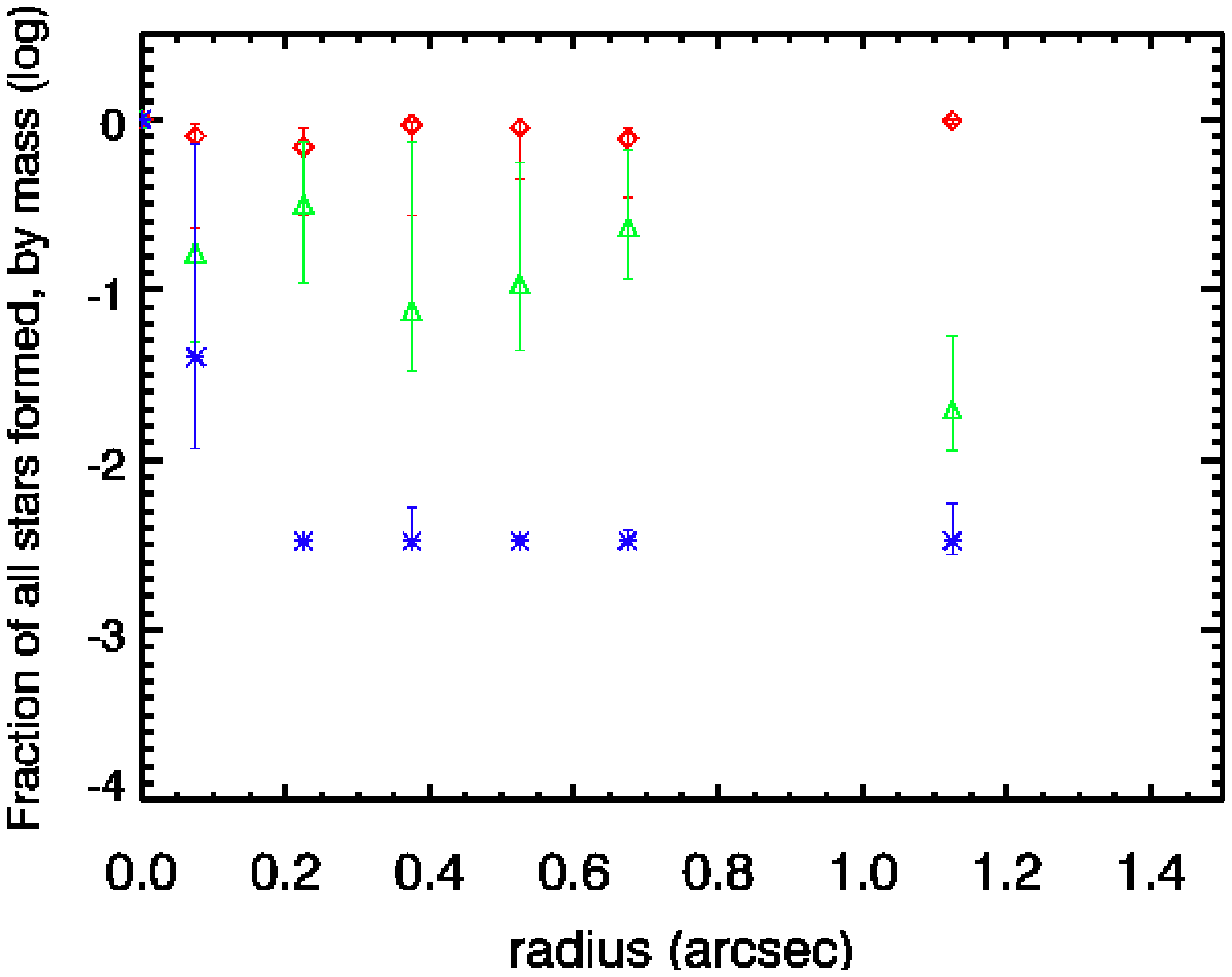}
  \end{center}
\end{figure}

\begin{figure}
  \caption{Plots of mean stellar populations statistics for three AGN populations:  Strong Type II (upper left), Strong Type I (upper right), and Weak Type I AGN (lower left).  The x-axis is the radial location in arcseconds and the y-axis is the fraction of stars in a given stellar population.  Colors denote different age ranges of stellar populations, as explained in \S 7.  }
  \begin{center}
    \includegraphics[width = 3in, height = 2.0in]{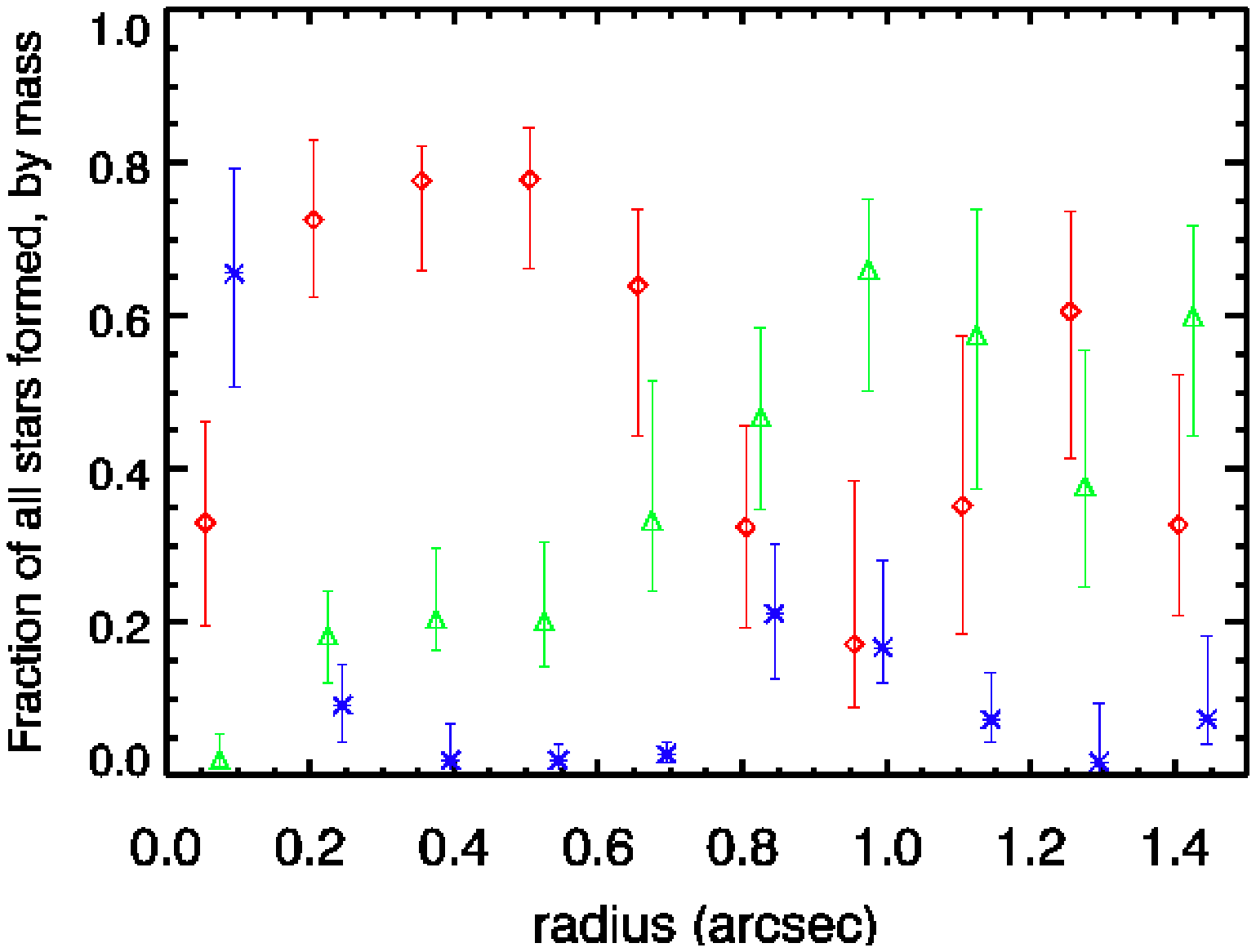}
    \includegraphics[width = 3in, height = 2.0in]{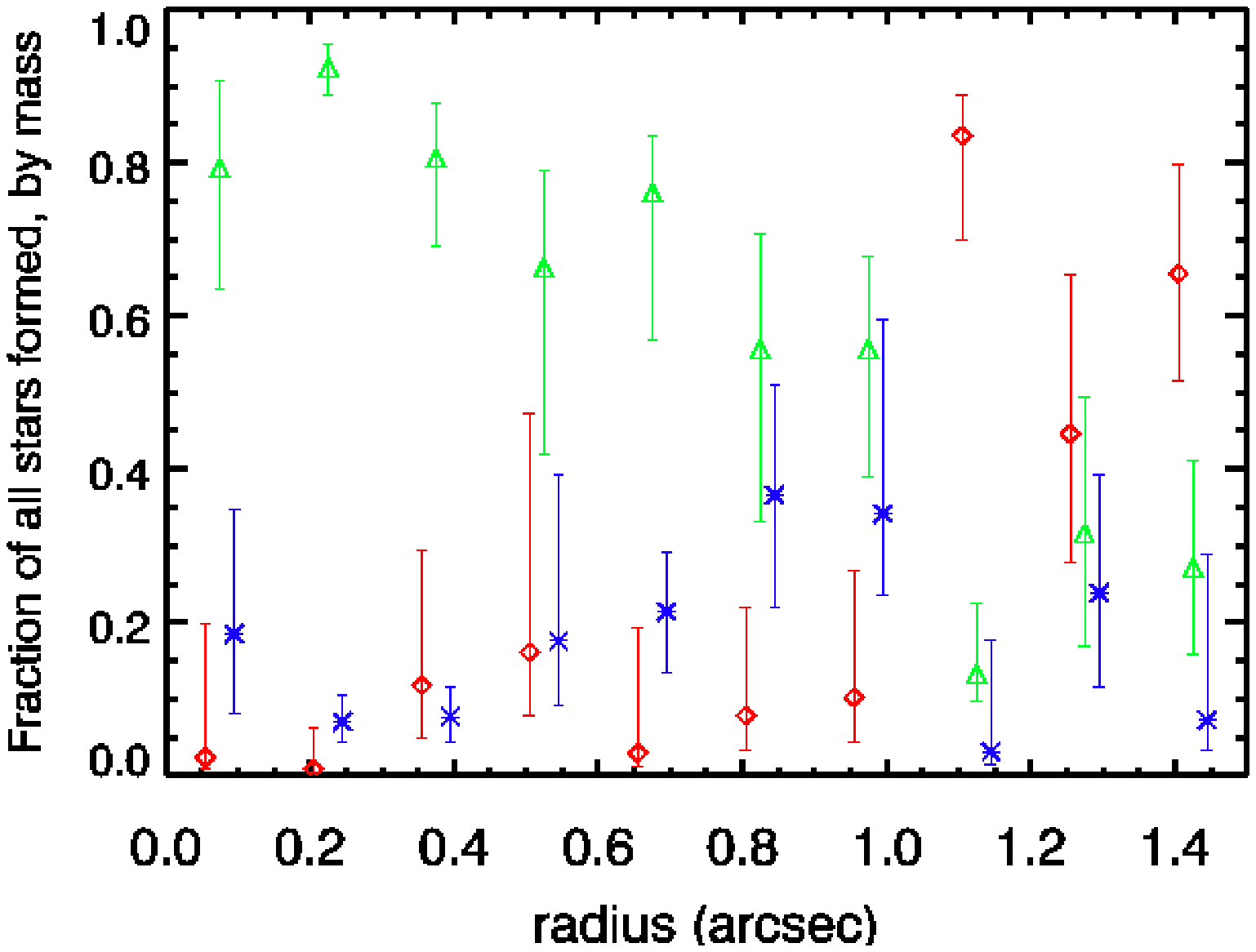}
        \includegraphics[width = 3in, height = 2.0in]{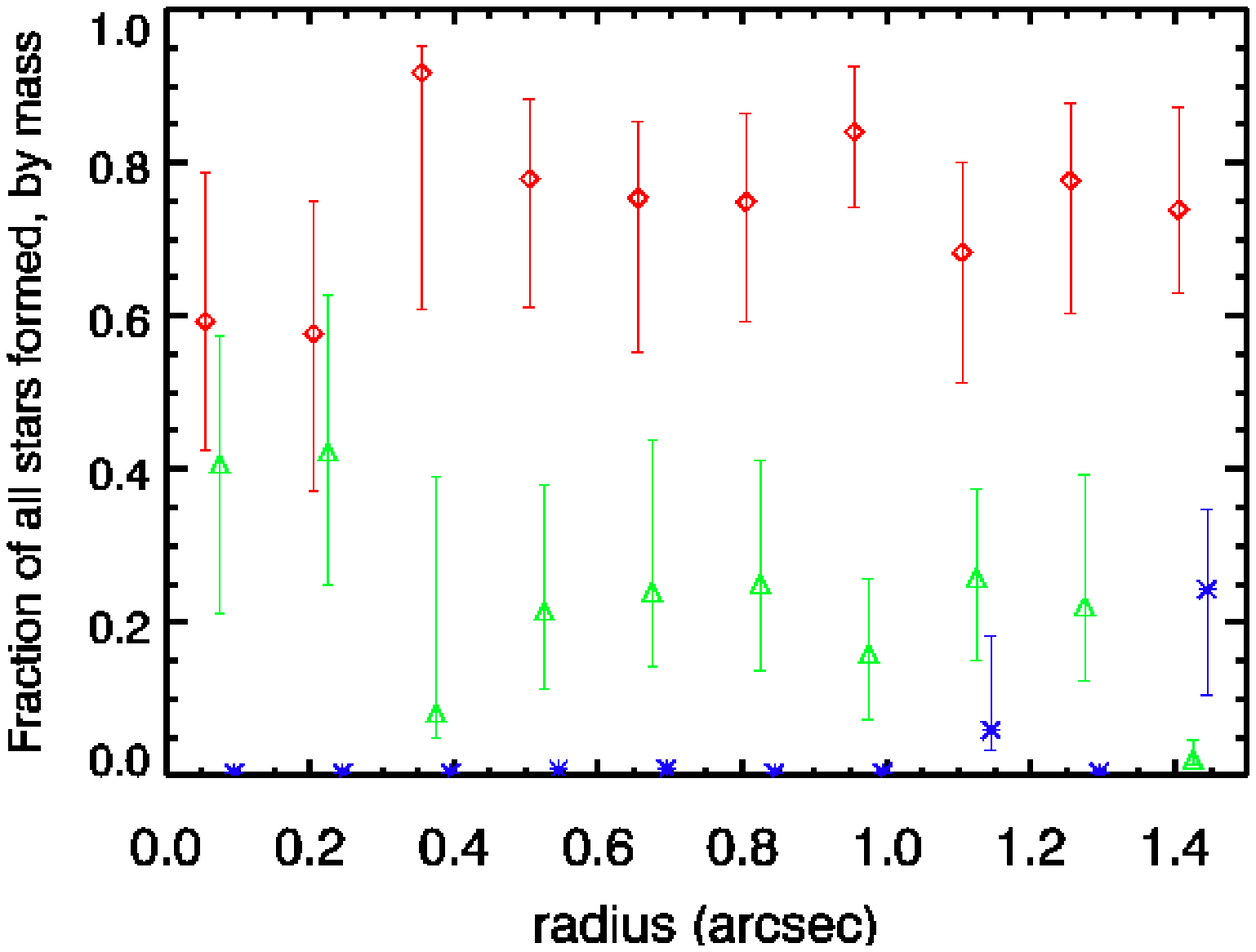}
  \end{center}
\end{figure}
\clearpage
\begin{figure}
  \caption{[OIII] line luminosities for all AGN sources plotted against the radially averaged fraction of young stellar populations (age $< 100$ Myr), not weighted by luminosity.}
  \begin{center}
    \includegraphics[width = 3in, height = 2.0in]{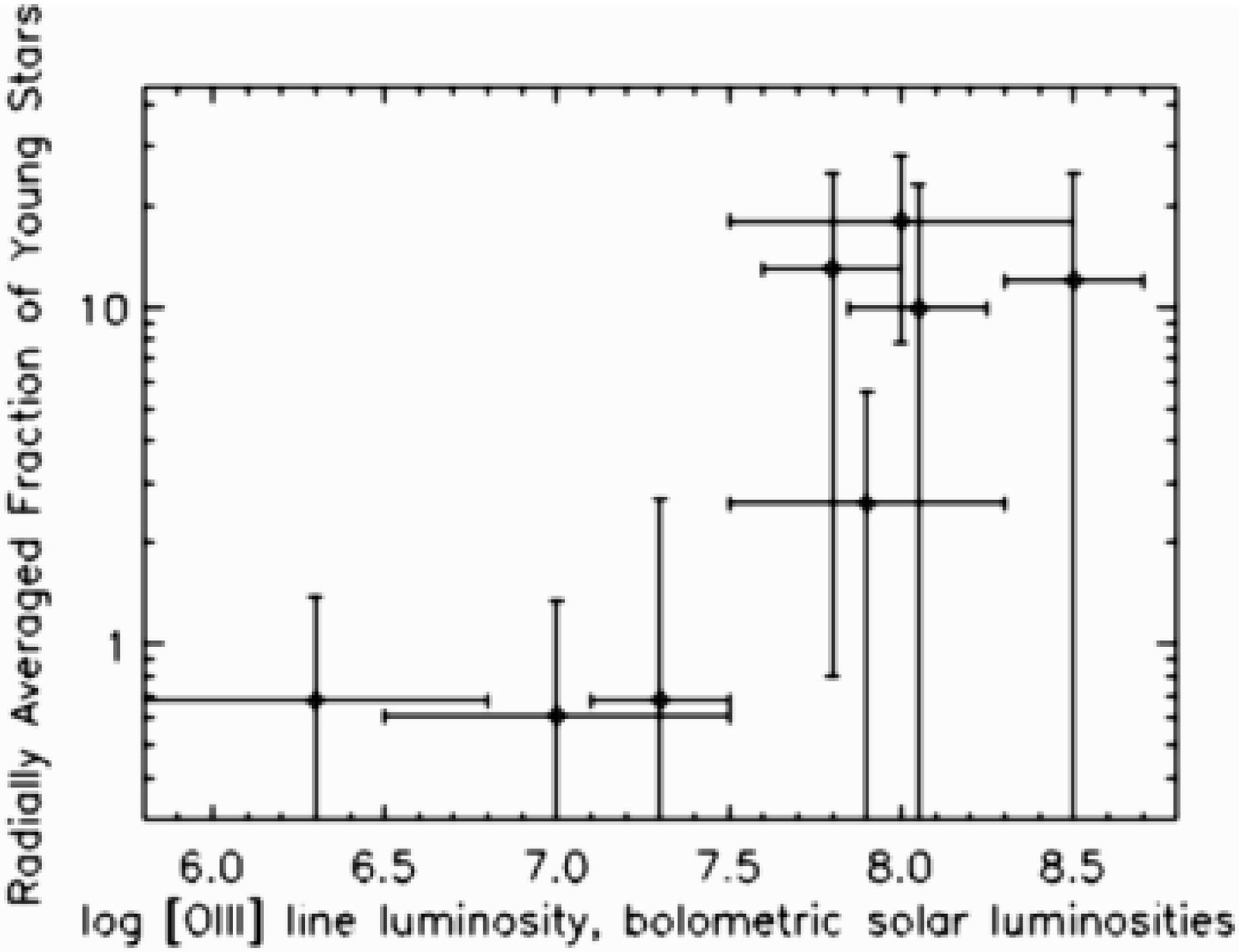}
  \end{center}
\end{figure}

\begin{figure}
  \caption{Tiled imagery and GALFIT models for Normal 1, as in Figure 3.}
  \begin{center}
    \includegraphics[width=5in,height=3.0in]{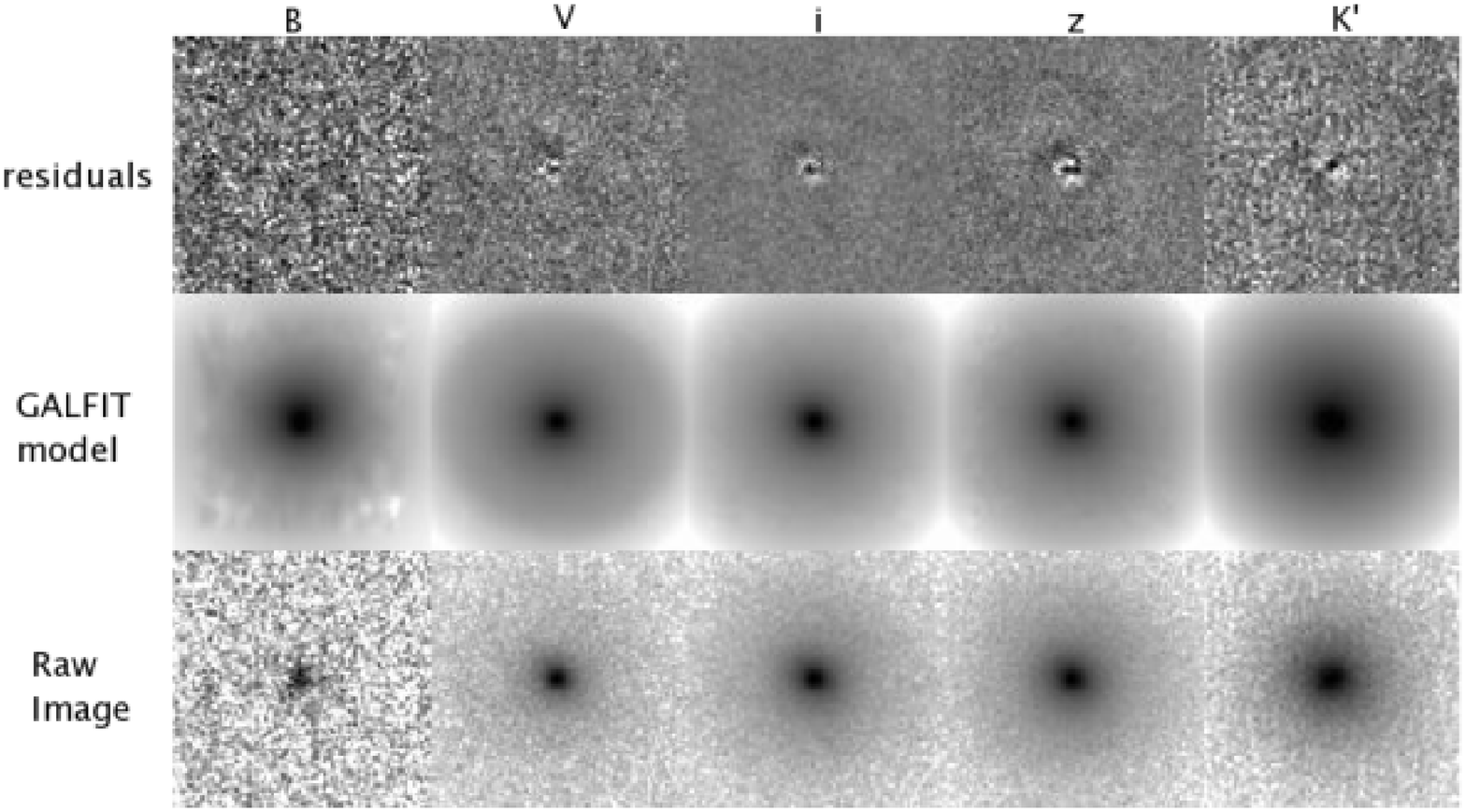}
  \end{center}
\end{figure}
\clearpage
\begin{figure}
  \caption{Tiled imagery and GALFIT models for Normal 2, as in Figure 3.}
  \begin{center}
    \includegraphics[width=5in,height=3.0in]{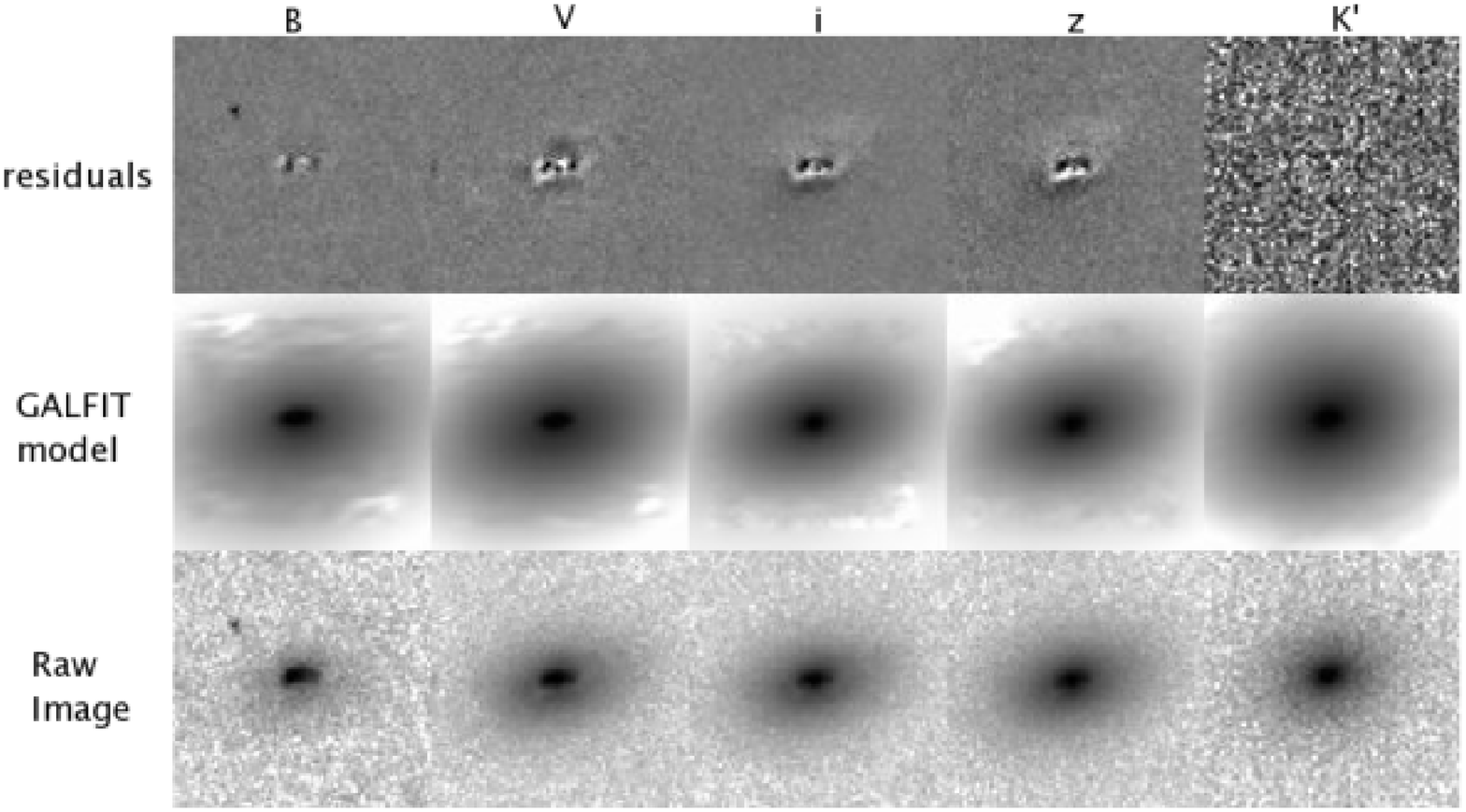}
  \end{center}
\end{figure}

\begin{figure}
  \caption{Tiled imagery and GALFIT models for Normal 3, as in Figure 3.}
  \begin{center}
    \includegraphics[width=5in,height=3.0in]{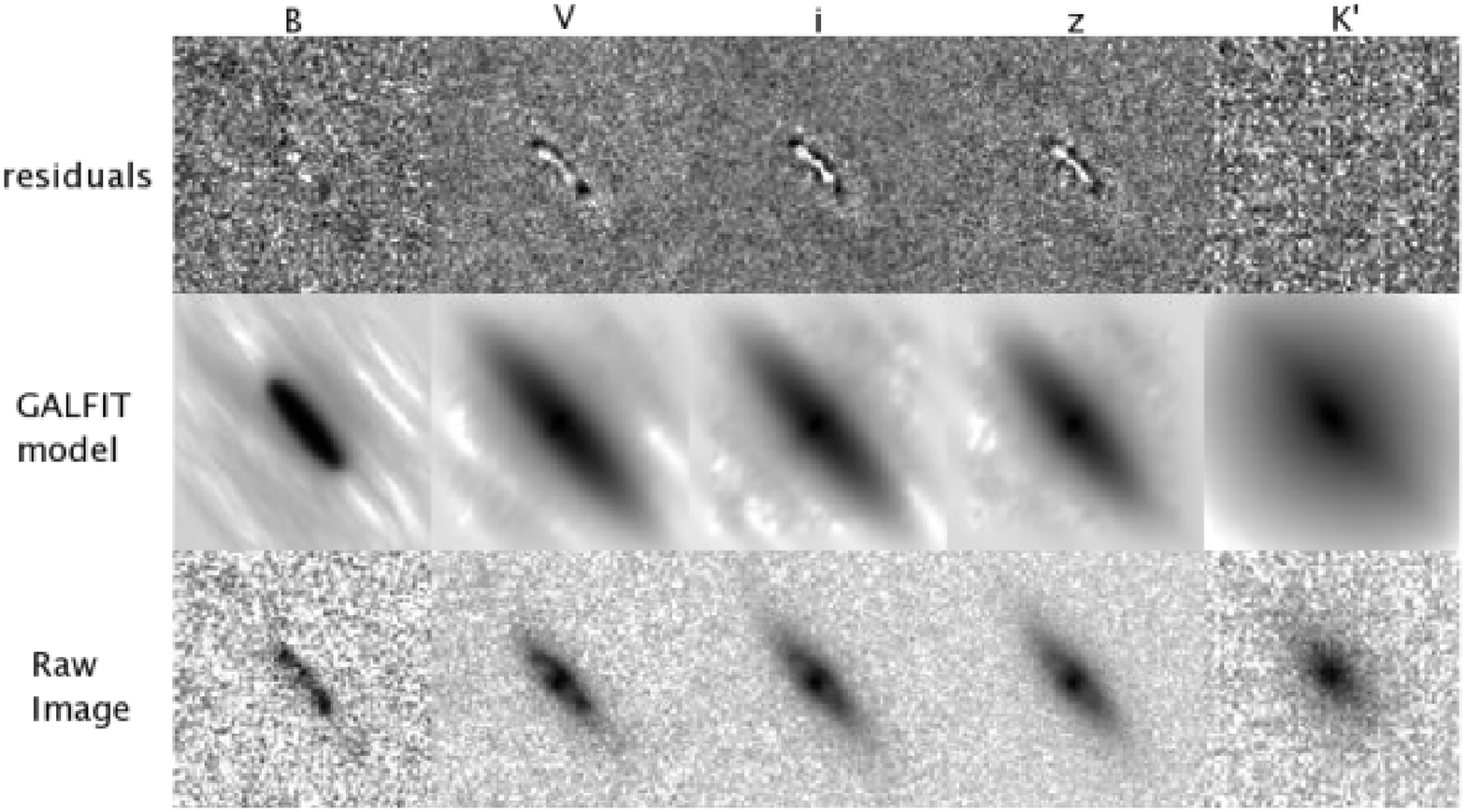}
  \end{center}
\end{figure}
\clearpage
\begin{figure}
  \caption{Tiled imagery and GALFIT models for Normal 4, as in Figure 3.}
  \begin{center}
    \includegraphics[width=5in,height=3.0in]{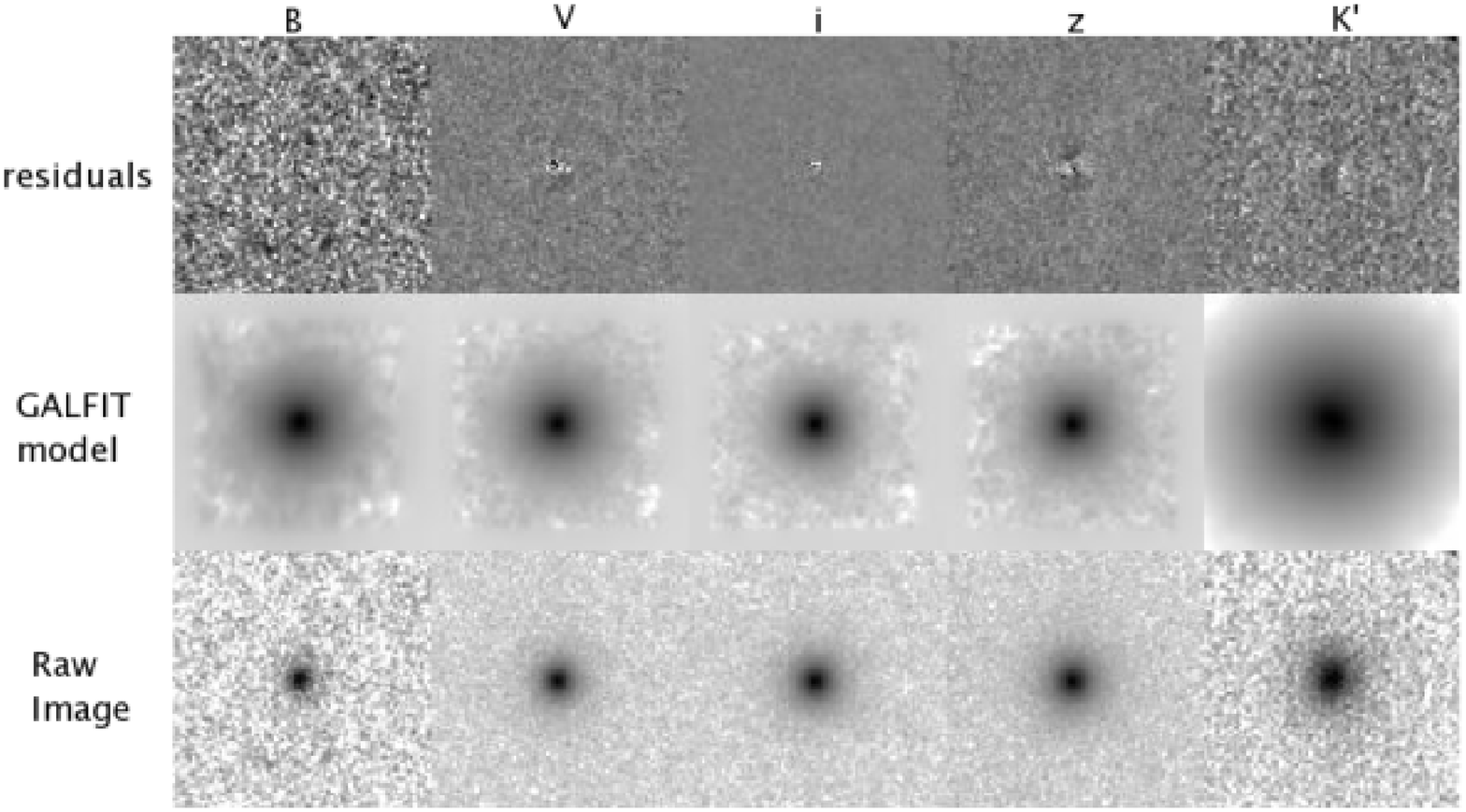}
  \end{center}
\end{figure}

\begin{figure}
  \caption{Tiled imagery and GALFIT models for Normal 5, as in Figure 3.}
  \begin{center}
    \includegraphics[width=5in,height=3.0in]{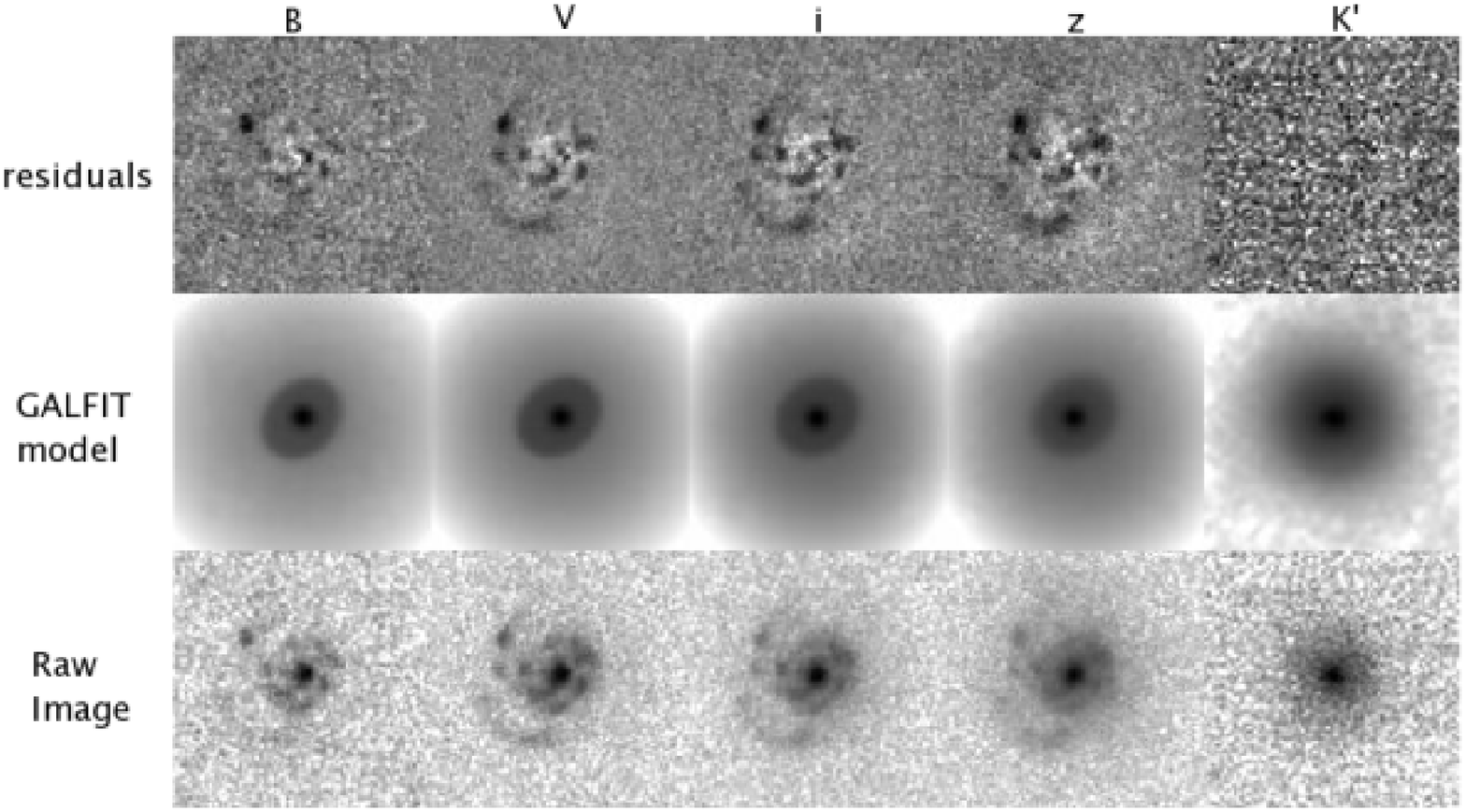}
  \end{center}
\end{figure}
\clearpage
\begin{figure}
  \caption{Tiled imagery and GALFIT models for Normal 6, as in Figure 3.}
  \begin{center}
    \includegraphics[width=5in,height=3.0in]{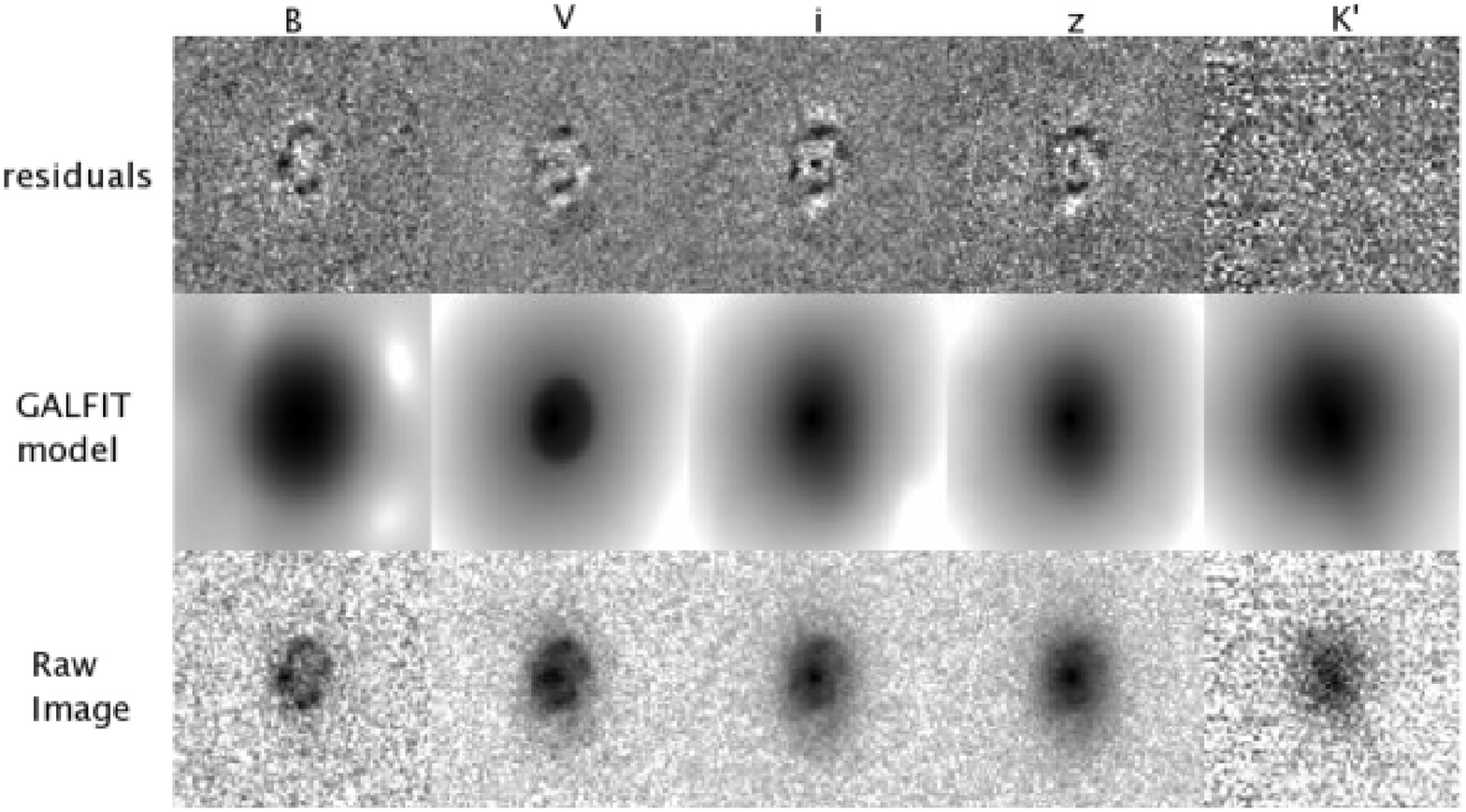}
  \end{center}
\end{figure}

\begin{figure}
  \caption{Tiled imagery and GALFIT models for Normal 7, as in Figure 3.}
  \begin{center}
    \includegraphics[width=5in,height=3.0in]{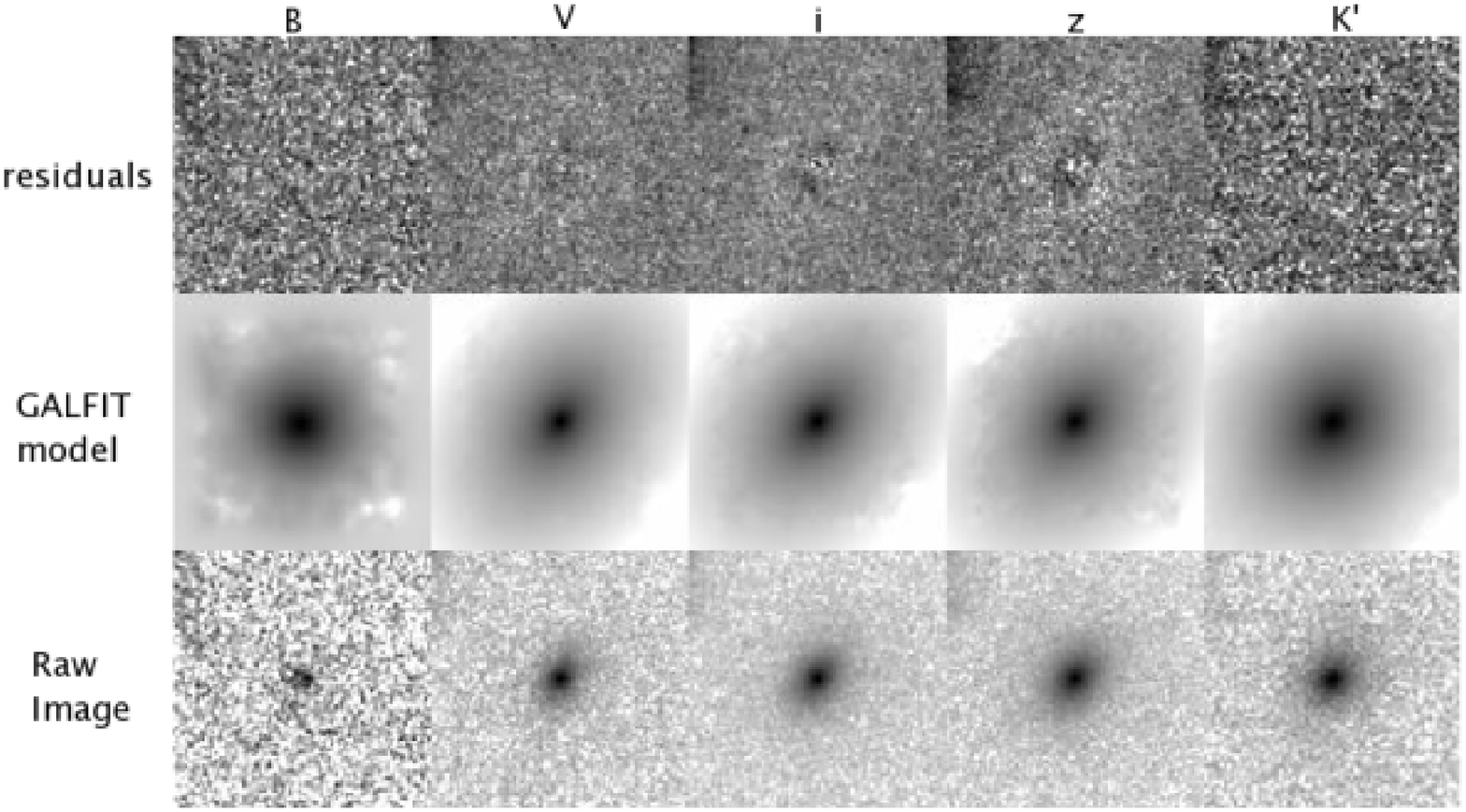}
  \end{center}
\end{figure}
\clearpage
\begin{figure}
  \caption{Tiled imagery and GALFIT models for Normal 8, as in Figure 3.}
  \begin{center}
    \includegraphics[width=5in,height=3.0in]{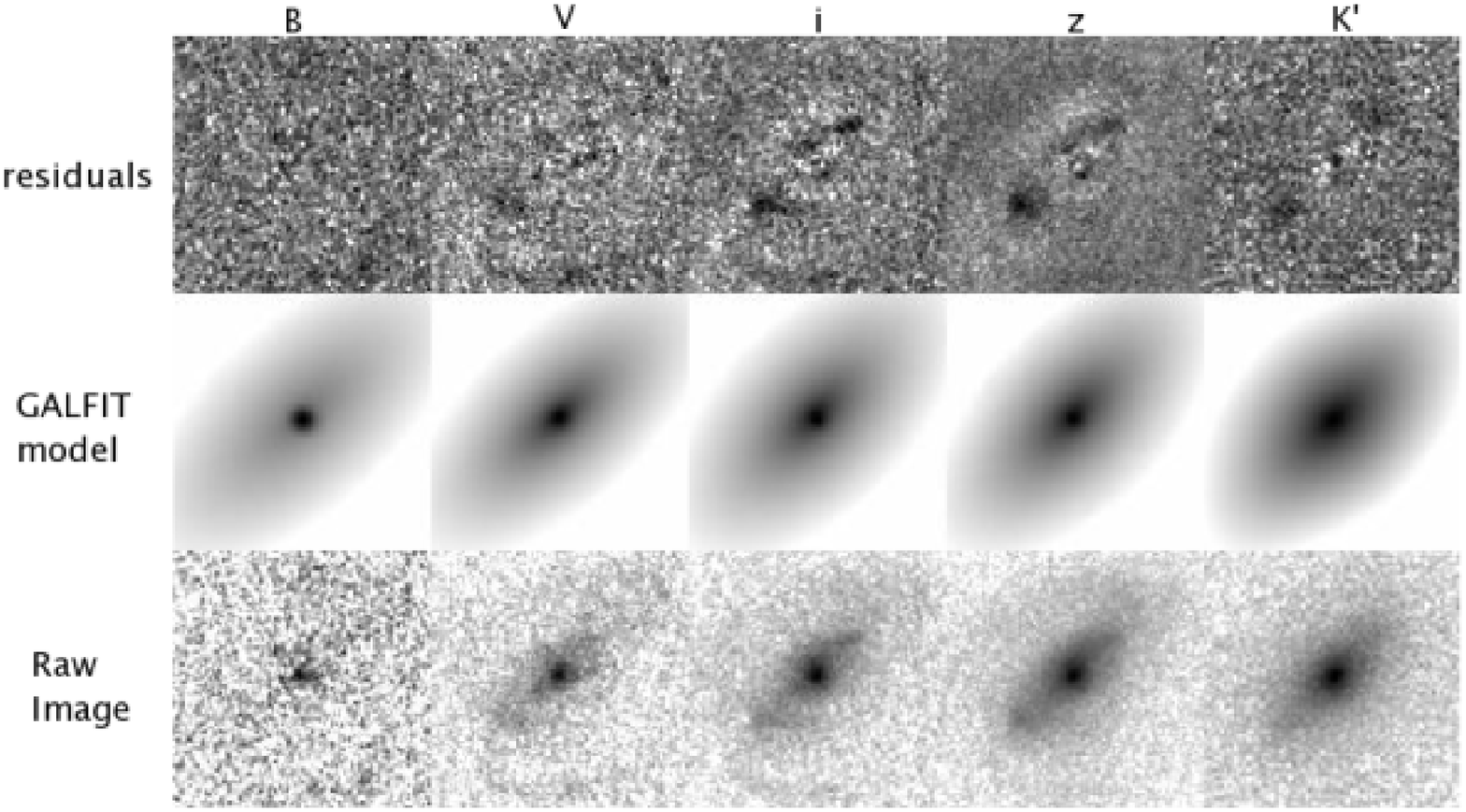}
  \end{center}
\end{figure}

\begin{figure}
  \caption{Tiled imagery and GALFIT models for Normal 9, as in Figure 3.}
  \begin{center}
    \includegraphics[width=5in,height=3.0in]{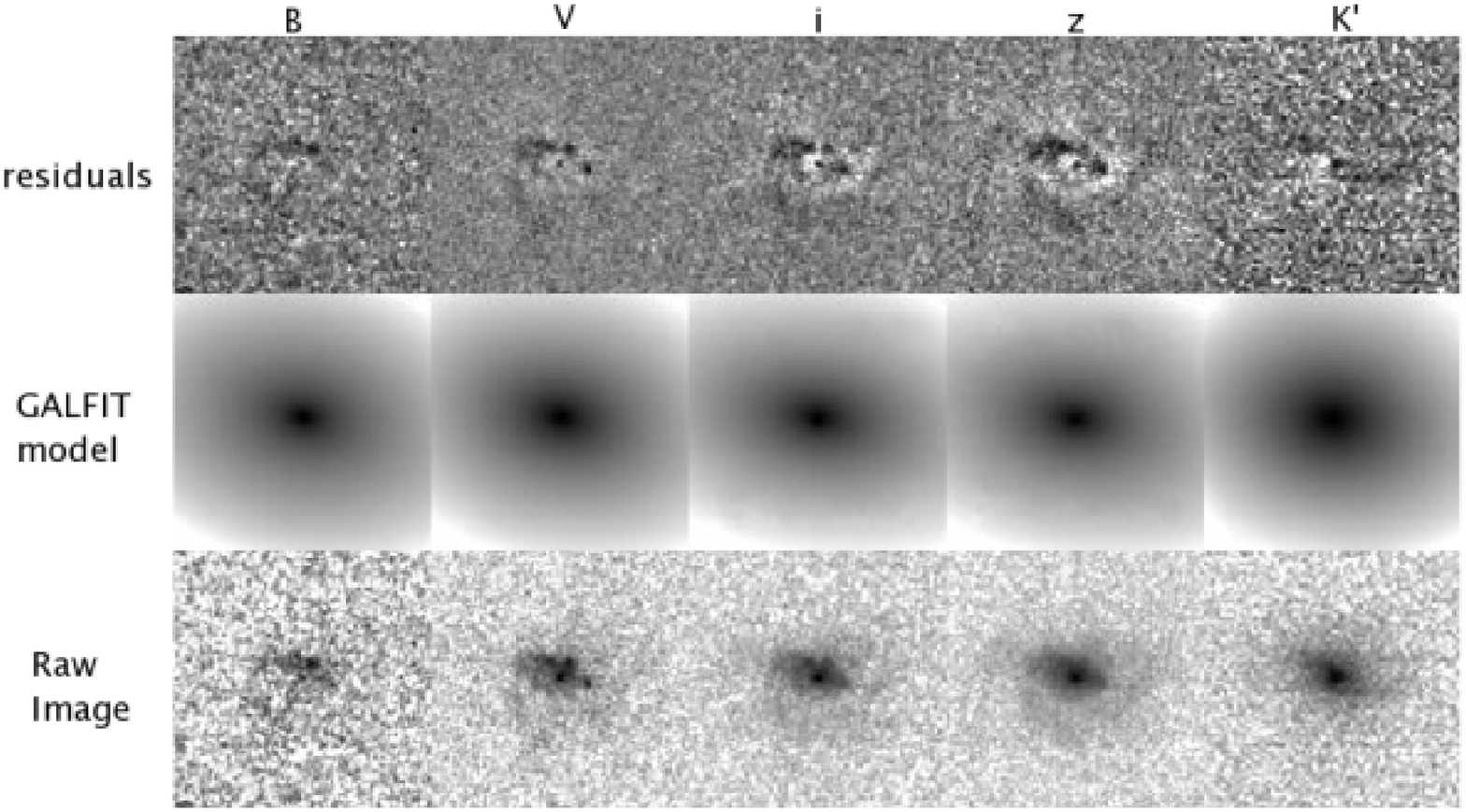}
  \end{center}
\end{figure}
\clearpage
\begin{figure}
  \caption{Radial profiles and fractions of different stellar populations plotted vs. radius for Normal 1, with plots and symbols as in Figure 12.}
  \begin{center}
    \includegraphics[width = 3in, height = 2.0in]{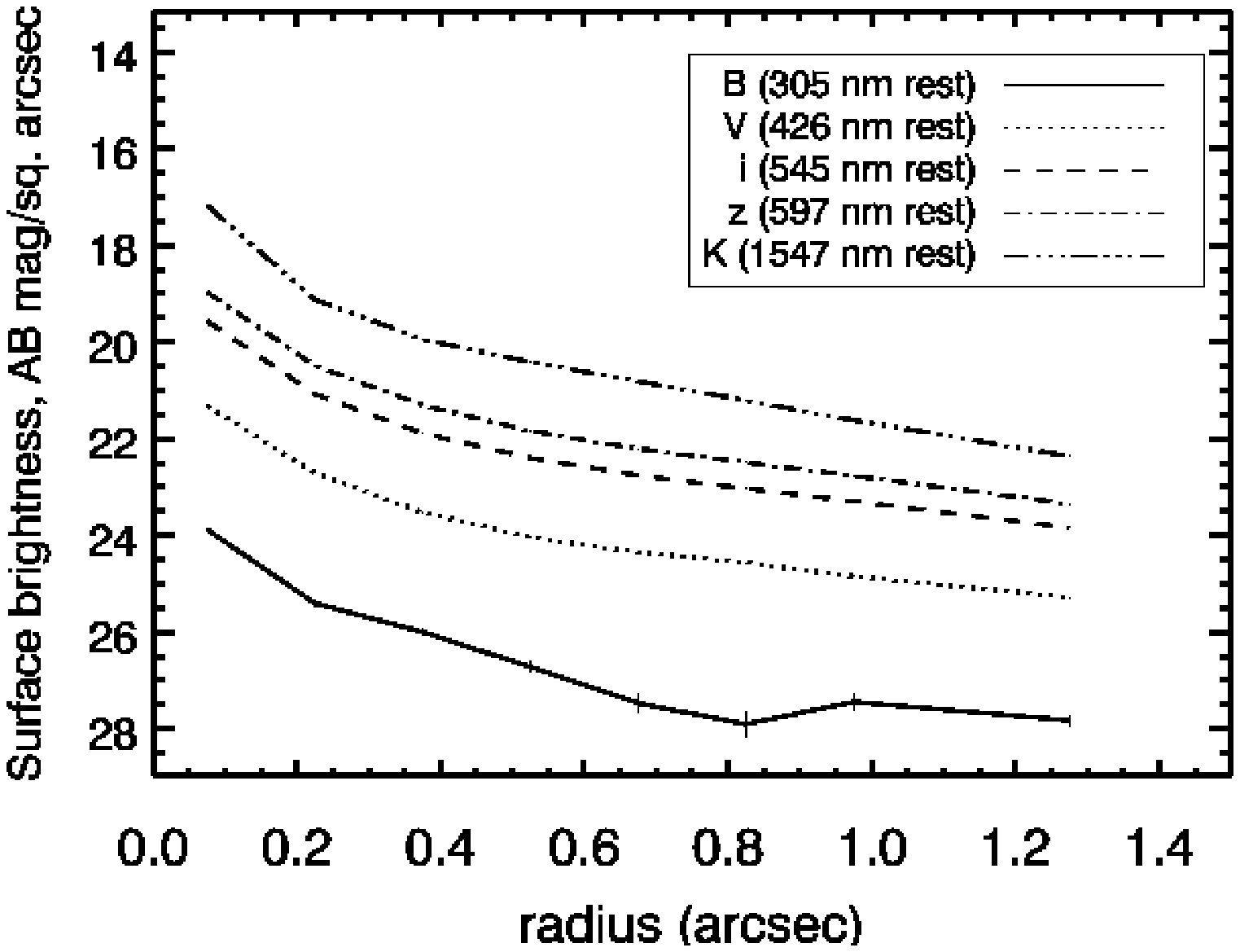}
    \includegraphics[width = 3in, height = 2.0in]{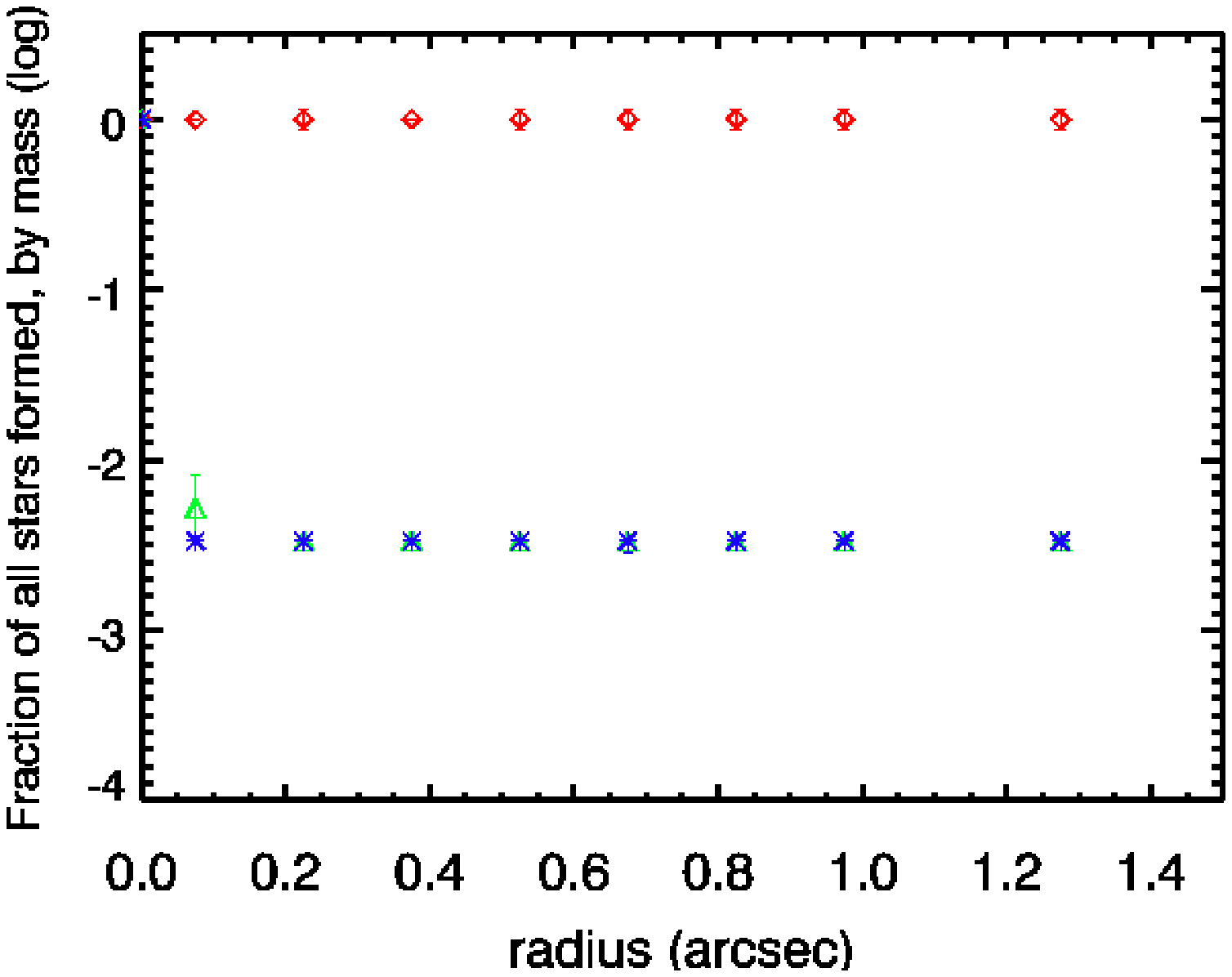}
  \end{center}
\end{figure}

\begin{figure}
  \caption{Radial profiles and fractions of different stellar populations plotted vs. radius for Normal 2, with plots and symbols as in Figure 12.}
  \begin{center}
    \includegraphics[width = 3in, height = 2.0in]{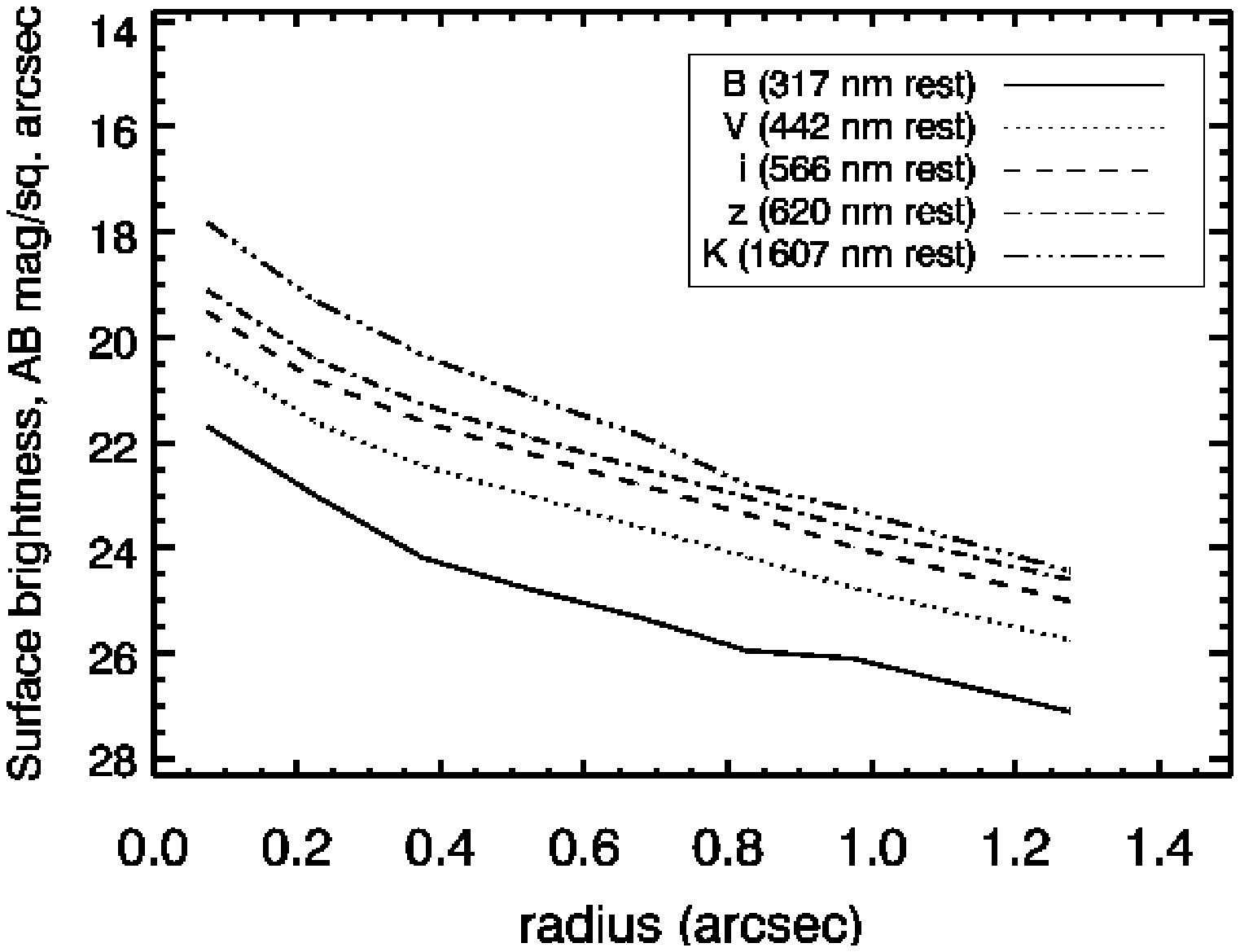}
    \includegraphics[width = 3in, height = 2.0in]{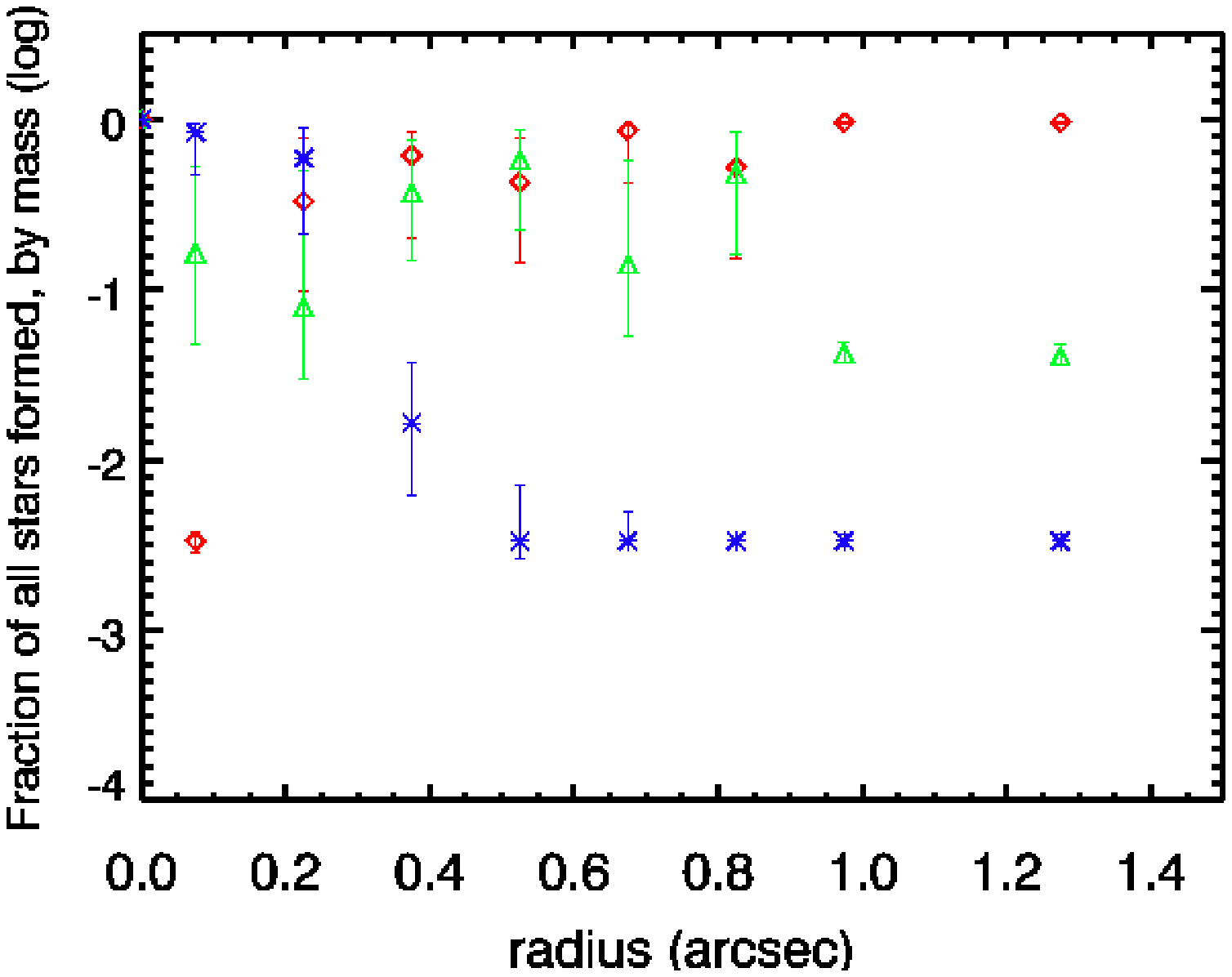}
  \end{center}
\end{figure}
\clearpage
\begin{figure}
  \caption{Radial profiles and fractions of different stellar populations plotted vs. radius for Normal 3, with plots and symbols as in Figure 12.}
  \begin{center}
    \includegraphics[width = 3in, height = 2.0in]{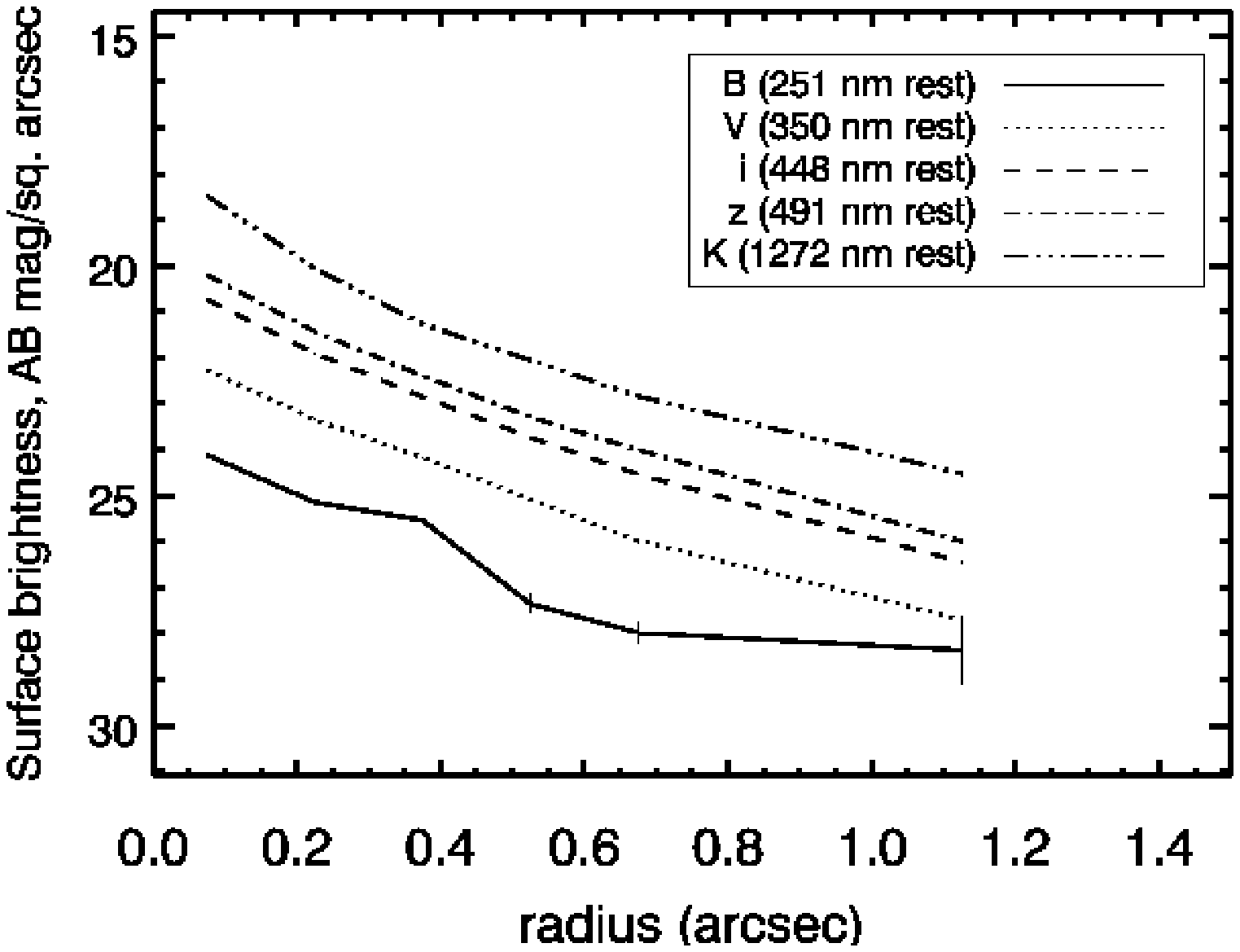}
    \includegraphics[width = 3in, height = 2.0in]{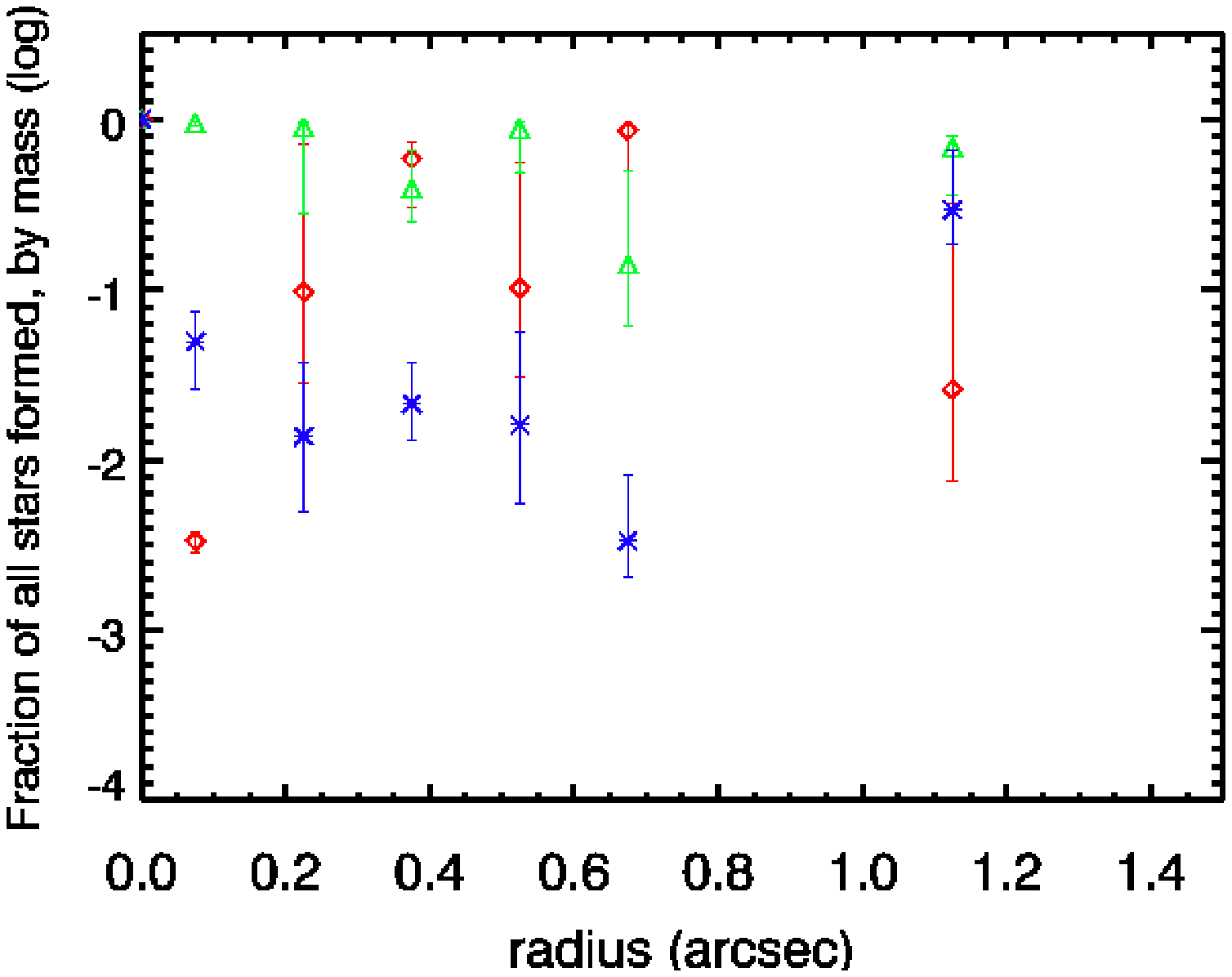}
  \end{center}
\end{figure}
    
\begin{figure}
  \caption{Radial profiles and fractions of different stellar populations plotted vs. radius for Normal 4, with plots and symbols as in Figure 12.}
  \begin{center}
    \includegraphics[width = 3in, height = 2.0in]{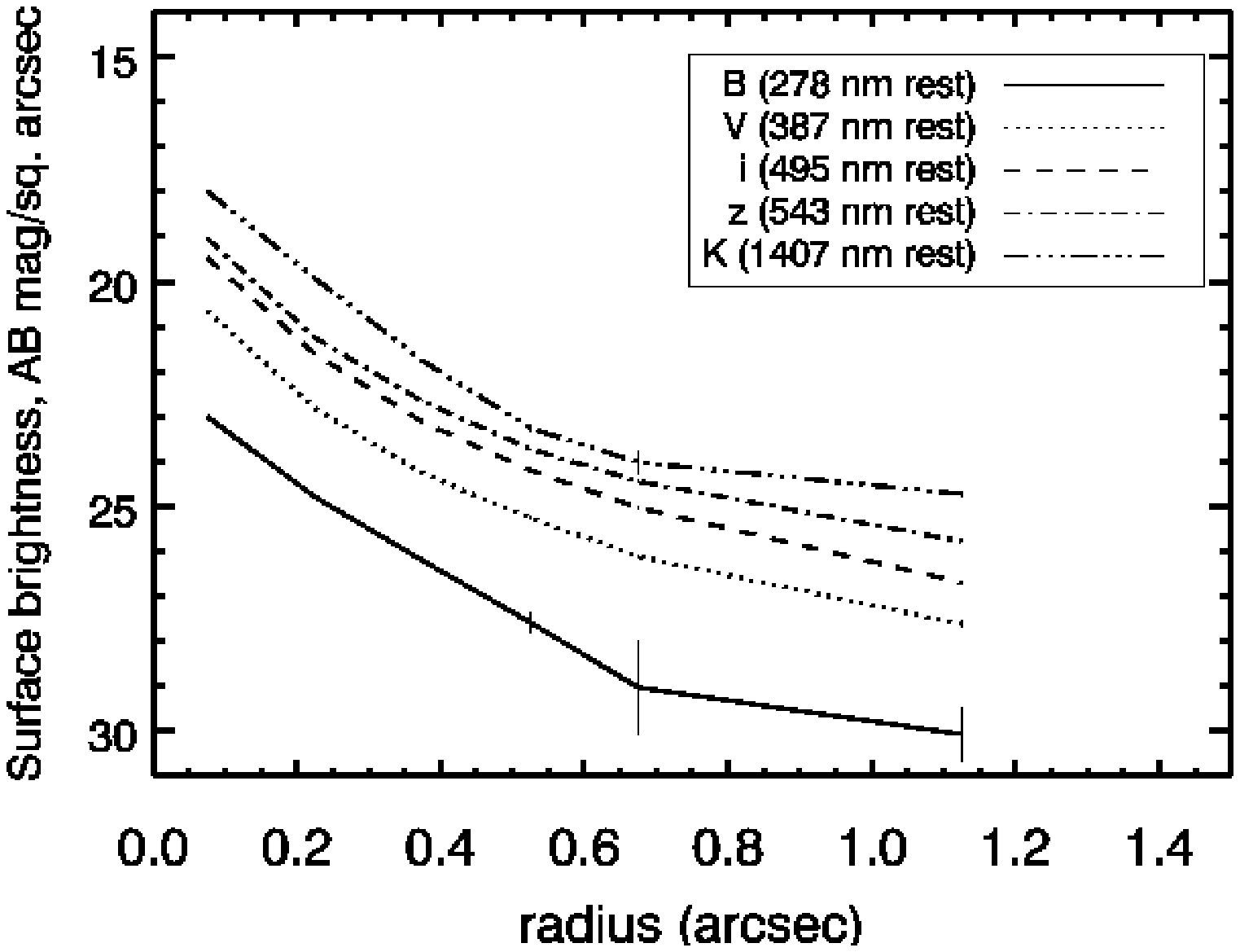}
    \includegraphics[width = 3in, height = 2.0in]{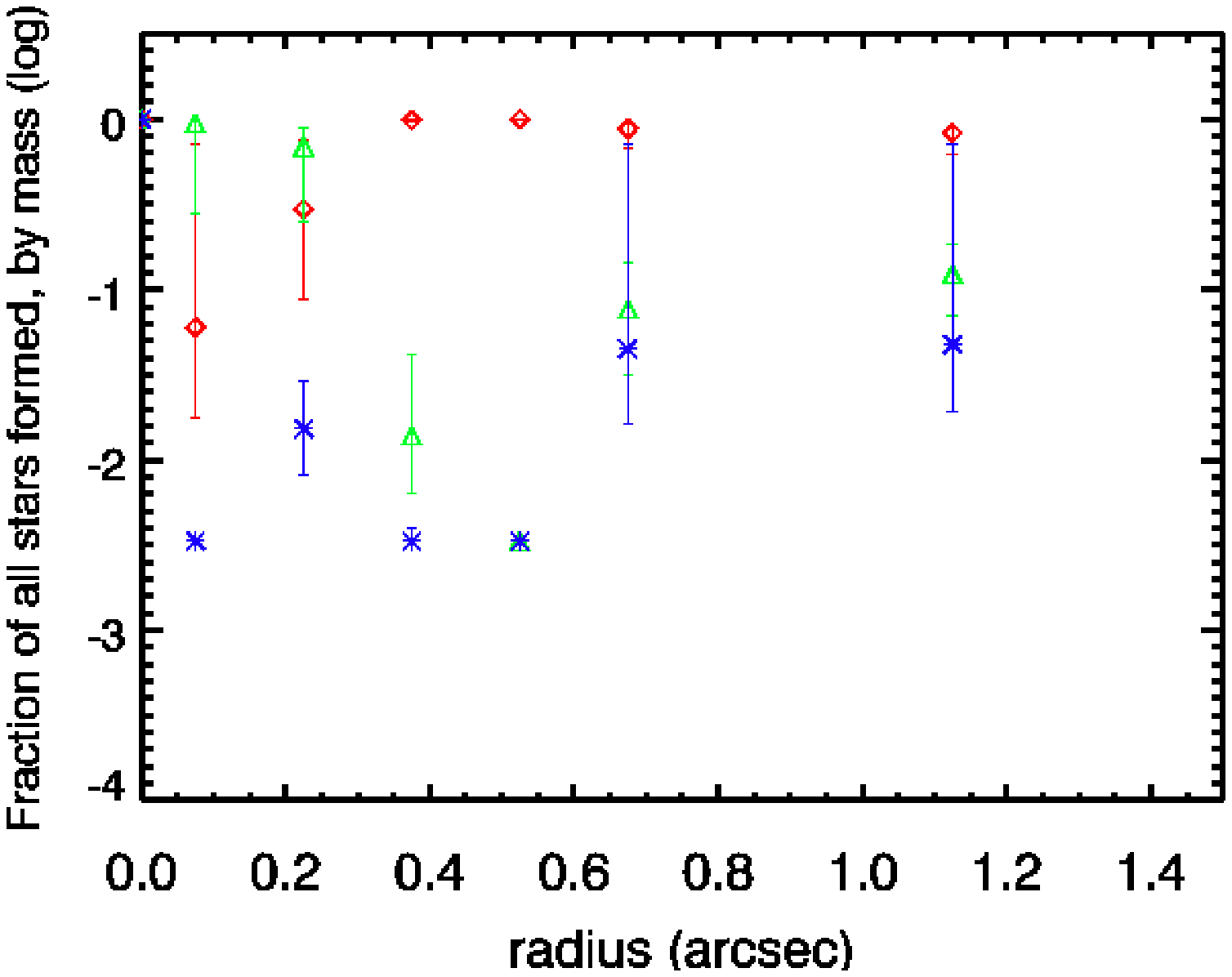}
  \end{center}
\end{figure}
 \clearpage
\begin{figure}
  \caption{Radial profiles and fractions of different stellar populations plotted vs. radius for Normal 5, with plots and symbols as in Figure 12.}
  \begin{center}
    \includegraphics[width = 3in, height = 2.0in]{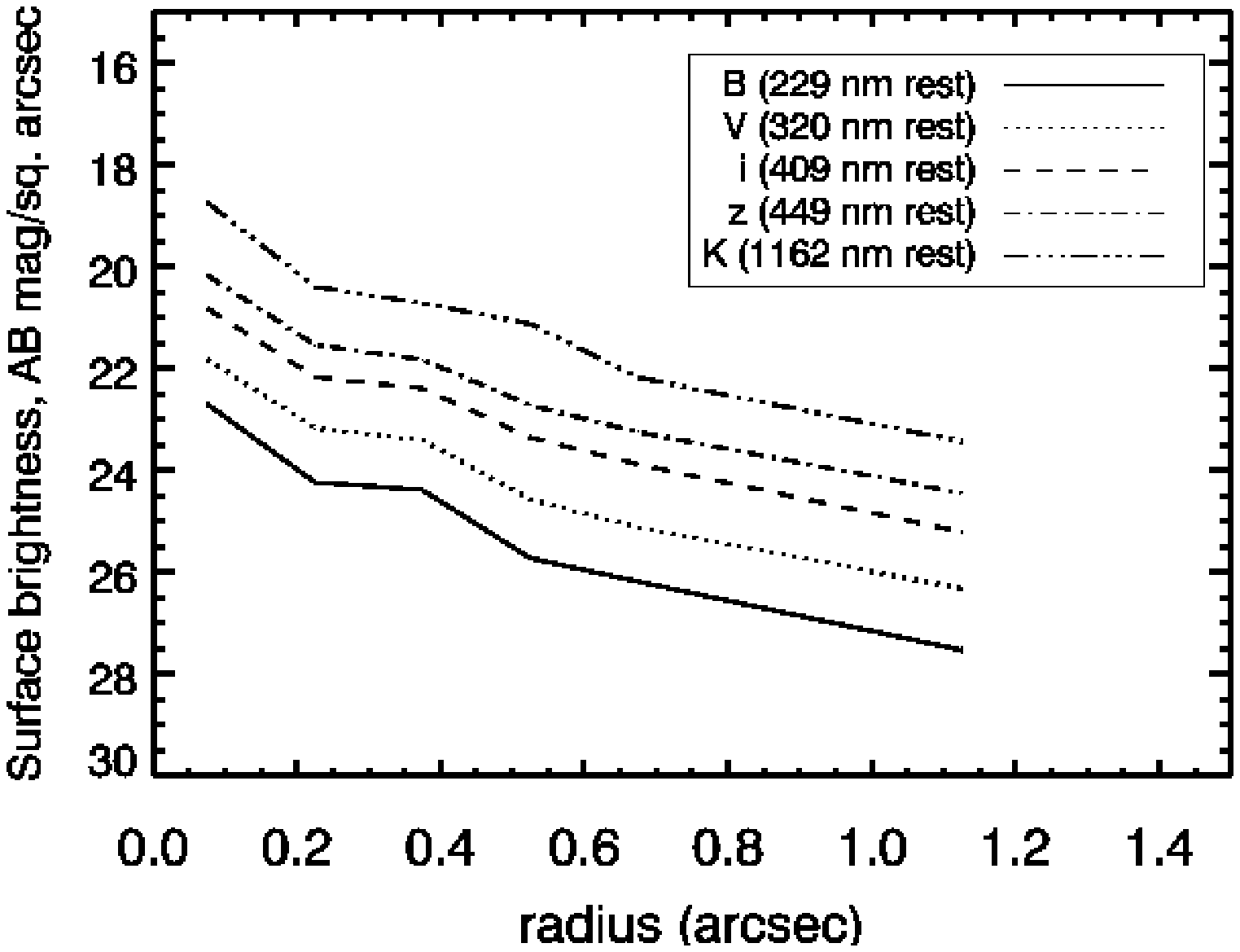}
    \includegraphics[width = 3in, height = 2.0in]{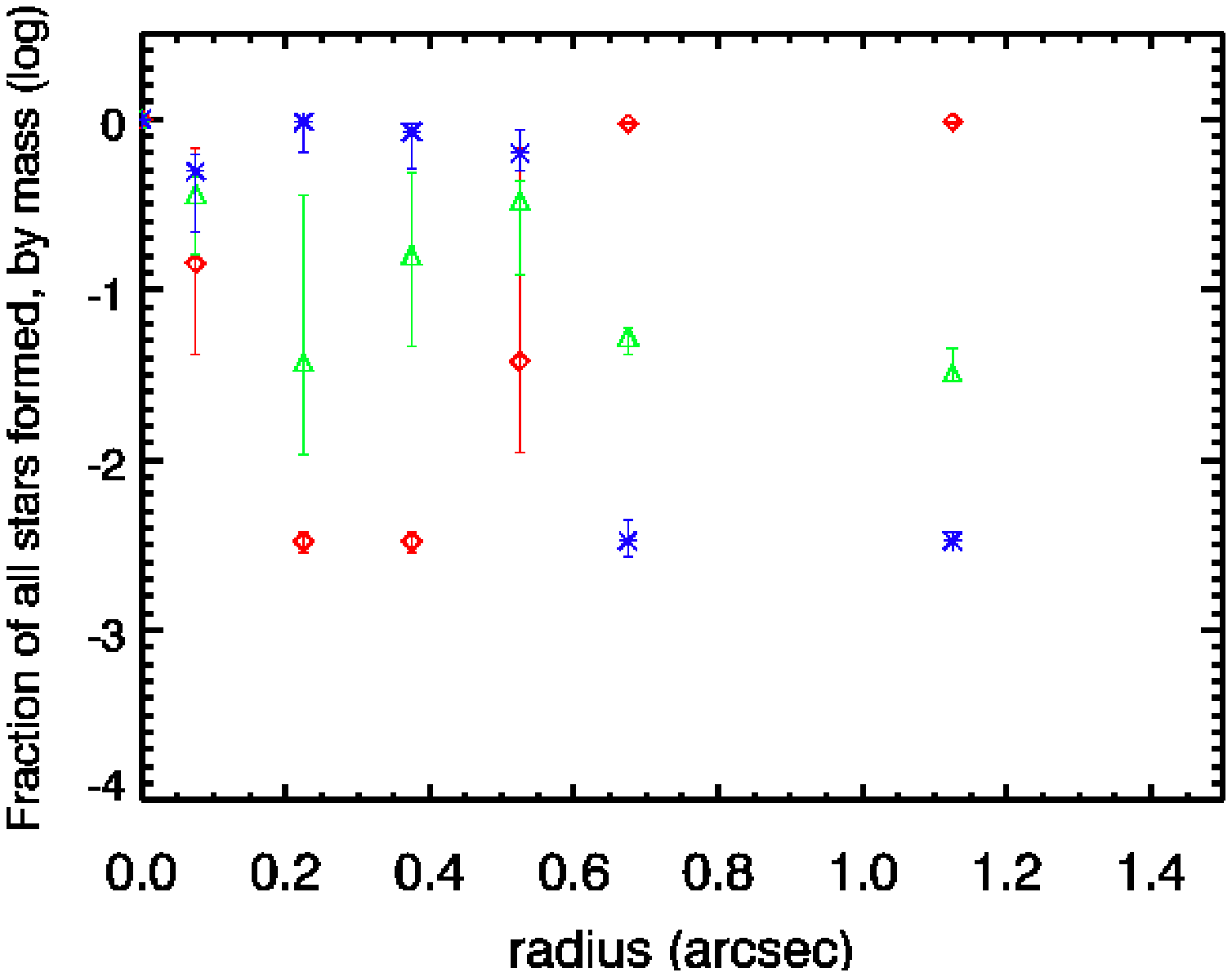}
  \end{center}
\end{figure}
    
\begin{figure}
  \caption{Radial profiles and fractions of different stellar populations plotted vs. radius for Normal 6, with plots and symbols as in Figure 12.}
  \begin{center}
    \includegraphics[width = 3in, height = 2.0in]{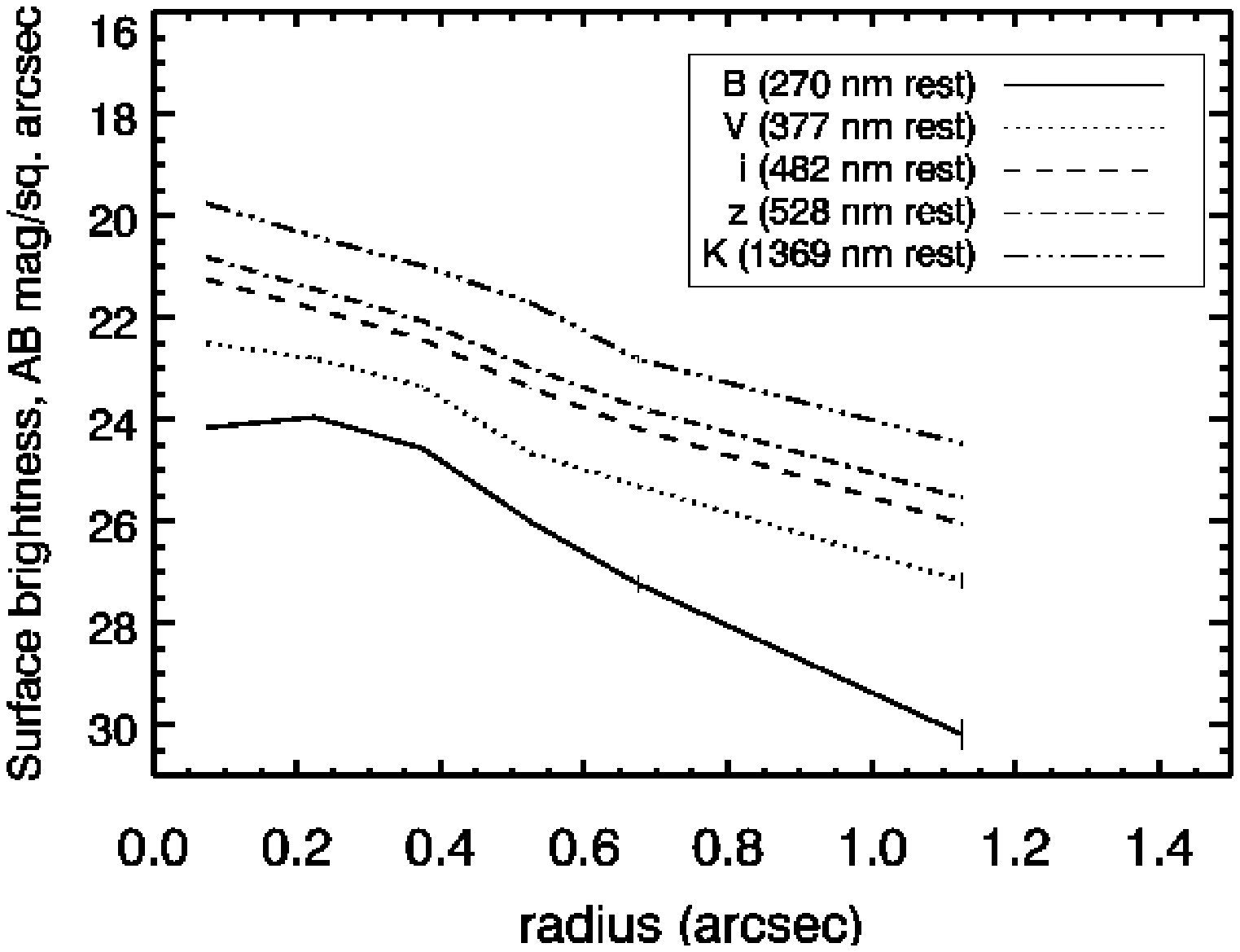}
    \includegraphics[width = 3in, height = 2.0in]{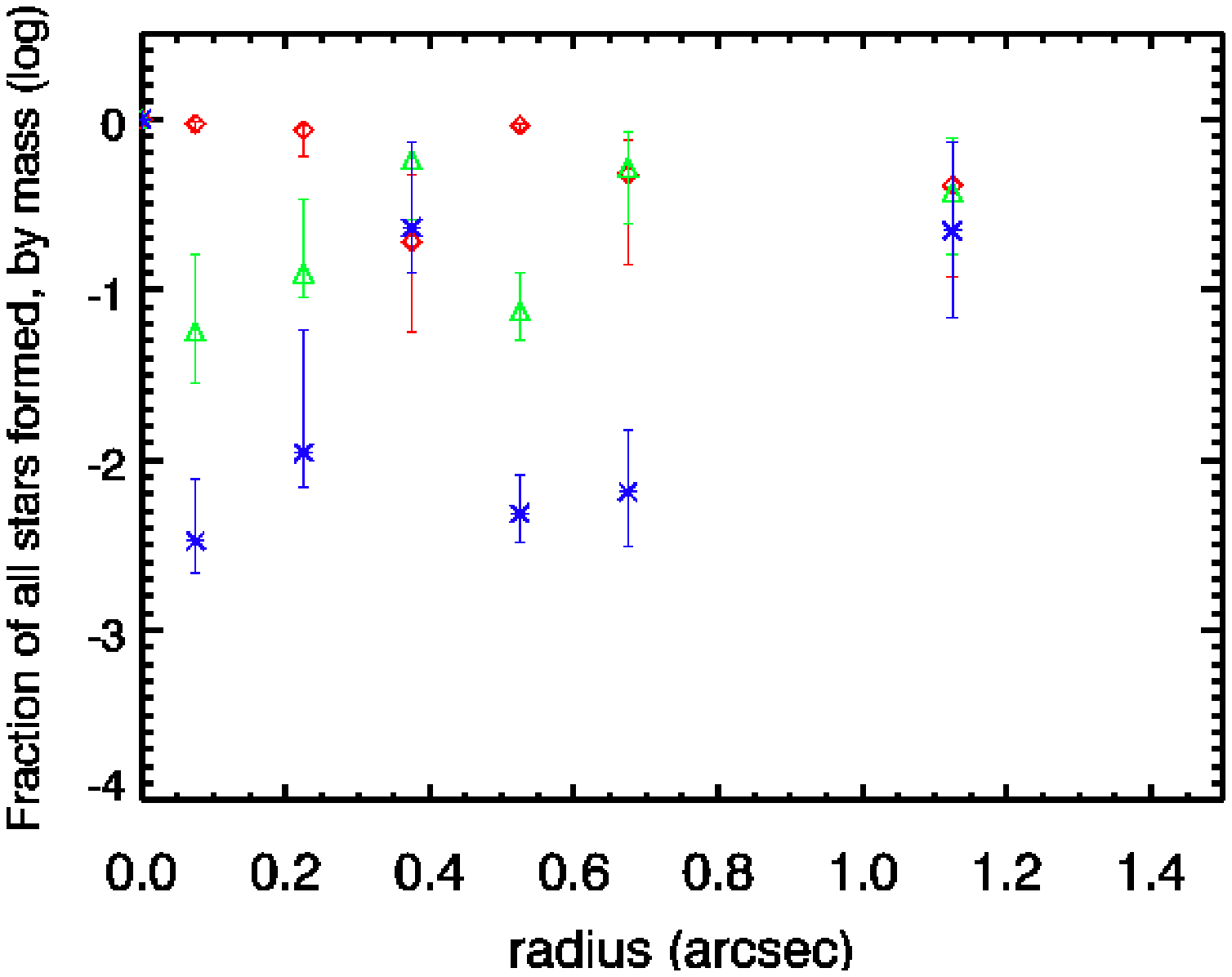}
  \end{center}
\end{figure}
    \clearpage
\begin{figure}
  \caption{Radial profiles and fractions of different stellar populations plotted vs. radius for Normal 7, with plots and symbols as in Figure 12.}
  \begin{center}
    \includegraphics[width = 3in, height = 2.0in]{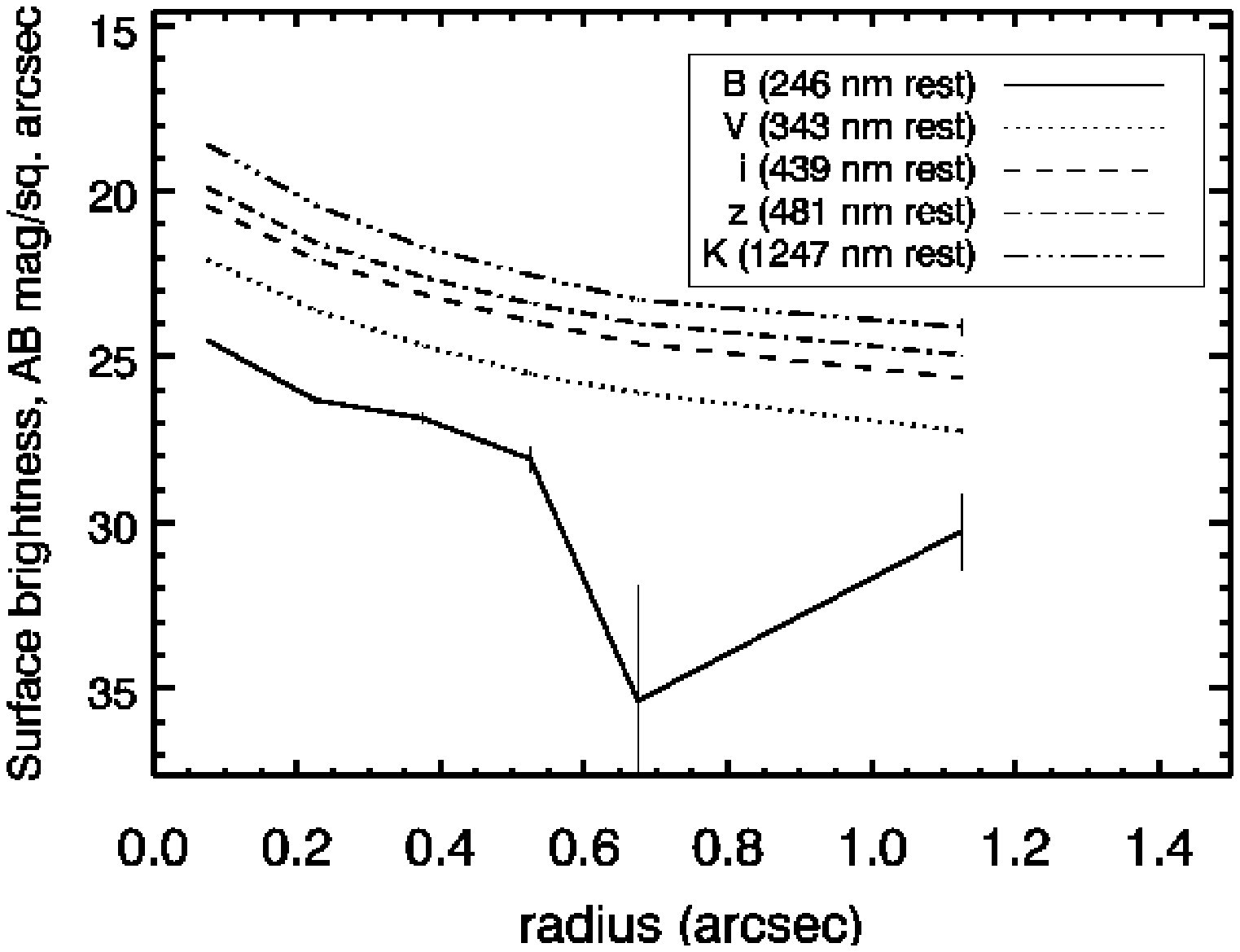}
    \includegraphics[width = 3in, height = 2.0in]{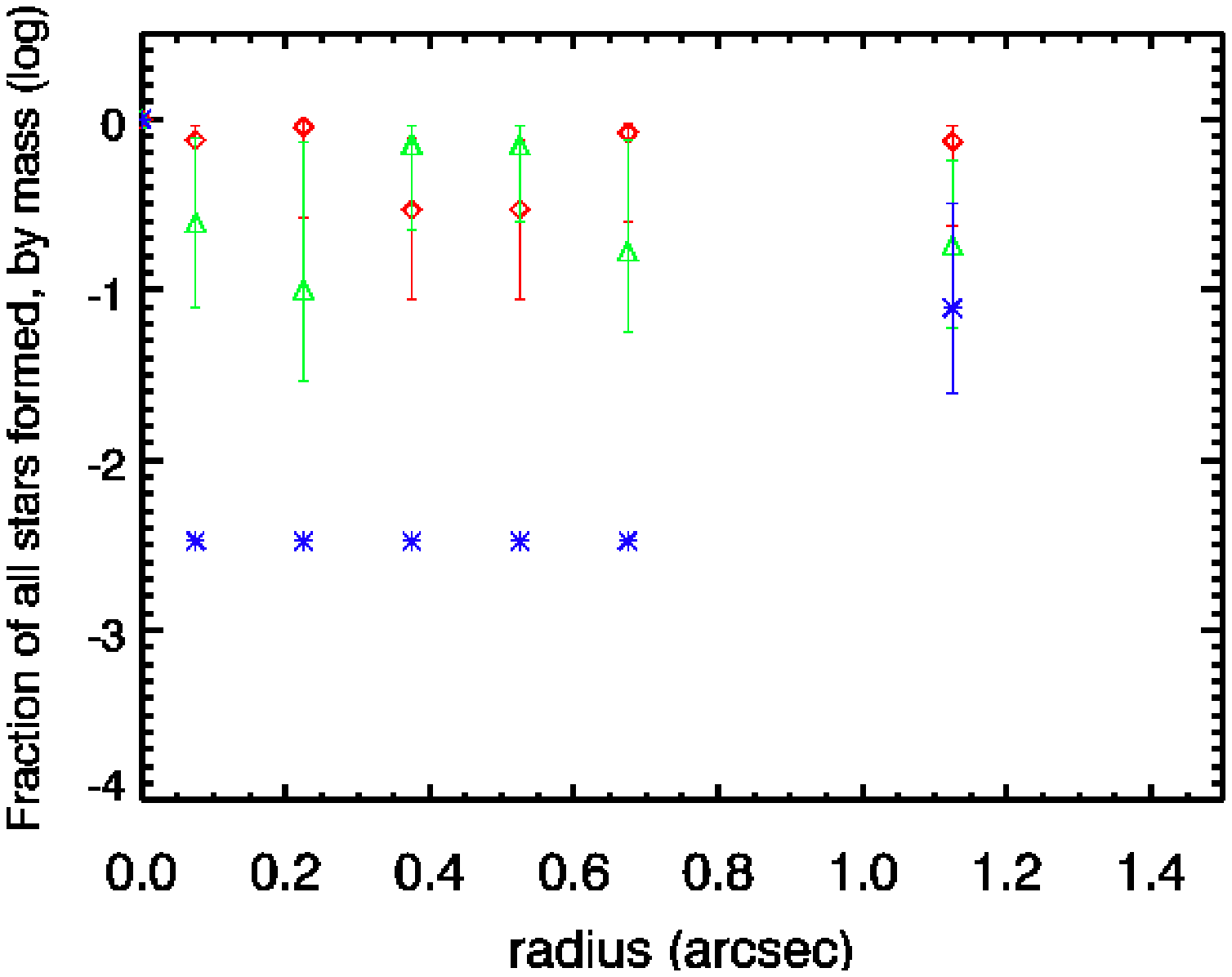}
  \end{center}
\end{figure}

\begin{figure}
  \caption{Radial profiles and fractions of different stellar populations plotted vs. radius for Normal 8, with plots and symbols as in Figure 12.}
  \begin{center}
    \includegraphics[width = 3in, height = 2.0in]{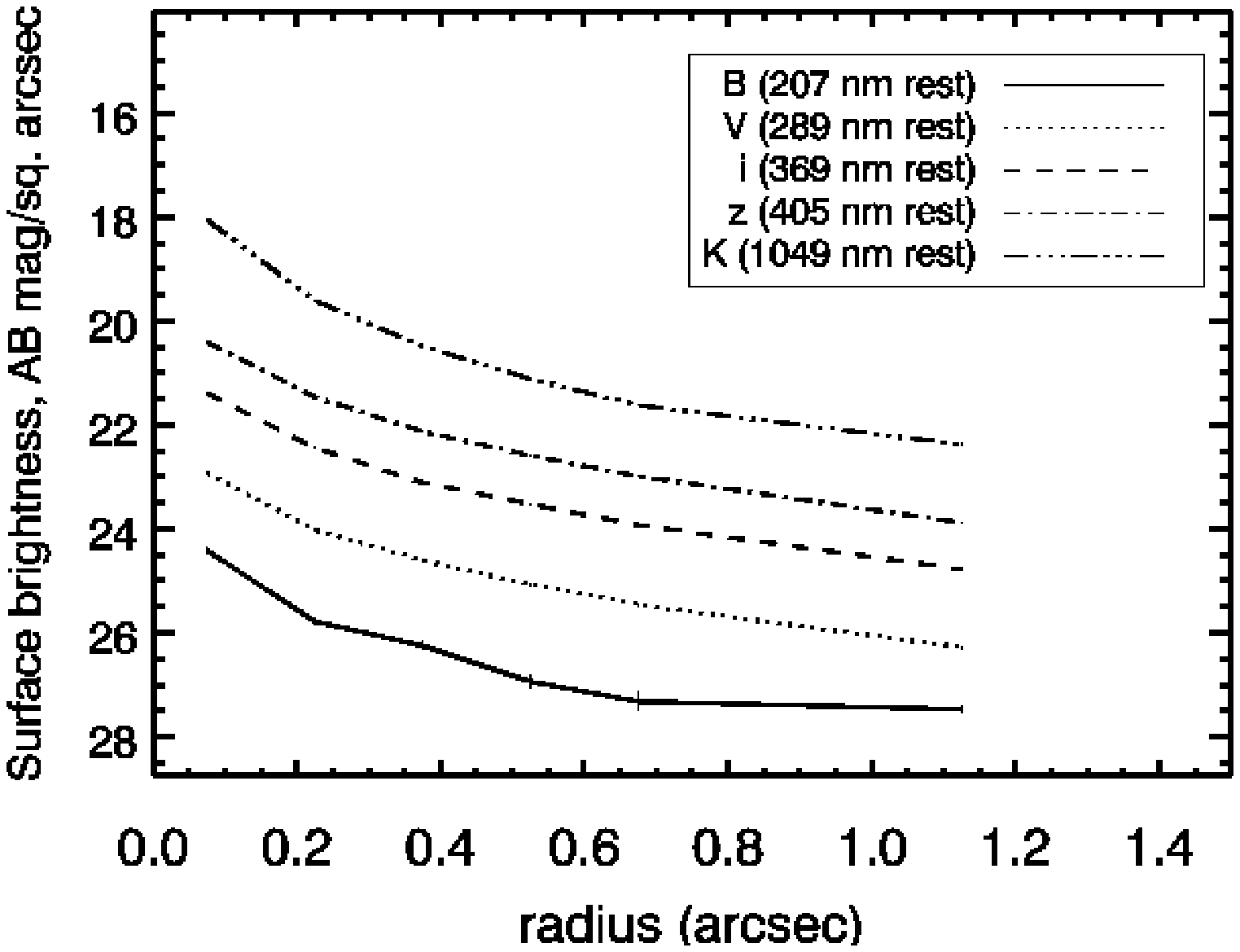}
    \includegraphics[width = 3in, height = 2.0in]{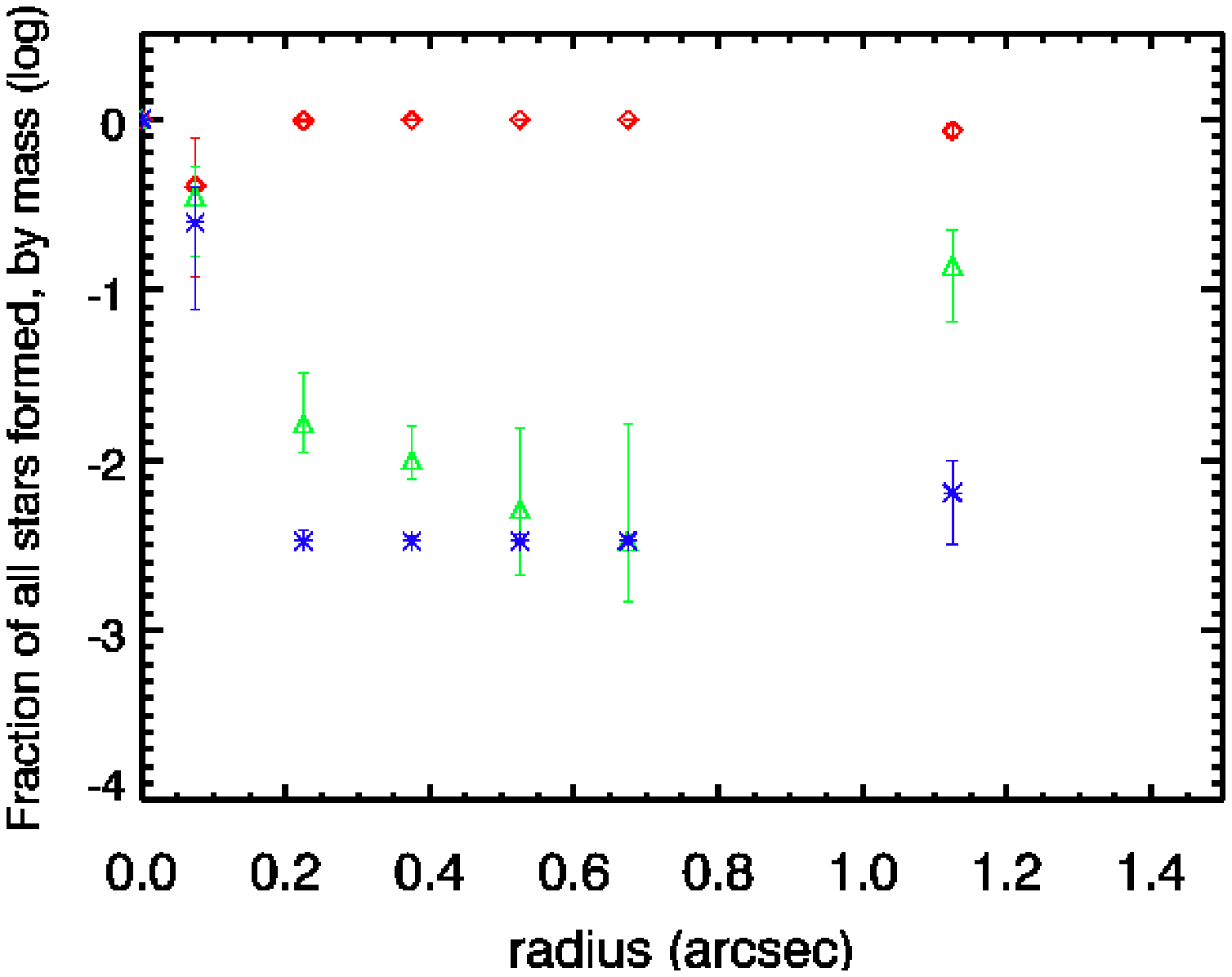}
  \end{center}
\end{figure}
\clearpage
\begin{figure}
  \caption{Radial profiles and fractions of different stellar populations plotted vs. radius for Normal 9, with plots and symbols as in Figure 12.}
  \begin{center}
    \includegraphics[width = 3in, height = 2.0in]{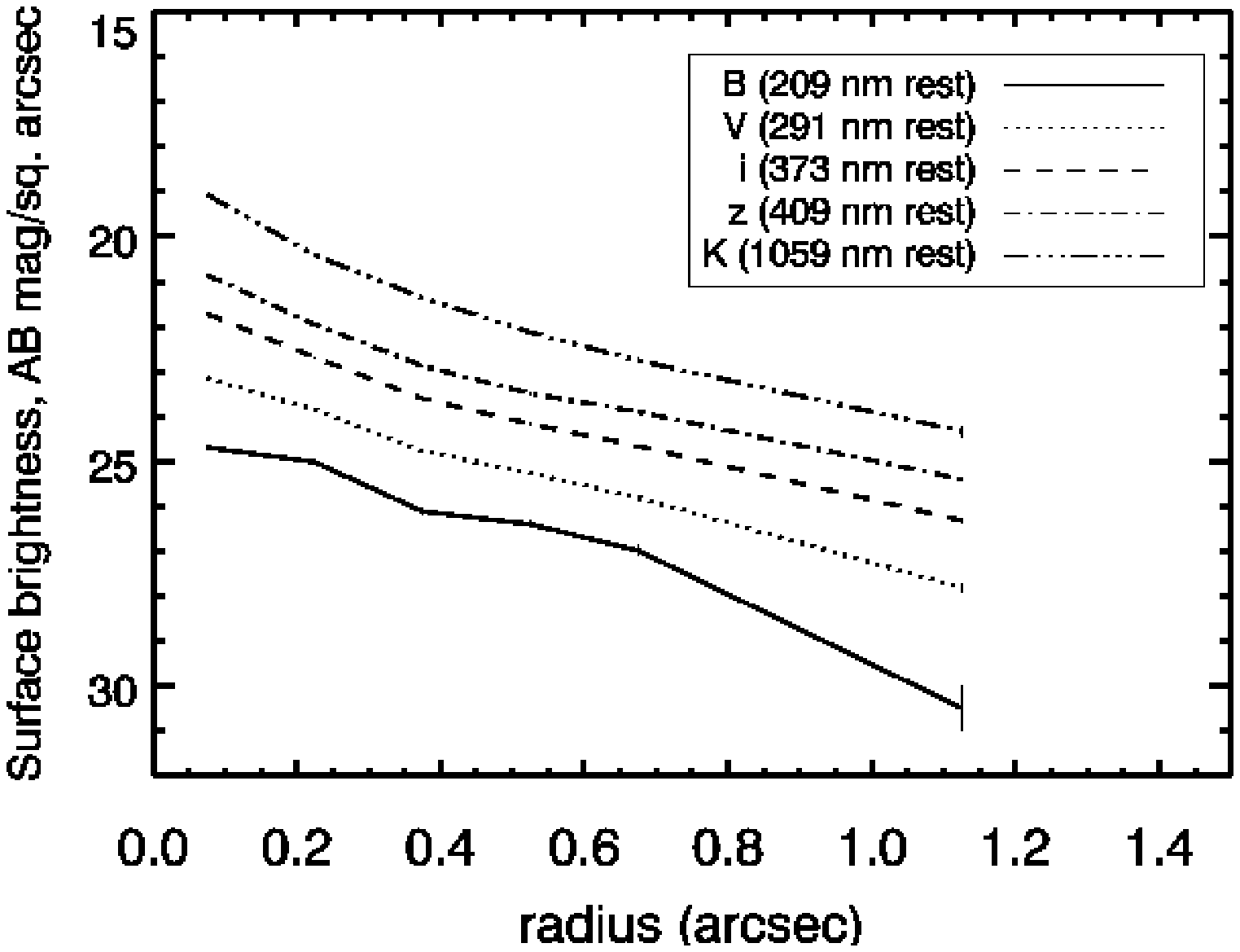}
    \includegraphics[width = 3in, height = 2.0in]{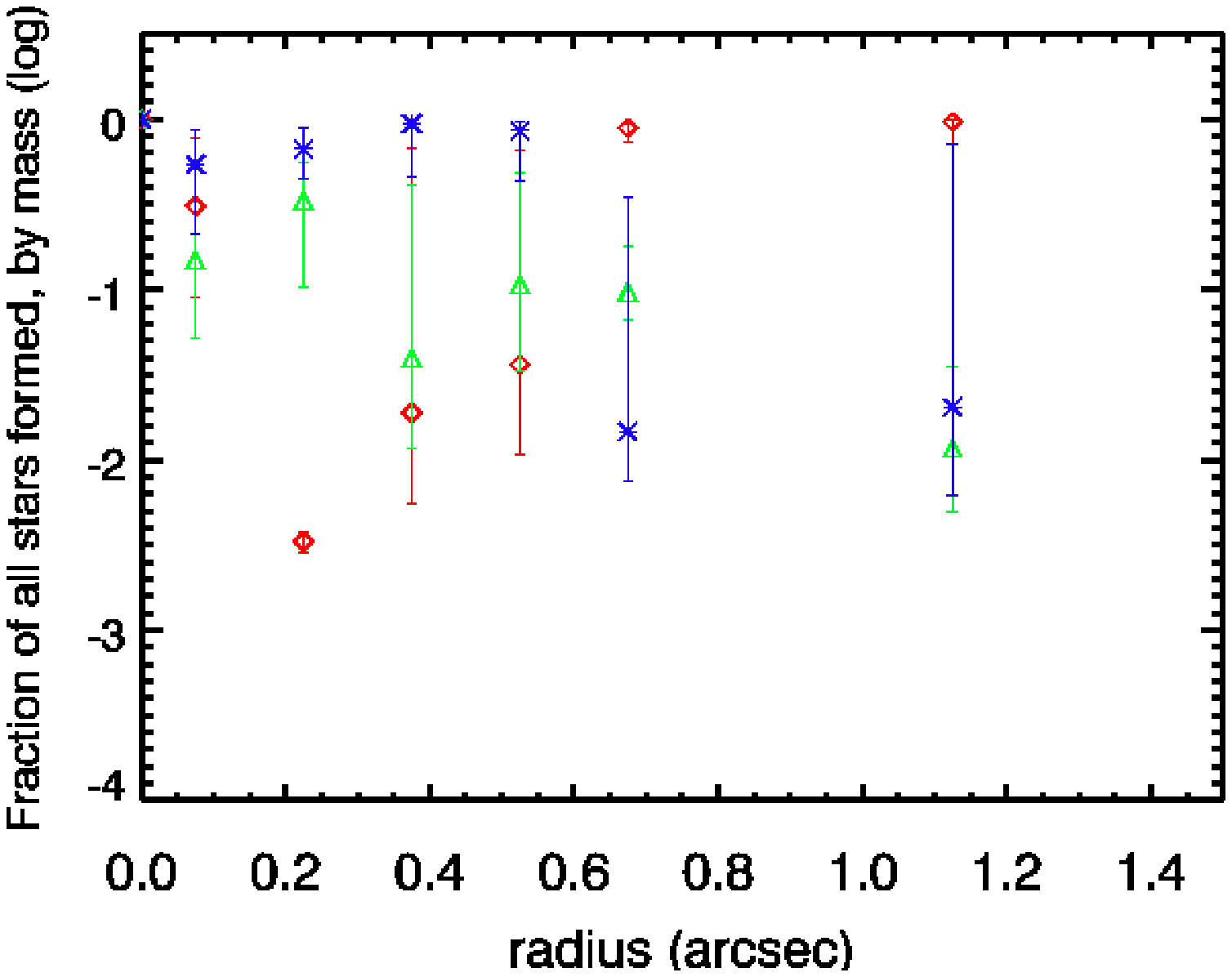}
  \end{center}
\end{figure}              

\begin{table}
\begin{center}
\caption{Table of fields imaged with Keck LGSAO and NGSAO.\label{tbl-1}}
\begin{tabular}[t]{lllllll}
\tableline\tableline
\multicolumn{1}{l}{Field} & \multicolumn{1}{l}{Tip-Tilt Star Coord.} & \multicolumn{1}{l} {Offset\tablenotemark{c}} & \multicolumn{1}{l}{R\tablenotemark{a}} & \multicolumn{1}{l}{D\tablenotemark{b}}  & \multicolumn{1}{l}{Date} & \multicolumn{1}{l}{Sources in Image}\\
\tableline

GS 40s&3\raisebox{0.5ex}{\scriptsize h}32\raisebox{0.5ex}{\scriptsize m}13.8\raisebox{0.5ex}{\scriptsize s} -27$^{\circ}$42$^\prime$13$\farcs$9 & S & 15.0 & 20.6& 2005 Sep 28 & XID 266,56,83 \\
GS 33&3\raisebox{0.5ex}{\scriptsize h}32\raisebox{0.5ex}{\scriptsize m}55.5\raisebox{0.5ex}{\scriptsize s} -27$^{\circ}$51$^\prime$6$\farcs$4 & W & 15.4 & 22.2 & 2005 Sep 28 & XID 15, Norm5-6\\
GS 68w&3\raisebox{0.5ex}{\scriptsize h}32\raisebox{0.5ex}{\scriptsize m}10.3\raisebox{0.5ex}{\scriptsize s} -27$^{\circ}$43$^\prime$7$\farcs$2&NW&16.0 & 21.1& 2005 Sep 27&XID 155,536,594,Norm1-4\\
GS 25&3\raisebox{0.5ex}{\scriptsize h}32\raisebox{0.5ex}{\scriptsize m}36.1\raisebox{0.5ex}{\scriptsize s} -27$^{\circ}$40$^\prime$5$\farcs$0& NE & 16.9 & 21.7& 2006 Dec 11  & XID 32\\
GS 71&3\raisebox{0.5ex}{\scriptsize h}32\raisebox{0.5ex}{\scriptsize m}31.9\raisebox{0.5ex}{\scriptsize s} -27$^{\circ}$41$^\prime$5$\farcs$6& W & 16.1 & 21.0 & 2006 Dec 11 & Norm7-9\\
\tableline
\end{tabular}
\tablenotetext{a}{R-band magnitude of guide star}
\tablenotetext{b}{distance between guide star and center of science field in arcseconds}
\tablenotetext{c}{Direction from guide star to center of science field}
\end{center}
\end{table}

\begin{table}
\begin{center}
\caption{Table of critical parameters of our AGN and comparison samples.\label{tbl-2}}
\begin{tabular}[t]{llllllll}
\tableline\tableline
\multicolumn{1}{l}{XID \#\tablenotemark{a}} & \multicolumn{1}{l}{Coordinates (J2000)\tablenotemark{a}} & \multicolumn{1}{l} {Redshift\tablenotemark{b}} & \multicolumn{1}{l}{z-type\tablenotemark{b}} &  \multicolumn{1}{l}{B (AB)\tablenotemark{c}} & \multicolumn{1}{l}{R (Vega)\tablenotemark{a}} & \multicolumn{1}{l}{X-ray class\tablenotemark{d}} \\
\tableline
83 & 3\raisebox{0.5ex}{\scriptsize h}32\raisebox{0.5ex}{\scriptsize m}14.98\raisebox{0.5ex}{\scriptsize s} -27$^{\circ}$42$^\prime$24$\farcs$9 & 1.76 & Photo & 25.13 & 23.08 & AGN I\\
155 & 3\raisebox{0.5ex}{\scriptsize h}32\raisebox{0.5ex}{\scriptsize m}7.98\raisebox{0.5ex}{\scriptsize s} -27$^{\circ}$42$^\prime$39$\farcs$4 &0.55 & Spect & 23.61 & 22.37 & AGN II\\
266 & 3\raisebox{0.5ex}{\scriptsize h}32\raisebox{0.5ex}{\scriptsize m}13.86\raisebox{0.5ex}{\scriptsize s} -27$^{\circ}$42$^\prime$49$\farcs$0 & 0.73 & Spect & 23.70 & 21.70 & AGN II\\
56 & 3\raisebox{0.5ex}{\scriptsize h}32\raisebox{0.5ex}{\scriptsize m}13.24\raisebox{0.5ex}{\scriptsize s} -27$^{\circ}$42$^\prime$41$\farcs$1 & 0.61 & Spect & 21.68 & 20.18 & AGN II\\
536 & 3\raisebox{0.5ex}{\scriptsize h}32\raisebox{0.5ex}{\scriptsize m}10.76\raisebox{0.5ex}{\scriptsize s} -27$^{\circ}$42$^\prime$34$\farcs$6 & 0.42 & Spect & 21.99 &  19.39 & galaxy\\
594 & 3\raisebox{0.5ex}{\scriptsize h}32\raisebox{0.5ex}{\scriptsize m}9.71\raisebox{0.5ex}{\scriptsize s} -27$^{\circ}$42$^\prime$48$\farcs$2 & 0.73 & Spect & 24.71 &  21.52 & AGN I\\
15 & 3\raisebox{0.5ex}{\scriptsize h}32\raisebox{0.5ex}{\scriptsize m}52.88\raisebox{0.5ex}{\scriptsize s} -27$^{\circ}$51$^\prime$20$\farcs$0 & 1.23 & Spect & 23.51 & 22.62 & AGN I\\
32 & 3\raisebox{0.5ex}{\scriptsize h}32\raisebox{0.5ex}{\scriptsize m}37.47\raisebox{0.5ex}{\scriptsize s} -27$^{\circ}$40$^\prime$0$\farcs$3 & 0.66 & Spect & 23.52 & 22.36 & AGN I\\
Norm1 & 3\raisebox{0.5ex}{\scriptsize h}32\raisebox{0.5ex}{\scriptsize m}10.38\raisebox{0.5ex}{\scriptsize s} -27$^{\circ}$43$^\prime$0$\farcs$9 & 0.422 & Photo & 23.12 & 22.1 & \ldots \\	
Norm2 & 3\raisebox{0.5ex}{\scriptsize h}32\raisebox{0.5ex}{\scriptsize m}9.23\raisebox{0.5ex}{\scriptsize s} -27$^{\circ}$42$^\prime$51$\farcs$1 & 0.369 & Photo & 24.64 & 21.4 & \ldots   \\	
Norm3 & 3\raisebox{0.5ex}{\scriptsize h}32\raisebox{0.5ex}{\scriptsize m}9.40\raisebox{0.5ex}{\scriptsize s} -27$^{\circ}$42$^\prime$36$\farcs$1 & 0.729 & Photo & 25.43 &  23.1  & \ldots \\	
Norm4 & 3\raisebox{0.5ex}{\scriptsize h}32\raisebox{0.5ex}{\scriptsize m}8.41\raisebox{0.5ex}{\scriptsize s} -27$^{\circ}$42$^\prime$31$\farcs$3 & 0.563 & Photo & 25.02 & 22.4 & \ldots \\	
Norm5 & 3\raisebox{0.5ex}{\scriptsize h}32\raisebox{0.5ex}{\scriptsize m}52.87\raisebox{0.5ex}{\scriptsize s} -27$^{\circ}$51$^\prime$14$\farcs$74 & 0.892 & Photo & 24.03 & 22.48 & \ldots \\
Norm6 & 3\raisebox{0.5ex}{\scriptsize h}32\raisebox{0.5ex}{\scriptsize m}55.01\raisebox{0.5ex}{\scriptsize s} -27$^{\circ}$50$^\prime$51$\farcs$64& 0.607 & Photo & 24.41 & 22.65 & \ldots \\
Norm7 & 3\raisebox{0.5ex}{\scriptsize h}32\raisebox{0.5ex}{\scriptsize m}31.65\raisebox{0.5ex}{\scriptsize s} -27$^{\circ}$41$^\prime$14$\farcs$28 &  0.764  & Photo &25.72&22.25 & \ldots \\ 
Norm8 & 3\raisebox{0.5ex}{\scriptsize h}32\raisebox{0.5ex}{\scriptsize m}32.99\raisebox{0.5ex}{\scriptsize s} -27$^{\circ}$41$^\prime$17$\farcs$06 &1.096& Photo & 24.46 & 22.57 & \ldots  \\
Norm9 & 3\raisebox{0.5ex}{\scriptsize h}32\raisebox{0.5ex}{\scriptsize m}32.90\raisebox{0.5ex}{\scriptsize s} -27$^{\circ}$41$^\prime$23$\farcs$93 &  1.076 & Photo&24.76 & 23.52 & \ldots  \\ 

\tableline
\end{tabular}
\tablenotetext{a}{XID numbers, coordinates, and ground-based R-band magnitudes (Vega) for AGN are from \citet{gia02}.}
\tablenotetext{b}{Spectroscopic redshifts are from \citet{szo04} and photometric redshifts are from \citet{zhe04}.}
\tablenotetext{c}{B-band magnitudes are measured directly from the ACS F435W (B) images, using a $3\farcs$ diameter aperture.}
\tablenotetext{d}{X-ray class is in \citet{szo04} and \citet{zhe04}.}
\end{center}
\end{table}

 \begin{table}
\begin{center}
\caption{Summary of X-ray, IR, and spectroscopic measurements for the AGN sample.  The X-ray class is as defined in \citet{szo04}, based on total 0.5 - 10 keV X-ray luminosity and hardness ratio.  The total 0.5 - 10 keV X-ray luminosities presented here, taken from \citet{toz06}, are in units of $10^{42}$ ergs s$^{-1}$.  The hardness ratio is defined as HR = (H-S) / (H+R) 
as in \citet{szo04} and \citet{gia02}, where H and S
are the net instrument count rates in the hard (2-10 keV) and the soft (0.5-2 keV) 
bands, respectively.  The neutral hydrogen column densities (in units of $10^{21}$ cm$^{-2}$) are also from \citet{toz06} and are derived from fits to X-ray spectra.  The X-ray to optical flux ratio (denoted X/O) is as defined originally in \citet{mac88} and modified in \citet{szo04}, with a correction for internal absorption in the Type II sources.  The 24 micron flux (denoted $f_{nu}(24)$) utilizes 24 micron fluxes published in catalog form on the Spitzer GOODS webpage (http://data.spitzer.caltech.edu/popular/goods/, Dickinson et al., in preparation).  The 24 micron-to-8 micron flux ratio (denoted as 24/8), defined as log($\nu f_{\nu}(24) / \nu f_{\nu}(8)$) in \citet{bra06}, additionally utilizes 8 micron fluxes extracted from the publically available images at the same website.  Logarithmic [OIII] line luminosities are in units of the bolometric solar luminosity ($4.0\times10^{33}$ ergs s$^{-1}$).  \label{tbl-3}}
\begin{tabular}[t]{llllllll}
\tableline\tableline
\multicolumn{1}{l}{XID \#} &\multicolumn{1}{l}{$L_{X,0.5-10}$}& \multicolumn{1}{l}{HR} & \multicolumn{1}{l}{$N_H$ } & \multicolumn{1}{l}{X/O}& \multicolumn{1}{l}{$f_{\nu}$(24)} & \multicolumn{1}{l}{24/8}  &
\multicolumn{1}{l}{log $L_{OIII}$} \\
\multicolumn{1}{l}{ } &\multicolumn{1}{l}{($\times10^{42}$ ergs s$^{-1}$)} &\multicolumn{1}{l}{} & \multicolumn{1}{l}{($\times10^{21}$ cm$^{-2}$)} & \multicolumn{1}{l}{(dex)} & \multicolumn{1}{l}{($\mu$Jy)} & \multicolumn{1}{l}{(dex)}  &
\multicolumn{1}{l}{} \\

\tableline
83 & 44.3 & -0.23 & $<$ 2.0 & -0.1 & 336 $\pm$ 7 &  -0.18 $\pm$ 0.12 & 7.9\tablenotemark{b}  $\pm$ 0.4  \\ 
155 &  3.07 & 0.16 & 35.9 & -0.6 & 522 $\pm$ 7 & 0.30 $\pm$ 0.08 & 8.0\tablenotemark{a}  \\
266 & 35 & 1.00 & 887 & -0.7 & 333 $\pm$ 6 & 0.30 $\pm$ 0.10 & 7.8\tablenotemark{a}  \\
56 & 26.2 & 0.11  & 16.2 & -0.1& 774 $\pm$ 9  & 0.37 $\pm$ 0.08 &  8.5\tablenotemark{a}   \\
536 & 0.3 & -0.25 & $<$ 3.0 & -2.3 & $<$ 90\tablenotemark{d} & -0.40 $\pm$ 0.15 & 6.3\tablenotemark{b} $\pm$ 0.5  \\
594 & 2.44 & -1.00 & $<$ 1.0 & -1.1 & $<$ 77\tablenotemark{d} & -0.22 $\pm$ 0.18 & 7.0\tablenotemark{b} $\pm$ 0.5,$\;< 6.8$\tablenotemark{c}  \\
15 & 63 & -0.45 & 2.6 & 0.2 & 140 $\pm$ 4  &  \ldots & 8.0\tablenotemark{b} $\pm$ 0.5 \\
32 & 11.1 & -0.49 & 1.3 &  -0.1 & 159 $\pm$ 4  & 0.15 & 7.3\tablenotemark{c} $\pm$ 0.2 \\
		
\tableline
\end{tabular}
\tablenotetext{a}{Computed from OIII line intensities in \citet{net06}.}
\tablenotetext{b}{$L_{OIII}$ estimated from X-ray luminosity, using fit for Type I AGN shown in figure 2 of \citet{net06}, with appreciable scatter} 
\tablenotetext{c}{$L_{OIII}$ estimated from spectra in \citet{szo04}}
\tablenotetext{d}{$f_{\nu}$(24) measured directly from MIPS images at the Spitzer GOODS website above}
\end{center}
\end{table}

 \begin{table}
\begin{center}
\caption{Summary of stellar populations analysis for AGN and comparison samples.  The dominant stellar populations are listed for two locations:  The inner region ($r < 0.4''$) and the outer region ($r > 0.7''$).  ``Young'' populations refer to those with an age less than 100 Myr, ``intermediate'' age populations have an age between 100 Myr and 1 Gyr, and ``old'' populations are older than 1 Gyr.  \label{tbl-4}}
\begin{tabular}[t]{lllll}
\tableline\tableline
\multicolumn{1}{l}{ID} & \multicolumn{1}{l}{X-ray class} & \multicolumn{1}{l}{Morphology} & \multicolumn{1}{l}{Inner Population, $r < 0.4''$} & \multicolumn{1}{l}{Outer Population, $r > 0.7''$}  \\
\tableline
56 & AGN II   & Irr & young & intermediate\\ 
155 &  AGN II  & Disk & young & old \\
266 & AGN II  & Sc & old & intermediate \\
83 & AGN I  & E & intermediate & old \\
536 & weak AGN I & E merger & old & old \\
594 & weak AGN I & E & intermediate & old \\
15 & BLAGN I\tablenotemark{a} & Disk & intermediate & intermediate \\
32 & weak BLAGN I\tablenotemark{a} & E & old & old \\
Norm1 & \ldots & E & intermediate & old \\		
Norm2 & \ldots & S0/a & young & old \\	
Norm3 & \ldots & Sa & intermediate & intermediate\\	
Norm4 & \ldots & E & old & old \\	
Norm5 & \ldots & Sc & young & old \\	
Norm6 & \ldots & Irr & old & equal mix \\	
Norm7 & \ldots & E & old & old \\	
Norm8 & \ldots & SBb & equal mix & old \\	
Norm9 & \ldots & Sd  & young & old \\

\tableline
\end{tabular}
\tablenotetext{a}{Broad line AGN from optical spectroscopy from \citet{szo04}.}
\end{center}
\end{table} 

\end{document}